\documentclass[fleqn,usenatbib]{mnras}

\usepackage{newtxtext,newtxmath}
\usepackage[T1]{fontenc}

\DeclareRobustCommand{\VAN}[3]{#2}
\let\VANthebibliography\thebibliography
\def\thebibliography{\DeclareRobustCommand{\VAN}[3]{##3}\VANthebibliography}

\usepackage{xcolor}

\usepackage{graphicx}	% Including figure files
\usepackage{amsmath}	% Advanced maths commands

\newcommand{\textbfake}{}
\newcommand{\omp}{\overline{\Omega}_{\rm env}}
\newcommand{\omg}{\overline{\Omega}_{\rm core}}
\newcommand{\mesa}{\fontfamily{qcr}{\selectfont{MESA}}\rm}

\title[Models of rotating sdB stars]{Asteroseismic rotation rates of hot subdwarf B stars hint at transient accretion from leftover common envelope matter}

\author[F.D. Moyano et al.]{Facundo D. Moyano$^{1,}$\thanks{e-mail: moyanofacu@hotmail.com},
Hongwei Ge$^{1}$,
Zhanwen Han$^{1}$,
Beatriz Bordad\'agua$^{2,3}$,
Murat Uzundag$^{4}$,
\newauthor{Philipp Podsiadlowski$^{2,5,6}$,
  Veronika Schaffenroth$^{7}$,
  Xuefei Chen$^{1}$,
  Zhengwei Liu$^{1}$
}
\\
$^{1}$ Yunnan Observatories, Chinese Academy of Sciences, Kunming 650216, China\\
$^{2}$ Heidelberger Institut für Theoretische Studien (HITS), Schloss-Wolfsbrunnenweg 35, 69118 Heidelberg, Germany\\
$^{3}$ Center for Astronomy (ZAH/LSW), Heidelberg University, Königstuhl 12, 69117 Heidelberg, Germany \\
$^{4}$ Institute of Astronomy, KU Leuven, Celestijnenlaan 200D, 3001, Leuven, Belgium \\
$^{5}$ London Centre for Stellar Astrophysics, Vauxhall, London, UK \\
$^{6}$ University of Oxford, St Edmund Hall, Oxford, OX1 4AR, UK \\
$^{7}$Thüringer Landessternwarte Tautenburg, Sternwarte 5, D-07778 Tautenburg, 
Germany
}

\date{Accepted 2026 April 13. Received 2026 April 13; in original form 2026 January 13.}

\pubyear{2026}

\begin{document}
\label{firstpage}
\pagerange{\pageref{firstpage}--\pageref{lastpage}}
\maketitle

\begin{abstract}
  Asteroseismology enabled measuring the rotation rate in the deep stellar interiors of stars across several evolutionary phases, advancing the theory of angular momentum transport in single stars from the main sequence to the white dwarf phase.
  However, binary stellar evolution products have not yet been studied in the context of angular momentum transport constrained by asteroseismology.
  Hot subdwarf B (sdB) stars can pulsate in non-radial modes, enabling probing of their internal rotation.
  Those in binary systems form through mass transfer, thus they can be used to probe theories of internal rotation in post-mass transfer stars.
  Here, we interpret observed asteroseismic core and envelope rotation rates of sdB stars in unsynchronised binary systems that formed through the common-envelope channel, using stellar evolution models of rotating sdB stars with internal magnetic fields.
  We find that when sdB stars form with the angular momentum content of red giant cores prior to common-envelope ejection, their predicted core rotation rates are two to ten times lower than measured asteroseismic rotation rates, and their envelope rotation rates are lower by two to five orders of magnitude.
  This suggests that the angular momentum content of sdB stars increases during their formation.
Since sdB stars in close binary systems may host circumstellar matter from a past common-envelope ejection, we show that if they accrete a small amount of matter, the combination of internal magnetic fields with angular momentum transfer through accretion spins up both the core and envelope to match their measured asteroseismic rotation rates.
\end{abstract}

\begin{keywords}
asteroseismology --
stars: rotation --
stars: binaries
stars: evolution --
stars: hot subdwarfs --
methods: numerical
\end{keywords}

%%%%%%%%%%%%%%%%%%%%%%%%%%%%%%%%%%%%%%%%%%%%%%%%%%

%%%%%%%%%%%%%%%%% BODY OF PAPER %%%%%%%%%%%%%%%%%%

   \section{Introduction}
   \label{intro}
   Asteroseismology is the discipline that probes the interior of stars through their oscillations.
   It has enabled measurements of internal rotation across multiple evolutionary phases, from the main sequence to the white dwarf stage.
   This can be done through various methods, such as identifying rotational splittings of non-radial gravity and pressure modes (g- and p-modes, respectively), the influence of rotation on the period spacing of g-modes, among others \citep[e.g.][]{Aerts2019,Aerts2021,Aerts2025}.
   Notably, in low- and intermediate-mass stars the rotation rate of the stellar cores was measured in thousands of stars such as main sequence stars \citep{Ouazzani2019,Li2020,Pedersen2022,Aerts2025}, subgiants \citep{Deheuvels2014,Deheuvels2020}, hydrogen-shell-burning red giants \citep{Mosser2012,Gehan2018,Li2024,Dhanpal2025}, core-helium burning red giants \citep{Mosser2012,Mosser2024}, and white dwarfs \citep[e.g.][]{Corsico2011,Kawaler2015,Hermes2017,Calcaferro2023,OliveiradaRosa2024,Romero2022,Romero2025}.
   All these measurements consistently showed that the transport of angular momentum (AM) in stellar interiors has to be highly efficient.
   This is contrary to the results obtained with standard theories of AM transport based on hydrodynamical processes \citep{Eggenberger2012,Marques2013}, as well as those obtained with different theories of internal magnetism or wave-induced transport \citep{Fuller2014, Cantiello2014, Takahashi2021, Moyano2023a,Bordadagua2025}.
   
   A mechanism efficient enough to partially address this issue is the generation of internal magnetic fields driven by a dynamo operating in radiative regions aided by the Tayler instability on azimuthal magnetic fields \citep{Tayler1973}, proposed originally by \citet{Spruit2002} and revised by \citet[][TSF dynamo hereafter]{Fuller2019}.
   In particular, this process can explain relatively well the core rotation rates of low-mass stars through their evolution on the lower red giant branch (RGB) to their core-helium burning phase \citep{Fuller2019,Eggenberger2022}.
   However, low-mass stars evolving through the red giant phase can follow a different evolutionary path if they are in binary systems tight enough to allow for mass transfer to a companion.
   If they begin transferring mass to a companion during the RGB they can lose most of their hydrogen-rich envelope mass, and if their cores are massive enough to ignite helium they can settle into the core-helium burning phase as a hot horizontal branch star.
   Stars in this phase are known as hot subdwarf B (sdB) stars \citep{Heber2016,Heber2026}; sdB stars in binary systems can form through stable mass transfer or unstable mass transfer leading to a common envelope ejection event \citep{Han2002,Han2003}.
   
   sdB stars are hot and compact, with effective temperatures in the range $20,000 \rm{K} \lesssim T_{\rm eff} \lesssim 40,000$K, surface gravities of  $5 \lesssim \log g \lesssim 6.2$, and radii of $R \simeq 0.1 - 0.3 R_{\odot}$ \citep{Heber2016,Heber2026}.
   A fraction of these sdB stars can pulsate in radial and non-radial modes through a $\kappa$ mechanism produced by an iron opacity bump due to the combined action of radiative levitation and gravitational settling  \citep{Charpinet1996,Charpinet1997,Charpinet2000,Charpinet2001,Charpinet2002a,Charpinet2002b}.
   Both g- and p-modes can be observed in pulsating sdB stars \citep[e.g.][]{Charpinet2008,Ostensen2010,Holdsworth2017,Uzundag2024}, which can be used to measure their core and envelope rotation rates if they are split by rotation as witnessed in several of them \citep[e.g.][ and references in Table \ref{table_singlesdb}]{Charpinet2018,Reed2021,Silvotti2022,Reed2025}.
   sdB stars are thought to uniquely form through binary interactions, thus they represent a useful probe of internal rotation and AM transport theory in post mass-transfer stars.
   Pulsating helium-core white dwarfs showing rotational splittings \citep[e.g. possibly TIC21187072,][]{Romero2022,Romero2025} or post-common envelope white dwarfs \citep{Hermes2015} are also useful probes of internal rotation in post mass-transfer stars, but their scarcity and additional uncertainties in their evolutionary histories render them less reliable for internal AM transport studies.   
   
   \textbfake{Moreover, space missions such as \textit{Kepler} \citep{Borucki2010} and TESS \citep{Ricker2015} have revolutionised the study of pulsating sdB stars.
     Before these space missions, reliable mode identifications were rare, whereas their uninterrupted, high-precision photometry made such identifications routine \citep[e.g.][]{Reed2011,Baran2012a}.
     Reliable mode identifications in asteroseismology are crucial to accurately probe the internal structure of stars, including their rotation.
     This enables linking observed periodicities to pulsation modes described by quantised spherical harmonics - characterised by the modal degree $\ell$, radial order $n$, and azimuthal order $m$.
     Mode identification typically relies on frequency multiplets caused by rotation (i.e.\ rotational splittings $\delta \nu_{n \ell m}$), or on asymptotic g-mode period sequences, where high-order g-modes of the same spherical degree exhibit nearly uniform period spacings \citep[e.g.][]{Reed2018,Uzundag2021}.
     Furthermore, rotational splittings enable measuring the rotation in both the sdB stars' deep helium-rich radiative regions and in their thin hydrogen-rich envelopes, through their g- and p-modes, respectively.
     This is because, to first order and slow uniform rotation, the rotational splittings are related to the rotation period via the relation $\delta \nu_{n \ell m} = (1 - C_{n \ell})/ P_{\rm rot}$ where $P_{\rm rot}$ is the mean rotation period in the regions probed by the modes and $C_{n \ell}$ is the Ledoux constant.
     For high radial-order p-modes the Ledoux constant tends to zero while for g-modes it tends to $C_{n \ell} \simeq [\ell (\ell + 1)]^{-1}$, thus providing a simple way to measure the rotational velocities in stellar interiors \citep[e.g.][]{Charpinet2018,Aerts2019}.
     These methods, combined with the large amount of data provided by space missions enabled detailed characterisations of sdB stars and measurements of rotation in their interiors.}
   
   However, asteroseismic rotation rates of sdB stars were not tested against stellar evolution models in the context of internal AM transport.
   Only those in tight binary systems were studied in the context of tidal interactions, for which it was found that sdB stars in binary systems with orbital periods below roughly one day are spun up by tidal forces \citep{Ma2024}.
   Furthermore, previous models of rotating sdB stars \citep{Sills2000,Kawaler2005} did not benefit from the recent advances in the theories of stellar rotation nor the large amount of asteroseismic rotation rates measured to date.

   In our work, we computed stellar evolution models of rotating sdB stars that are not affected by tidal synchronisation (i.e. they are not spun up by tides), taking into account the constraints provided by asteroseismic rotation rates in the evolutionary phases previous to the tip of the RGB to compute accurate rotating RGB models of the progenitors of sdB stars.
   Specifically, we model sdB stars in binary systems that may have formed through a common-envelope ejection event near the tip of the RGB \citep{Han2002,Han2003}, for which we computed models calibrated to surface and asteroseismic rotation rates from the main sequence to the RGB.
   We then use these models to construct sdB models with rotation and compare them to  their asteroseismic rotation rates, which we outline in the following sections.

   %==========================================================================================
   \section{Input physics of stellar models}
   \label{methods}
   \subsection{Main sequence to red giants}
   \label{inputs_msrgb}
   %Initial masses and velocities
   We computed stellar evolution models using the open-source software Modules for Experiments in Stellar Astrophysics \citep[\mesa; ][]{Paxton2011,Paxton2013,Paxton2015,Paxton2018,Jermyn2023} version r24.08.1, for which we only mention the most relevant input parameters in our simulations, and we refer the reader to the input files for details\footnote{All of the models computed and presented in this work, along with the files necessary to recompute them are publicly available at \url{https://zenodo.org/records/17332652}.}.
   All of our models are computed from the zero age main sequence (ZAMS) assuming initial rigid rotation, initial masses in the range $M= 0.8 - 2.2 M_{\odot}$ with an step of $0.1 M_{\odot}$, and an initial metallicity of $Z=0.02$.
   \textbfake{We choose initial rotation periods $P_{\rm rot}= 0.3,1,5,10,50$ days for stars in the mass range $M=0.8 - 1.3 M_{\odot}$ and $P_{\rm rot}=0.39, 0.58, 1.16, 2.31$ days for stars in the mass range $M=1.4 - 2.2 M_{\odot}$.
     The latter rotation periods correspond to rotation rates of $\Omega/2\pi = 5,10,20,30\ \mu$Hz, with $\Omega$ the angular velocity.
     The rotation periods for the lower-mass range ($M=0.8 - 1.3 M_{\odot}$) are chosen based on the range of values measured in low-mass stars that develop convective envelopes during the main sequence, and whose rotation periods can be measured by photometric modulation of their light-curves due to surface spots \citep[e.g.][]{Santos2021}.
     Meanwhile, the rotation periods for the higher-mass range ($M=1.4 - 2.2 M_{\odot}$) are chosen based on the range of near-core rotation rates of main sequence g-mode pulsators \citep[e.g.][]{Ouazzani2019,Li2020,Pedersen2022,Aerts2025}, which are often found to rotate nearly-rigidly across the main sequence \citep{VanReeth2018,Li2020,Saio2021,Moyano2023b,Moyano2024}, providing thus a good estimate of their surface rotation periods at the ZAMS.
   The whole range of initial rotation periods corresponds to surface rotational velocities in the range $V_{\rm rot,ZAMS} \simeq 1 - 250$ km/s, depending on the initial mass of the star \footnote{\textbfake{The initial surface rotational velocities can be estimated as $V_{\rm rot} [\rm{km/s}] = 50.6 \left ( \frac{P_{\rm rot}}{\rm {day}} \right )^{-1} \left ( \frac{R}{R_{\odot}} \right ) $}}.}

   %Overshooting
   We include convective overshooting following the exponentially decaying parametrisation of the diffusion coefficient \citep{Freytag1996,Herwig1997}, for which we choose an overshooting strength ($f_{\rm ov}$, the associated free parameter) increasing with stellar mass in the form \citep[e.g.][]{Paxton2011}
   \begin{equation}
     f_{\rm ov}(M)= \frac{f_{\rm ov,0}}{2} \left [ 1 - \cos \left ( \pi \frac{M-M_{\rm min}}{M_{\rm max} - M_{\rm min}} \right ) \right ],
   \end{equation}
   where we choose $f_{\rm ov,0}=0.016$, $M_{\rm min}=1.1 M_{\odot}$, $M_{\rm max}=2.1 M_{\odot}$.
   This parametric form enables a gradual increase of the overshooting parameter in the mass range of models that we present in this work, and is favoured by previous works on detailed modelling of eclipsing binary stars \citep{Claret2016,Claret2017,Claret2018,Claret2019}.
   We limit the extent of the overshooting regions to those where the diffusion coefficient is above $D_{\rm min} = 10^{-10}$ cm$^{2}$/s.
   
   %AM transport
   We include internal AM transport via the Eddington-Sweet circulation and the secular shear instability as hydrodynamical processes \citep[e.g.][]{Heger2000}, and we include internal magnetic fields via the TSF dynamo \citep{Fuller2019}.
   In the framework of the TSF dynamo, the azimuthal magnetic fields are amplified through radial differential rotation until they become unstable to the Tayler instability, which destabilises the azimuthal fields and allows for the amplification of radial magnetic fields \citep{Spruit2002,Fuller2019,Eggenberger2022}.
   To trigger the Tayler instability and saturate according to the TSF dynamo, a minimum value of the azimuthal magnetic field is needed, which is related to the local shear ($q \equiv \partial \log \Omega / \partial \log r$), and whose minimum value is given by
   \begin{equation}
     q_{\rm min} \equiv \left | \frac{\partial \log \Omega}{\partial \log r} \right | _{\rm min} = \alpha^{-3} \left ( \frac{N_{\rm eff}}{\Omega} \right )^{5/2} \left ( \frac{\eta}{r^2 \Omega} \right )^{3/4} ,
     \label{eq_qmin}
     \end{equation}
   where $\alpha$ is a free parameter of order unity, $\Omega$ is the angular velocity, $\eta$ is the magnetic diffusivity, $r$ is the radial coordinate, $N_{\rm eff} = (\eta/K) N^{2}_{\rm T} + N^{2}_{\mu}$ is the effective Brunt-V\"ais\"al\"a frequency as enabled by thermal diffusivity ($K$), and $N^{2}_{T}$ and $N^{2}_{\mu}$ are the thermal and chemical composition components of the Brunt-V\"ais\"al\"a frequency, respectively.
   The associated viscosity which sets the efficiency of AM transport by the TSF dynamo is given by \citep{Fuller2019}
   \begin{equation}
     \nu_{\rm mag} = \alpha^{3} r^{2} \Omega \left ( \frac{\Omega}{N_{\rm eff}} \right )^{2}.
     \label{eq_magvisc}
   \end{equation}
   \textbfake{The magnetic viscosity ($\nu_{\rm mag)}$ enters as an additional diffusion coefficient into the equation of AM transport, given by \citep{Paxton2013}}
   \begin{equation}
     \begin{split}
       \left( \frac{\partial \Omega}{\partial t}\right)_m= & \frac{1}{i}\left( \frac{\partial}{\partial m}\right)_t \left[ (4\pi r^2 \rho)^2 i (D_{\rm hydro}+\nu_{\rm mag}) \left( \frac{\partial \Omega}{\partial m}\right)_{t} \right] \\
       & - \frac{\Omega}{r}\left(\frac{\partial r}{\partial t}\right)_m\left(\frac{\rm{d} \log i}{\rm{d} \log r} \right) , 
     \end{split}
     \label{eq_amt_mesa}
   \end{equation}
   \textbfake{where $i$ is the specific moment of inertia of a shell at mass coordinate $m$, and $D_{\rm hydro}$ is the diffusion coefficient associated to the hydrodynamical processes included in our models, namely the Eddington-Sweet circulation and the secular shear instability; the rest of the variables have their usual meaning.
   The net effect of introducing $\nu_{\rm mag}$ in Eq. \ref{eq_amt_mesa} is to decrease the gradients of angular velocity in radiative regions, leading to rigid rotation if the AM transport efficiency is high enough since the first term of this equation accounts for diffusion of AM while the second one accounts for contraction or expansion.}
   \textbfake{Equations \ref{eq_qmin} \& \ref{eq_magvisc} determine the efficiency of AM transport by internal magnetic fields} and depend  sensitively on the value of $\alpha$, which we calibrate to $\alpha=1.5$ by requiring that a $1.3 M_{\odot}$ model reaches a mean core rotation rate of $\Omega/2\pi \simeq 650$ nHz during the lower part of the RGB, which correspond to the mean mass and core rotation rates in the largest sample of RGB stars analysed to date \citep{Li2024}.
   
   %Rotational mixing
   Although our work is not focused on reproducing any signature of chemical composition in sdB stars, chemical mixing induced by rotation in main sequence stars could affect the mass and size of their convective cores, as well as their chemical stratification which in turn could change the AM budget of the red giants' cores which are used to construct the sdB models.
   Therefore, to account for rotational mixing we employ the Eddington-Sweet circulation, and the secular shear instability \citep[e.g.][]{Heger2000} as in the case of AM transport.
   Currently these prescriptions allow for a free parameter to vary the efficiency of transport of chemical elements with respect to AM, denoted by $f_{\rm c} \equiv D_{\rm chem}/D_{\Omega}$ with $D_{\rm chem}$ and $D_{\Omega}$ the diffusion coefficients for the transport of chemical elements and AM, respectively.
   We set this parameter to $f_{\rm c} = 0.017$ since it was shown to provide a good fit to the surface boron depletion in B type stars \citep{Jin2024} and is therefore a calibrated value of the rotational mixing efficiency.
   Another free parameter regarding the rotational mixing is the parameter $f_{\mu}$ which is a multiplicative factor that changes the chemical composition gradient in the computation of the diffusion coefficients of transport of chemical elements.
   We choose $f_{\mu} = 0.1$ for this parameter following  the work of \citet{Jin2024} as well.

   %AM loss: Magnetic braking
   Magnetic braking due to magnetised winds during the main sequence can change the AM content of the stars and thus affect the rotation rate of our red giants' models, which are then needed to construct the sdB models.
   Therefore, we include surface magnetic braking for stars with masses low enough to develop convective envelopes during the main sequence ($M \le 1.3 M_{\odot}$ in our models); although subgiants with masses just above the Kraft break mass ($\simeq 1.3 M_{\odot}$) can potentially experience magnetic braking \citep{vanSaders2013} we do not explore its effect in this work.
   We specifically include the torque produced by the magnetised winds following the prescription of \citet{Matt2015} as
   \begin{equation}
     \frac{dJ}{dt} = - T_{0} \left ( \frac{\tau_{\rm c}}{\tau_{\rm c, \odot}} \right ) ^{p} \left ( \frac{\Omega}{\Omega_{\odot}} \right ) ^{p+1} ,
   \end{equation}
   in the unsaturated regime where the Rossby number is subcritical, and
   \begin{equation}
     \frac{dJ}{dt}= - T_{0} \chi^{p} \left ( \frac{\Omega}{\Omega_{\odot}} \right ) ,
   \end{equation}
   in the saturated regime where the Rossby number is critical \citep{Matt2015}.
   In both cases $J$ is the AM, $\tau_{\rm c}$ is the convective turnover timescale, while $T_{\rm 0}$ is given by
   \begin{equation}
     T_{0} = K \left ( \frac{R}{R_{\odot}} \right )^{3.1} \left ( \frac{M}{M_{\odot}} \right ) ^ {0.5} \gamma^{-2m} ,
   \end{equation}
   where $\gamma = \sqrt{1+ (u/0.072)^{2}}$ with $u \equiv V_{\rm rot}/V_{\rm rot,crit}$ the ratio of surface rotational velocity to its critical value and the rest of the variables have their usual meaning.
   This formulation has four free parameters: $K,m,p,\chi$, which can change the efficiency of the magnetic torques and thus the resulting rotation rates.
   To calibrate these parameters we follow the work of \citet{Gossage2021,Gossage2023}, who calibrated them to reproduce the surface rotation period of stars in open clusters at different ages.
   We chose, $m=0.22, p=2.6, \chi=14$ \citep[similarly to][]{Gossage2023}, and $K = 3 \times 10^{29}$ which is different from the value calibrated by \citet{Gossage2023}.
   This is due to the different input physics of our models and theirs, nonetheless we note that our re-calibrated value of $K$ leads to a good enough agreement between our models and the surface rotation periods of open clusters as done by \citet{Gossage2023}.
   With this approach, we can give a relatively accurate estimate of the rotation rate of the red giants by the time they reach the RGB tip since no significant AM loss due to magnetic braking is expected after the end of the main sequence, although AM loss due to mass-loss can still be significant in upper RGB stars and thus leads to an additional uncertainty.
   These in turn can lead to a moderately accurate estimate of the rotation rates expected in sdB stars.
   
   \textbfake{These rotational constraints enable us calibrating the internal distribution and global content of AM in red giants' models, which we then used to construct  sdB models.
     Nevertheless, we emphasise that currently asteroseismology cannot probe the rotation rate of the core in RGB tip stars, because in upper RGB stars gravity-dominated mixed modes have small amplitudes and thus the core properties cannot be measured \citep[e.g.][]{Grosjean2014}.
     All asteroseismic measurements of both core and envelope rotation rates of post-main sequence stars from the main sequence to the RGB tip are currently only possible for stars up until the RGB bump \citep[e.g.][]{Mosser2012,Gehan2018,Deheuvels2020,Li2024,Dhanpal2025}.
     Therefore, in this work we assume that no significant AM is lost and that there are no significant changes in the redistribution of the internal AM after the RGB bump.
     Although comparisons of models with measurements of core rotation rates in low-mass single core-helium burning stars suggest that the TSF dynamo gives a relatively accurate core rotation rate near the RGB tip \citep[e.g.][]{Fuller2019,Eggenberger2022}.
     This is because the core rotation rates of low-mass core-helium burning stars (i.e.\ those that experience a helium flash at the RGB tip) depend directly on the core rotation at the RGB tip since during the helium flash the timescales are short enough such that local conservation of AM dictates the evolution of the core rotation rates and thus its initial rate during the core-helium burning phase.}
   
   %------------------------------------------------------------------------------------------

               %
   %
   \begin{figure} \includegraphics[width=\columnwidth]{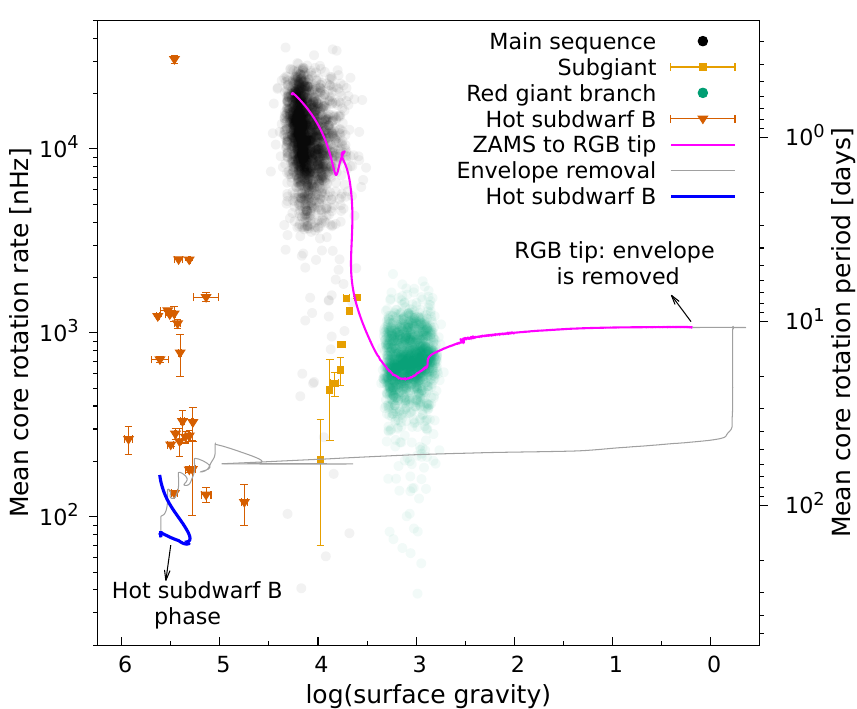}
     \caption{Mean core rotation rate (or equivalently period on the right axis) as a function of surface gravity.
       The data points are asteroseismic measurements of mean core rotation rates taken from the literature \citep[][and references in Tables \ref{table_singlesdb} and \ref{table_binarysdb}]{Deheuvels2014,Deheuvels2020,Li2020,Li2024,Aerts2025} of stars in different phases as indicated in the legend.
       The line is a stellar evolution model computed from the ZAMS ($\log g \simeq 4.3$) to the sdB phase.
     We indicate the location of the RGB tip where we remove the hydrogen-rich envelope to construct our sdB models, then the thinner part of the line shows the evolution during the envelope removal, and the sdB phase is shown with thicker lines as indicated in the figure.
     The initial mass of this model is $1.5 M_{\odot}$ and has an initial rotation rate of $\Omega /2 \pi= 20 \mu$Hz, equivalent to an initial rotational velocity of $V_{\rm rot,ZAMS} = 135$ km/s at the ZAMS.}
     \label{omegac_gsurf_zamstosdb}
   \end{figure}

   \subsection{Input physics: hot subdwarf B models}
   \label{inputs_sdb}
      %Envelope removal
   To construct our rotating sdB models, we use our models presented in Sect. \ref{inputs_msrgb} at the RGB tip as a starting point, where we define the RGB tip as the time the models begin burning helium with an energy generation rate of $\epsilon_{\rm nuc} = 10$ erg/(g s).
   We then use an arbitrarily high mass-loss rate ($\dot{M} \simeq 10^{-6} - 10^{-3} M_{\odot}$/yr) to remove most of the hydrogen-rich envelope before the central helium abundance changes due to core-helium burning.
   In our RGB tip models that start burning helium in degenerate conditions ($M_{\rm ZAMS} \le 1.8 M_{\odot}$), this is achieved by simply using high-enough mass-loss rates such that most of the envelope is removed before the helium flash occurs.
   Whereas for our RGB tip models that can ignite helium in partial- or non-degenerate conditions ($M_{\rm ZAMS} > 1.8 M_{\odot}$) we artificially stop the change in chemical composition due to nuclear reactions to prevent any change in the central abundance of helium by the time the RGB model contracts towards the sdB regime.
   This way we are essentially simulating a common-envelope ejection event at the tip of the RGB, as is expected for most sdB stars in short period binary systems \citep{Han2002,Han2003}.
   We then switch off the mass-loss rate once the hydrogen-rich envelope mass reaches values of $M_{\rm env}/M_{\odot} = 5 \times 10^{-5}, 10^{-4}, 5 \times 10^{-4}, 10^{-3}, 5 \times 10^{-3}$, \textbfake{where we define the hydrogen-rich envelope mass $M_{\rm env}$ as the total mass of hydrogen in the whole interior.
     This definition is valid for our sdB models since the hydrogen is uniquely present in their outer regions (see Fig. \ref{propdiag_chemprof}), thus forming a hydrogen-rich envelope.
     We do not include any mass-loss by winds nor any other process during the sdB phase, which leads to an almost constant $M_{\rm env}$ in sdB models with progenitor masses $M_{\rm ZAMS} \le 1.9 M_{\odot}$.
     Above this mass, sdB models can burn some hydrogen and thus decrease their $M_{\rm env}$.}

   During the envelope removal process, our rotating models not only lose mass but also AM.
   To treat the AM loss due to mass-loss, we remove the AM contained only in the mass lost in each timestep.
   The AM loss ($\Delta J$) at each timestep is then given by
   \begin{equation}
     \Delta J= \int_{M-\Delta M}^{M} j(m) dm \ ,
   \end{equation}
   where $m$ is the mass coordinate, $j(m)$ is the specific AM contained in each layer, and $\Delta M$ is the mass lost in each timestep.

   Additionally, in our sdB models we include overshooting from the boundary of the helium-burning convective core following the exponential scheme \citep{Freytag1996,Herwig1997} and adopting $f_{\rm ov}=0.04$ for all of them.
   Although, this overshoot strength is larger than that of our main sequence models, we choose this value because it reproduces the mean period spacing of dipole g-modes in our observational sample of sdB stars (see Sect. \ref{sect_deltapi1}).
   Nevertheless, its effect on the mean core rotation rates of the sdB models is negligible.
   Rotational mixing and AM transport by hydrodynamical processes and internal magnetic fields are treated in the same way as for our models computed from the ZAMS to the red giant phase (Sect. \ref{inputs_msrgb}).
   
     The resulting sdB models have surface gravities in the range $\log g \simeq 4.8 - 6.1$, total masses in the range $M_{\rm sdB} \simeq 0.35 - 0.48 M_{\odot}$, and surface rotation periods in the range $P_{\rm rot} \simeq 10 - 10^{8}$ days, which we present in the next section.

   %==========================================================================================

   \section{Stellar evolution models and asteroseismic indicators}
   \label{models_astero_methods}
   In this section we show only a representative model of the evolution from the ZAMS to the sdB phase, to qualitatively explain their behaviour (Fig. \ref{omegac_gsurf_zamstosdb}).
   \textbfake{This representative model has an initial mass of $M_{\rm ZAMS}=1.5 M_{\odot}$, an initial velocity of $V_{\rm rot,ZAMS}=135$ km/s, and metallicity $Z=0.02$.}
   At the tip of the RGB, we remove the envelope as described in Sect. \ref{inputs_sdb} to construct the rotating sdB models, which we show for a few representative models only until the central helium mass fraction decreases to $Y=10^{-6}$ (Sect. \ref{sect_sdbmodels}).
   After this point our sdB models would evolve towards the white dwarf phase producing CO white dwarfs of $M \lesssim 0.5 M_{\odot}$, however we limit this work to the sdB phase only.
   The whole grid of models is publicly available for the reader to inspect at Zenodo\footnote{\url{https://zenodo.org/records/17332652}.}.

%------------------------------------------------------------------------------------------   
   \subsection{Evolution from the ZAMS to the RGB tip}
   In Fig. \ref{omegac_gsurf_zamstosdb} we show the evolution of the mean core rotation rate ($\omg/2\pi$) as a function of the surface gravity of a representative model of our grid.
   The data points are measurements of asteroseismic core rotation rates of main sequence Gamma Doradus stars \citep{Li2020,Aerts2025}, subgiants \citep{Deheuvels2014,Deheuvels2020}, RGB stars \citep{Li2024}, and sdB stars (see Tables \ref{table_singlesdb} and \ref{table_binarysdb}).
   We compute $\omg$ in our models as an average value of the rotation rate in the g-mode cavity over the time the g-modes spend in the cavity in a local asymptotic analysis as \citep{Goupil2013}
   \begin{equation}
     \frac{\omg}{2\pi} = \frac{1}{2\pi} \frac{\int_{g_{1}}^{g_{2}} \Omega N \frac{dr}{r}}{\int_{g_{1}}^{g_{2}} N \frac{dr}{r}} ,
     \label{eq_omg}
     \end{equation}
   where $N$ is the Brunt-V\"ais\"al\"a frequency, and $g_{1}, g_{2}$ are the radial coordinates of the boundaries of the g-mode cavity.
   We take the g-mode cavity as the regions where $\nu_{\rm g} < N,S_{1} $, where $\nu_{\rm g}$ is the pulsation frequency and $S_{1}$ is the Lamb frequency of dipole modes.
   In our models until the RGB tip we choose $\nu_{\rm g}=\nu_{\rm max}$, that is, the maximum oscillation frequency as given by scaling relations \citep{Kjeldsen2011}.

   The evolution of the model in Fig.\ \ref{omegac_gsurf_zamstosdb} begins at the ZAMS ($\log g \simeq 4.3$) and proceeds as a single star until it reaches the RGB tip ($\log g \simeq 0.25$).
   During this whole phase the evolution of the core rotation rate is dominated by the AM transport driven by the internal magnetic fields generated by the TSF dynamo.
   During the main sequence phase, the star rotates rigidly.
   However, during the subgiant phase, the core begins to decouple from the envelope as it contracts.
   Despite the rapid core contraction, the AM transport enforced by the TSF dynamo is efficient enough to extract AM from the core and deposit it into the expanding envelope, thereby slowing the core down.
   Therefore, $\omg/2\pi$ decreases by roughly 1 dex at the end of the main sequence, located at $\log g \simeq 3.75$ where the loop characteristic of stars with convective cores in HR diagrams is usually seen.
   After the core spins down, the star expands towards lower surface gravities while its core contracts.
   At this stage ($\log g \simeq 3.5$) the core can spin up slightly because the barrier of chemical composition around the hydrogen-burning shell decreases the efficiency of the TSF dynamo, enabling the helium core to partially retain its AM.
   This evolutionary scenario is similar for stars at different initial velocities in the mass range that we study in this work, namely $M = 0.8 - 2.2 M_{\odot}$ \citep{Fuller2019,Eggenberger2022}.
   Once the models reach the RGB tip we stop the computation and proceed to remove the envelope as explained in Sect. \ref{inputs_sdb}.

   \begin{figure} \includegraphics[width=\columnwidth]{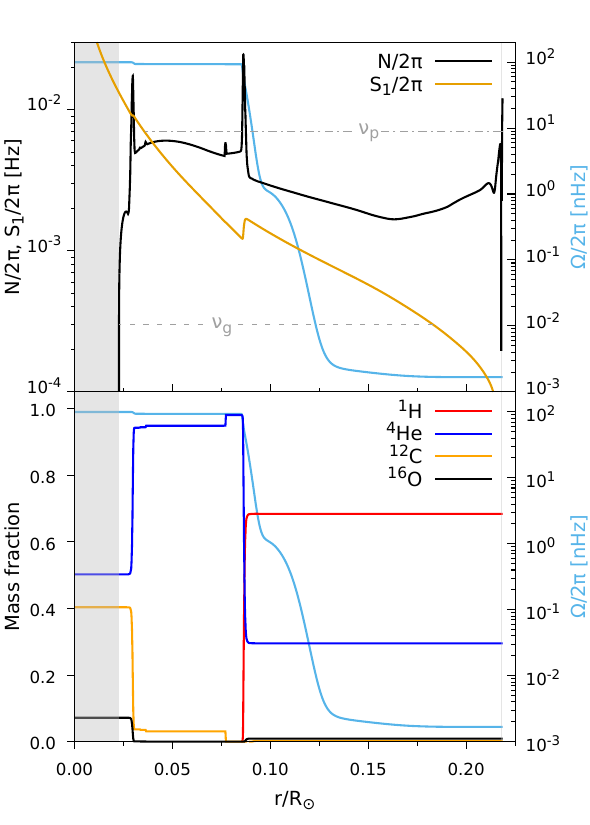}
     \caption{Propagation diagram (top) and chemical composition profile (bottom) of a sdB model at the middle of the core-helium burning phase ($Y=0.5$).
       The propagation diagram shows the Brunt-V\"ais\"al\"a and Lamb (for $\ell =1 $) frequencies, along with the typical g- and p-mode frequencies that we adopted in this work, $\nu_{\rm g}$ (dashed) and $\nu_{\rm p}$ (dashed-dot), respectively.
       The chemical composition profile at the bottom shows the mass fraction of hydrogen, helium, carbon, and oxygen, as indicated in the legend.
       In both panels, the grey shaded regions are convective and the rotation profile is also shown, with its values on the right hand-side axis.
     }
     \label{propdiag_chemprof}
   \end{figure}
   %
   %

%------------------------------------------------------------------------------------------
      \subsection{Evolution towards the sdB phase}
   Shortly after the envelope removal begins, the model presented in Fig. \ref{omegac_gsurf_zamstosdb} experiences a helium flash enabling the helium core to expand as it gradually becomes non-degenerate.
   Henceforth, $\omg/2\pi$ decreases by roughly 0.7 dex at $\log g \simeq 0.25$.
   After the helium flash the envelope contracts making the star evolve towards higher surface gravities, eventually experiencing some sub-flashes at $\log g \simeq 4.5 - 5.5$ before the core starts burning helium in non-degenerate conditions which in turn defines the beginning of the sdB phase.
   \textbfake{These and previous different evolutionary phases occur on different timescales, which we illustrate in Fig. \ref{omegac_age} for further clarity.}
   However, we emphasise that the evolution towards the sdB phase that we compute with hydrostatic models is likely not realistic and it is just a numerical way to construct the sdB models.
   Other evolutionary sequences can be obtained depending on the effective temperature and helium-core mass at which the helium flash occurs \citep{Lanz2004,MillerBertolami2008,Arancibia-Rojas2024}.
   Nonetheless, the rotational evolution of the core is likely robust against any changes in the behaviour of the envelope, such as its effective temperature or size.

%------------------------------------------------------------------------------------------   
   \subsection{Mean core rotation rate of sdB models}
      We compute $\omg/2\pi$ in our sdB models in a similar way as for the models evolving from the ZAMS to the RGB tip (i.e. using Eq. \ref{eq_omg}), but we determine the extent of the g-mode cavity by choosing a typical g-mode pulsation frequency in sdB stars of $\nu_{\rm g} = 3 \times 10^{-4}$Hz \citep[e.g.][]{Charpinet2018}.
      In Fig. \ref{propdiag_chemprof} we show a propagation diagram of one of our sdB models in the middle of the core-helium burning phase ($Y=0.5$) which illustrates the location and extension of its pulsation cavities.
      The leftmost peak in Brunt-V\"ais\"al\"a frequency is due to the gradient of chemical composition in the transition region from the carbon-oxygen rich core to the helium rich radiative interior (CO-He transition hereafter), while the second peak from left to right is due to the transition from the helium-rich radiative interior to the hydrogen-rich envelope (He-H transition hereafter).
      The CO-He and He-H transitions are illustrated in the bottom panel of Fig. \ref{propdiag_chemprof} where the chemical composition of hydrogen, helium, carbon and oxygen of the same model are shown.
   In both figures, the chemically homogeneous regions extend farther away from the boundary of the convective core (shown as grey shaded regions) because the models include convective-core overshooting and we assume that the overshoot regions remain non-adiabatic; this explains why the peak of Brunt-V\"ais\"al\"a frequency due to the CO-He transition is displaced from the boundary of the convective core.
   Near the surface of the model, a sharp decrease and a third peak of Brunt-V\"ais\"al\"a frequency appear because of a thin convective zone that arises in the HeII/HeIII partial ionisation zone \citep[e.g.][]{Charpinet2000}.

\begin{figure} \includegraphics[width=\columnwidth]{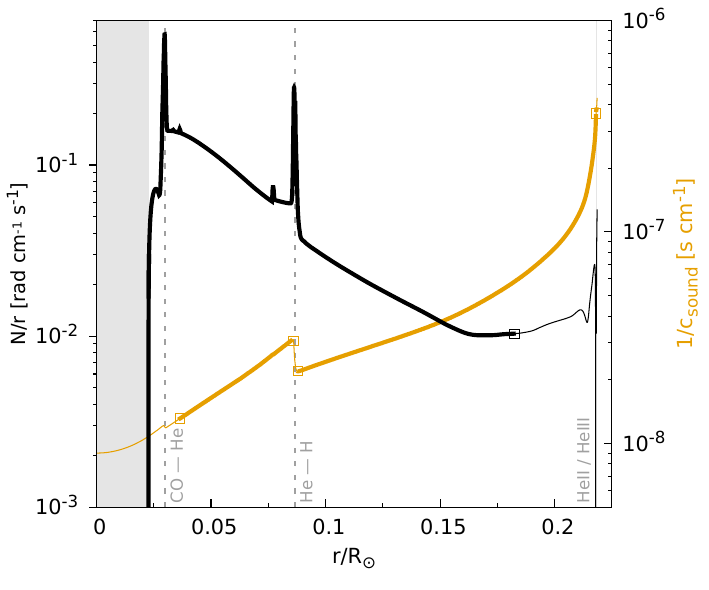}
     \caption{Weight functions used to compute the mean envelope and core rotation rates by using Eqs. \ref{eq_omg} and \ref{eq_omp}.
       The thick part of the lines show the g- and p-mode cavities of the same model shown in Fig. \ref{propdiag_chemprof}.
       \textbfake{The locations of the CO-He and He-H transitions are indicated by the dashed lines, the location of the HeII/HeIII partial ionisation zone near the surface is also indicated.
       For clarity, the turning points of the inner and outer p-mode cavities are highlighted with squared symbols, as well as the outer turning point of the gravity-mode cavity.}
     }
     \label{weight_functions}
      \end{figure}

   In our framework, the g-mode cavity can also extend to the hydrogen-rich envelope, and therefore the value of $\omg/2\pi$ can be affected by the angular velocity in the hydrogen-rich envelope.
   However, we show in Fig.\ \ref{weight_functions} that the weight function used in Eq.\ \ref{eq_omg} to compute $\omg$ has the largest values in the CO-He transition, which corresponds to the peak seen at $r/R_{\odot} \simeq 0.03$.
   Above this region, the contribution to the $\omg/2\pi$ mainly comes from the helium-rich radiative interior with an additional localised contribution at the He-H transition (at $r/R_{\odot} \simeq 0.08$).
   In the hydrogen-rich envelope, the weight function $N/r$ drops below $\simeq 0.05$, and since in these regions the angular velocity decreases by at least 4 orders of magnitude (see Fig. \ref{propdiag_chemprof}), its contribution to the value of $\omg$ is negligible.
   Moreover, in our models the change in rotation rate from the helium-rich radiative regions to the convective core is relatively small ($\lesssim 1 \%$ difference), therefore the core rotation rate computed with Eq. \ref{eq_omg} is a robust estimate of the mean rotation rate in the core of a sdB as sensed by g-modes.
   %
   %

%------------------------------------------------------------------------------------------
   \subsection{Mean envelope rotation rate of sdB models}
   \label{method_omp}
   Rotational splittings of p-modes in sdB stars can also be used to measure the mean rotation rate in the p-mode cavity, where they propagate.
   Since in sdB stars p-modes have larger amplitudes in the hydrogen-rich envelope \citep{Charpinet2000} with little to no contamination from the helium-rich radiative regions, they can be considered to probe the envelope rotation rate of sdB stars.
   Thus, they serve as an additional source of information on the AM evolution of post mass-transfer stars in binary sdB stars, which we present in Sect. \ref{astero_envrot}.
   To compare these measurements to our rotating sdB models we first compute the mean envelope rotation rates ($\omp /2\pi$) of our models as \citep{Goupil2013}
   \begin{equation}
     \frac{\omp}{2\pi} = \frac{1}{2\pi} \frac{\int_{p_{1}}^{p_{2}} \Omega \frac{dr}{c_{\rm s}}}{\int_{p_{1}}^{p_{2}} \frac{dr}{c_{\rm s}}} ,
     \label{eq_omp}
   \end{equation}
   where $p_{1},p_{2}$ are the radial coordinates of the boundaries of the p-mode cavity, defined as the regions where $\nu_{\rm p} > N,S_{1}$.
   To define the boundaries of the p-mode cavity we chose $\nu_{\rm p} = 7 \times 10^{-3}$ Hz which is a typical pulsation frequency of p-modes as observed in sdB stars \citep[e.g.][]{Charpinet2008}.
   This approach is similar to the one we followed to compute $\omg$ in our models by using Eq. \ref{eq_omg}.

   \begin{figure} \includegraphics[width=\columnwidth]{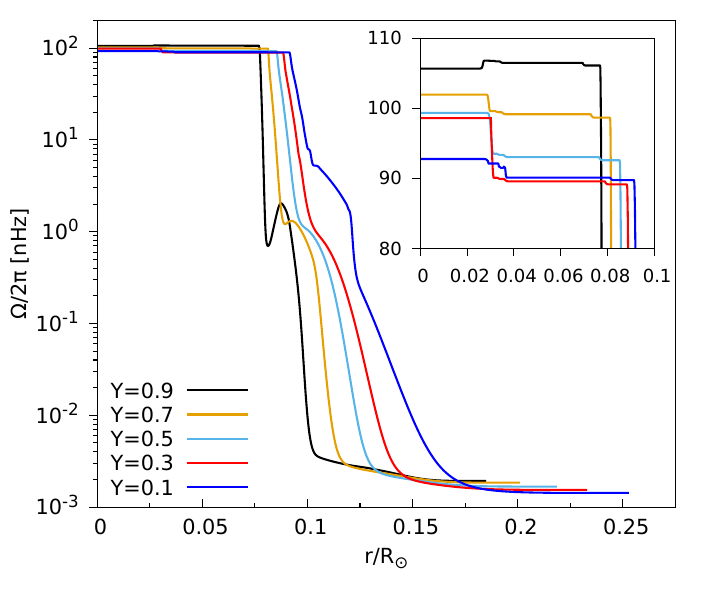}
     \caption{Rotation rate as a function of the radial coordinate of a representative sdB model shown at different ages as given by its central helium mass fraction ($Y$).
       The inset shows a zoomed view into the inner regions below the hydrogen-rich envelope.
     }
     \label{profile_rotation}
   \end{figure}
   \begin{figure*}
     \centering     \includegraphics[width=18cm]{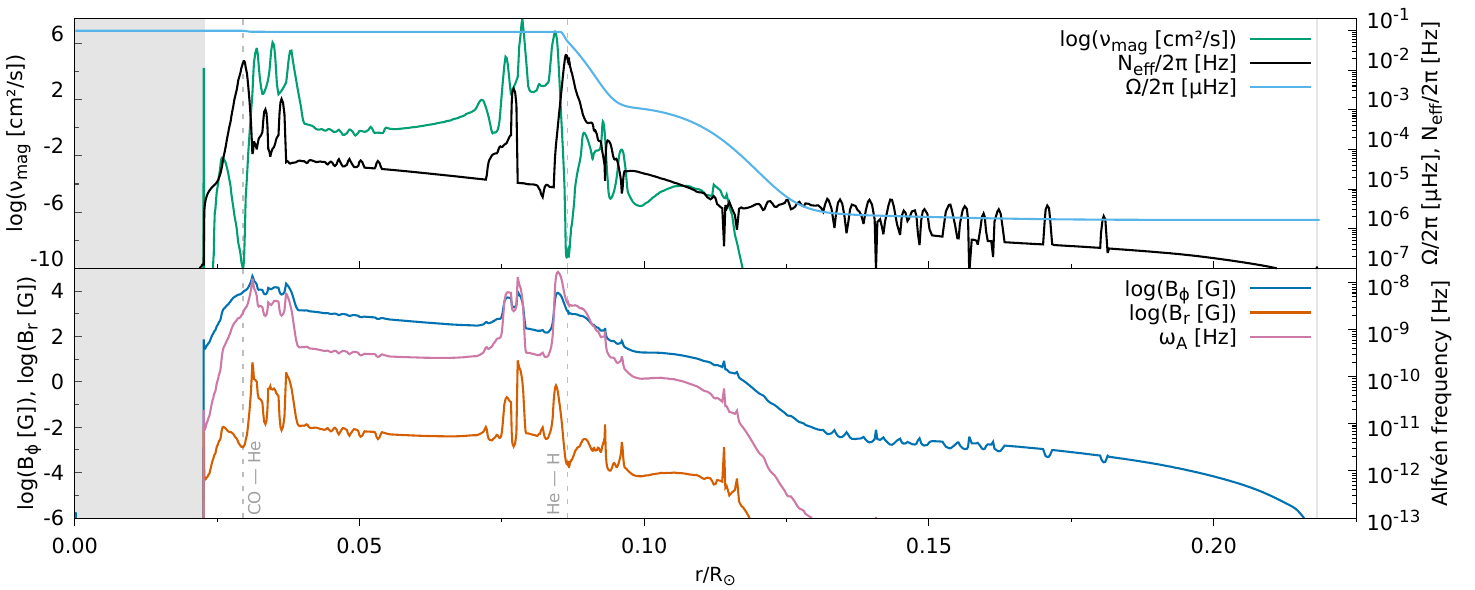}
     \caption{Magnetic viscosity ($\nu_{\rm mag}$), rotation rate ($\Omega /2\pi$), and effective Brunt-V\"ais\"al\"a frequency ($N_{\rm eff}$) are shown in the top panel as a function of the radial coordinate for the same sdB model shown in Fig. \ref{propdiag_chemprof}.
      Azimuthal ($B_{\phi}$), and radial ($B_{\rm r}$) components of the magnetic field, and Alfv\'en frequency ($\omega_{\rm A}$) are shown in the bottom panel.
      The grey shaded regions are convective, \textbfake{while the dashed lines indicate the transitions from the carbon-oxygen core to the helium-rich radiative interior and from the helium-rich radiative interior to the hydrogen-rich envelope.}
     }
     \label{magvisc_bfield}
   \end{figure*}

   In Fig. \ref{propdiag_chemprof} we show the p-mode frequency that we chose compared to the Lamb (for $\ell=1$) and Brunt-V\"ais\"al\"a frequencies in one of our sdB models.
   In our models, there can be two p-mode cavities: one in the helium-rich radiative interior starting just below the He-H transition (located at the peak in Brunt-V\"ais\"al\"a frequency), and one in the hydrogen-rich envelope just above the He-H transition.
   However, since p-modes have larger amplitudes in the hydrogen-rich envelope \citep[e.g. Fig.\ 11 of ][]{Charpinet2000}, we only take the cavity in the hydrogen-rich envelope to compute $\omp$ in our models.
   Although the sharp increase of the rotation rate near the He-H transition can potentially affect $\omp$, we show in Fig. \ref{weight_functions} that according to the weight function used in Eq. \ref{eq_omp} the contribution of the He-H transition is below $\simeq 10 \%$.
   Therefore, the value of $\omp$  of our sdB models is a weighted mean rotation rate of their hydrogen-rich envelope with the near-surface layers contributing the most to it.

%------------------------------------------------------------------------------------------
   \subsection{Angular momentum transport in sdB stars}
   During most of its lifetime, the sdB model expands towards lower surface gravities due to the expansion of its envelope and the inner helium-rich radiative regions, which explains why $\omg/2\pi$ decreases (see Fig. \ref{omegac_gsurf_zamstosdb}).
   Only once the central helium mass fraction drops below $Y \simeq 0.1$, the envelope and its core contract, spinning up while the star reaches higher surface gravities.
   In Fig. \ref{propdiag_chemprof} we show a rotation profile (i.e. the profile of angular velocity as a function of a spatial coordinate) at the middle of the core-helium burning phase of one representative sdB model, where the core rotates roughly five orders of magnitude faster than its surface.
   This kind of rotation profile is characteristic of all sdB models where the envelope is removed along with the AM contained in the mass lost, and no additional processes of AM transport nor external torques are included during the envelope ejection or the ensuing evolution.

   Moreover, in Fig. \ref{profile_rotation} we show the rotation profile of a sdB model at different times during the core-helium burning phase which illustrates the evolution of the rotation rate in the interior of our sdB models.
   The regions below the He-H transition rotate roughly as a solid body, whereas the regions above it (i.e.\ the hydrogen-rich envelope) can develop strong differential rotation, with a change in angular velocities spanning 3 to 5 orders of magnitude.
   This strong differential rotation can develop because the AM transport by the TSF dynamo is inhibited by chemical composition gradients since the turbulent motions of the gas triggered by the Tayler instability are stabilised by buoyancy.
   Additionally, the efficiency of AM transport by the TSF dynamo also decreases in the hydrogen-rich envelope due to its relatively low angular velocity. This increases the timescale required to amplify azimuthal magnetic fields through radial differential rotation. Consequently, the evolution of the rotation rate of the layers above the He-H transition mostly depend on their contraction and expansion.

   In Fig. \ref{magvisc_bfield} we show the magnetic viscosity ($\nu_{\rm mag}$) associated to the TSF dynamo, which sets the efficiency of AM transport.
   The magnetic viscosity in the helium-rich radiative interior can reach values in the range $\nu_{\rm mag} \simeq 10^{2} - 10^{6}$ cm$^2$/s due to its high angular velocity, however, it decreases abruptly to negligible values ($\nu_{\rm mag} \lesssim 10^{-6}$ cm$^2$/s) in the He-H transition due to the strong buoyancy forces exerted by the change in chemical composition around the peak of effective Brunt-V\"ais\"al\"a frequency.
   Above these regions, $\nu_{\rm mag}$ decreases because the rotation rates are much smaller (see Eq. \ref{eq_magvisc}).

   In the bottom panel of Fig. \ref{magvisc_bfield} we show the amplitude of the magnetic fields and the Alfv\'en frequency in our representative model.
   The Alfv\'en frequency indicates where the azimuthal fields are most likely to become unstable and hence increase the amplitude of the radial magnetic fields which explains why their maxima coincide.
   The azimuthal magnetic fields can reach amplitudes of $B_{\phi} \simeq 1- 10$kG in the helium-rich radiative interior, whilst in the regions close to the surface they drop below $10^{-4}$ G.
   This is because the angular velocities near the surface are too small to amplify the azimuthal fields by radial differential rotation.
   The radial component of the magnetic field has lower amplitudes, reaching at most $B_{\rm r} \simeq 1$G in the helium-rich radiative interior.
   These magnetic field amplitudes are typical of our models, in which the radial magnetic fields are always much smaller than the azimuthal fields, as is characteristic of the TSF dynamo.

   \begin{figure} \includegraphics[width=\columnwidth]{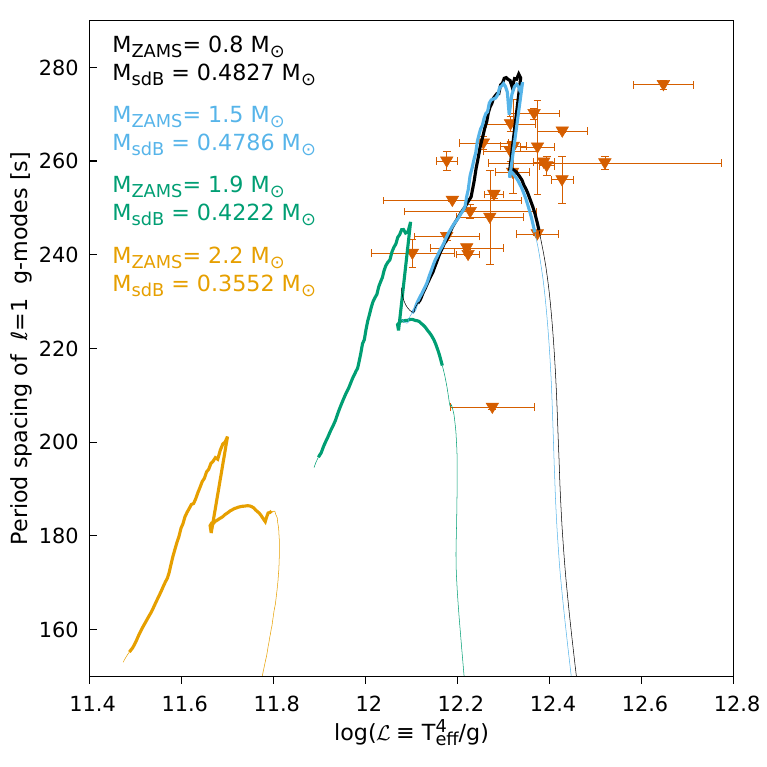}
     \caption{Period spacing of $\ell = 1$ gravity modes as a function of the inverse of the flux-weighted surface gravity ($\mathcal L$), which is a proxy for the luminosity.
       The data points are measurements taken from the literature (see Tables \ref{table_singlesdb} and \ref{table_binarysdb}) while the lines show four of our sdB evolutionary tracks where their progenitor-mass at the ZAMS ($M_{\rm ZAMS}$) and mass during the sdB phase ($M_{\rm sdB}$) are indicated in the figure.
       The evolution begins at the leftmost point of each track, and is only shown when the period spacing is above $150$ seconds.
       The thick part of the tracks highlight the region where the central helium mass fraction is in the range $0.9 > Y > 0.1$.
       The steep decrease in period spacing seen in the models towards the end of the core-helium burning phase (e.g. $\Delta \Pi_{1} \simeq 200$ s in the $M_{\rm ZAMS} = 2.2 M_{\odot}$ model) is due to a breathing pulse.
       The models have a hydrogen-rich envelope mass of $M_{\rm env}= 5 \times 10^{-4} M_{\odot}$ and an initial rotation period of $P_{\rm rot}=10$ days for the $M_{\rm ZAMS}=0.8 M_{\odot}$ model, and an initial rotation rate of $\Omega/2\pi = 20 \mu$Hz (or equivalently $P_{\rm rot} = 0.58$ days) for the rest of the models.
     }
     \label{deltapi1_specl}
      \end{figure}
   %
   %

%==========================================================================================
   \section{Comparison to asteroseismic measurements of sdB stars}
   \label{sect_sdbmodels}

%------------------------------------------------------------------------------------------ 
\subsection{Period spacing of dipole gravity modes}
\label{sect_deltapi1}
Before proceeding with a detailed comparison of our sdB models and their observed asteroseismic rotation rates we first compare their effective temperature and surface gravities as provided by spectroscopic measurements, and the period spacing of their dipole g-modes (see Tables \ref{table_singlesdb} and \ref{table_binarysdb}).
   The period spacing of dipole g-modes of the observational sample is that usually obtained by searching for vertical ridges in the \'echelle diagram, usually referred to as $\Delta P_{1}$ \citep[e.g.][]{Uzundag2021}, whereas in our models we compute it in the asymptotic limit as \citep[][]{Charpinet2008}
   \begin{equation}
     \Delta \Pi_{\ell} = \frac{2 \pi^{2}}{\sqrt{\ell (\ell +1)}} \left ( \int_{g_{1}}^{g_{2}} N \frac{dr}{r} \right ) ^{-1} ,
     \label{eq_deltapi}
   \end{equation}
   where we take $\ell=1$ to account for dipole modes, with $g_{1},g_{2}$ the radial coordinates of the boundaries of the g-mode cavity as in Eq. \ref{eq_omg} now using a typical g-mode pulsation frequency in sdB stars ($\nu_{\rm g} = 3 \times 10^{-4}$Hz) to define them.
   
   In Fig. \ref{deltapi1_specl} we show that the period spacing of our models can only reproduce the data if the initial mass of the sdB's progenitor is below 1.9 $M_{\odot}$.
   This occurs because sdB models with progenitor masses above this value ($2 M_{\odot} \le M_{\rm ZAMS} \le 2.2 M_{\odot}$) ignite helium in non-degenerate conditions, which leads to sdB stars with lower masses \citep[e.g.][]{Arancibia-Rojas2024}.
   This in turn leads to lower period spacings because $\Delta \Pi_{1}$ is proportional to the extension of the helium-burning core, which in turn scales with its mass and hence the mass of the sdB.
   We also compare the effective temperature and the surface gravities of the observational sample to our models via the inverse of the flux-weighted surface gravity, defined as $\mathcal L \equiv T^{4}_{\rm eff}/g $.
   This variable is proportional to the luminosity-to-mass ratio \citep{Langer2014} and highlights whether both spectroscopic effective temperatures and surface gravities can be reproduced by the models.   
   Our models can only reproduce $\mathcal L$ if the mass of the sdB progenitor is below $1.9 M_{\odot}$, which occurs because our sdB models with more massive progenitors ($2 M_{\odot} \le M_{\rm ZAMS} \le 2.2 M_{\odot}$) lead to sdB models with lower masses and hence lower luminosities.
   We therefore only use our sdB models with progenitor masses below $1.9 M_{\odot}$ to compare against the asteroseismic rotation rates because the current observational sample suggest most of them formed from such progenitors.

   \begin{figure}   \includegraphics[width=\columnwidth]{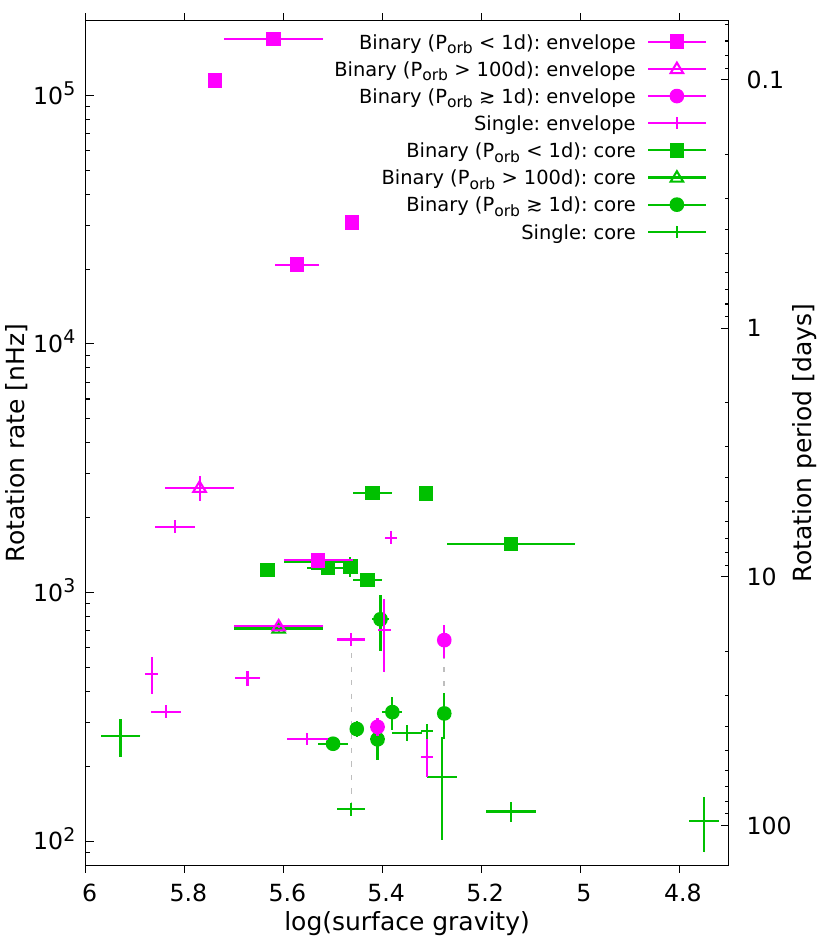}
     \caption{Asteroseismic mean core- and envelope-rotation rates (or equivalently rotation periods on the right-hand side axis) of sdB stars as a function of their surface gravity.
       The rotation rates were taken from the literature (see Tables \ref{table_singlesdb} and \ref{table_binarysdb}), and represent the rotation rates as measured by the splittings of gravity- and pressure-modes.
        The symbol types show whether the stars are in binary systems and whether they experience tidal synchronisation \citep[those with $P_{\rm orb} <1$ day likely experience tidal synchronisation;][] {Ma2024}.
       The colours show either the mean core rotation rates or the mean envelope rotation rates as indicated in the legend.
       The only two stars with significant radial differential rotation have their rotation rates connected by dashed lines.
     }
     \label{omegac_omegas_data}
      \end{figure}
   %
%

%------------------------------------------------------------------------------------------
   \subsection{Asteroseismic core rotation rates}
   Before comparing the asteroseismic rotation rates to our models, we divide the observational sample into different categories depending on whether they are single or in a binary system, and on their orbital periods ($P_{\rm orb}$) if they are in a binary system.
   In Fig. \ref{omegac_omegas_data} we show the whole sample of sdB stars with measurements of asteroseismic rotation rates used in this work (see also Tables \ref{table_singlesdb} \& \ref{table_binarysdb}).
   \textbfake{To compare against our stellar evolution models, we only selected sdB stars from the literature whose either core or envelope rotation rates were measured through rotational splittings of either g- or p-modes, respectively.
     This is because it enables a consistent comparison with other stars for which asteroseismic rotation rates were measured, such as main sequence stars, subgiants, and red giants (see Fig. \ref{omegac_gsurf_zamstosdb}), from both a theoretical and empirical point of view (see also Sect. \ref{models_astero_methods}).
     Moreover, we only select those sdB stars which have at least one clear detection of a rotational splitting with its azimuthal components well detected.
     That is, we do not select those sdB stars that have lower boundaries on their asteroseismic rotation rates based on the non-detection of rotational splittings \citep[e.g.][]{Reed2016}, or where only incomplete multiplets are detected \citep[e.g.][]{Ketzer2017}.
     Although this decreases the size of our observational sample, we ensure that it is not affected by potential biases due to limitations on the data available.
     More complete compilations of asteroseismic data and further interpretations including the above mentioned cases are presented by \citet{Reed2021,Silvotti2022,Reed2025}.}
   
   \textbfake{From the sample presented in Fig.\ \ref{omegac_omegas_data} (or Tables \ref{table_singlesdb} \& \ref{table_binarysdb}), we note that only a few sdB stars have asteroseismic measurements of both their core- and envelope-rotation rates \citep[e.g.][]{Kern2017,Kern2018,Reed2019,Su2024,Reed2025}, with only two of them showing significant core-to-envelope radial differential rotation, namely TIC441725813 \citep{Su2024} and LTCnC \citep{Reed2025}.
     The fact that most of these sdB stars only have measurable core or envelope rotation rates partially occurs because not all sdB stars are observed to pulsate in both g- and p-modes, likely because the expected number and kind of excited modes depends on their surface properties such as effective temperature and surface \citep[e.g.][]{Bloemen2014}.
     Moreover, if the pulsation axis is seen pole on or the rotation period is longer than the length of the photometric data, rotational splittings may not be detectable at all.
     Further limitations on the data such as a low frequency resolution, or the nature of individual cases such as frequency or amplitude modulations \cite[e.g. EPIC211823779;][]{Reed2018}, may affect the detectability of rotational splittings.}
   
   Those sdB stars in binary systems with $P_{\rm orb} < 1$ day experience tidal synchronisation which affects both their core and envelope rotation rates \citep{Ma2024}.
   Since we do not study tidal interactions in this work we exclude these stars from our further comparisons and only focus on those sdB stars where tidal effects can be safely neglected, which corresponds to sdB stars in binary systems with $P_{\rm orb} > 1$ day.
   Nevertheless, most of these sdB stars have orbital periods $P_{\rm orb} < 15$ days, short enough to consider that they formed through the common-envelope channel \textbfake{ \citep{Han2002,Han2003,Nelemans2010,Clausen2012,Ge2022,Ge2024,Rodriguez-Segovia2025a,Rodriguez-Segovia2025b}.}
   
   There are only two sdB stars in binary systems with long orbital periods ($P_{\rm orb} > 100$ days) in our sample, one of which is rotating nearly rigidly (PG1315-123).
   These stars have detected companions but their orbital periods were not yet measured, although a lower boundary was estimated \citep{Reed2019}.
   We do not include these sdB stars in our further comparisons because they have probably formed through stable mass-transfer rather than from a common-envelope ejection, and thus do not fit in our theoretical framework where we simulate a common-envelope ejection at the RGB tip.
   Thus, in our further comparisons to our models we only include sdB stars in binary systems with $P_{\rm orb} \gtrsim 1$ day, and single sdB stars because these could have non-detected companions.

\begin{figure*}
     \centering     \includegraphics[width=18cm]{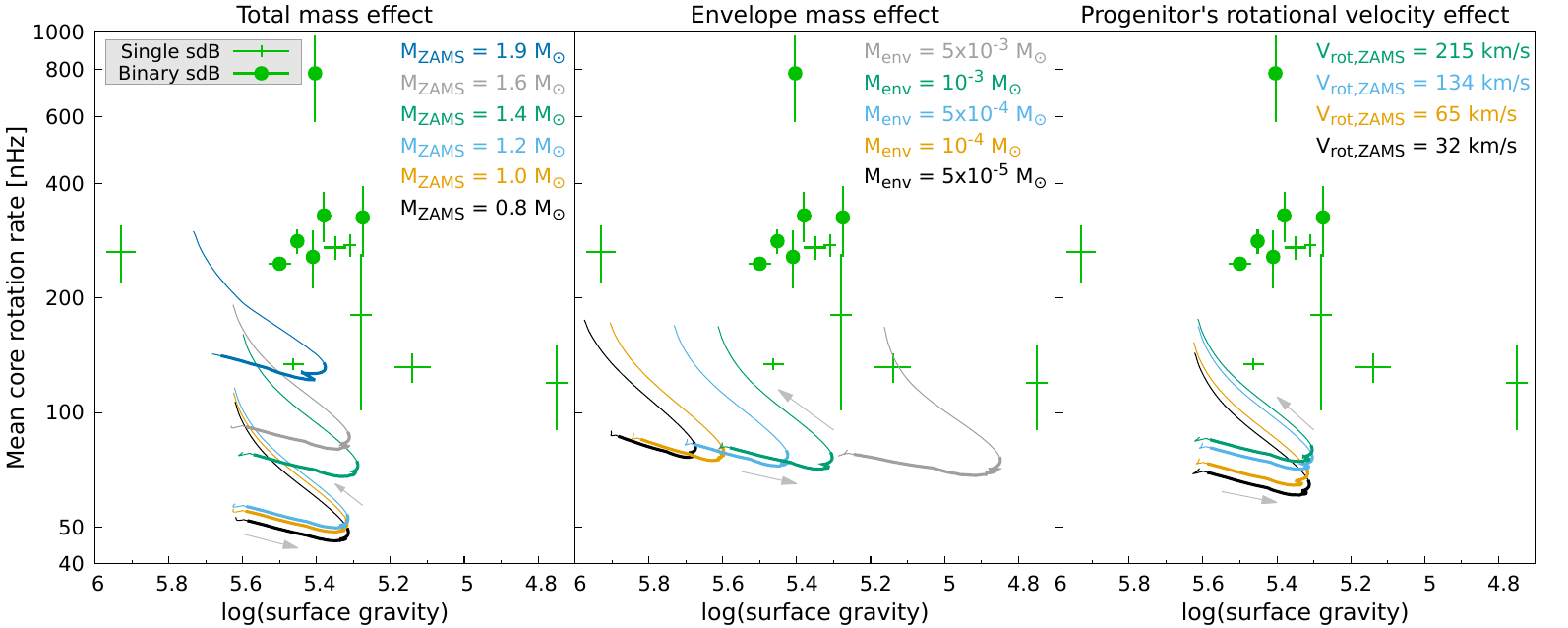}
     \caption{Mean core rotation rate as a function of surface gravity.
       The data points are measurements of mean core rotation rates as given by the splitting of g-modes, taken from the literature (see Tables \ref{table_singlesdb} and \ref{table_binarysdb}).
       Both single (crosses) and binary (circles) sdB stars in unsynchronised short orbital period systems ($P_{\rm orb} \gtrsim 1$ day) are shown.
       The lines are stellar evolution models of sdB stars shown only during the core-helium burning phase, where the thick lines highlight the parts where their central helium mass fraction is in the range $0.9 > Y > 0.1$.
       The evolutionary sense is indicated by gray arrows.
       From left to right, the three panels show the effect of the progenitor's mass at the ZAMS ($M_{\rm ZAMS}$), the effect of the envelope mass of the sdB ($M_{\rm env}$), and the effect of the progenitor's initial rotational velocity at the ZAMS ($V_{\rm rot,ZAMS}$).
       Unless otherwise indicated in the panels, the progenitor's models has an initial mass of $M_{\rm ZAMS}=1.5 M_{\odot}$ and initial rotational velocity of $V_{\rm rot,ZAMS}=134$ km/s at the ZAMS, and an envelope mass of $M_{\rm env}=10^{-3} M_{\odot}$ during the sdB phase.
       In the leftmost panel, starting from the lowest $M_{\rm ZAMS}$ value, the mass of the sdB models are: 0.4767, 0.4751, 0.4749, 0.4744, 0.4655, and 0.4164 $M_{\odot}$.
     }
     \label{omegac_gsurf_sdb}
   \end{figure*}

   After narrowing down the range of models and asteroseismic rotation rates that should be interpreted in our framework, we proceed to compare them and illustrate the effects of the models' parameters on their $\omg$ values in Fig. \ref{omegac_gsurf_sdb}.
   In particular, we note that the mass of the sdB progenitor has the largest impact on $\omg$ in our sdB models, which occurs because the value of $\omg$ at the RGB tip scales with the initial mass of the model.
   Since at the RGB tip the envelope is removed, the AM content of the core determines $\omg$ in the sdB phase.
   The initial mass at the ZAMS also determines the sdB mass, which implies then that $\omg$ scales inversely with the sdB mass.
   The mass of the hydrogen-rich envelope does not have any impact on $\omg$, affecting only the extension of the hydrogen-rich envelope and hence the surface gravity.
   And the initial rotational velocity of the models at the ZAMS has only a minor effect, affecting $\omg$ by a factor two at most.
   This happens because the internal magnetic fields generated by the TSF dynamo lead to a narrow range in $\omg$ for models with different initial velocities by the time they reach the RGB \citep[see Fig. 2 of][]{Eggenberger2022}.

   These are just some models that illustrate the effects of their properties on  $\omg$, although our whole grid of sdB models in the mass range $M_{\rm ZAMS}=0.8 - 1.9 M_{\odot}$ contains 270 evolutionary tracks which we do not fully show here for the sake of clarity.
   However, independently of the changes in the models' properties, the models are always in slight disagreement with the data.
   Only the models of the least massive sdB stars (most massive progenitors) can mildly match the $\omg$ of those sdB stars in binary systems.
   But this is only possible during the late stages of these models, when the central helium abundance drops below $Y=0.01$ in mass fraction (i.e. less than $1\%$).
   In this phase the models are evolving in rather short timescales and should not represent the majority of the observations.
   As for the single sdB stars, they can be well explained by these same models.
   However, given that they could be really single and result from a merger of two white dwarfs \citep[e.g.][]{Schwab2018}, their interpretation in our framework remains questionable.

   %
   %

%------------------------------------------------------------------------------------------
   \subsection{Asteroseismic envelope rotation rates}
   \label{astero_envrot}
\begin{figure*}
    \centering     \includegraphics[width=18cm]{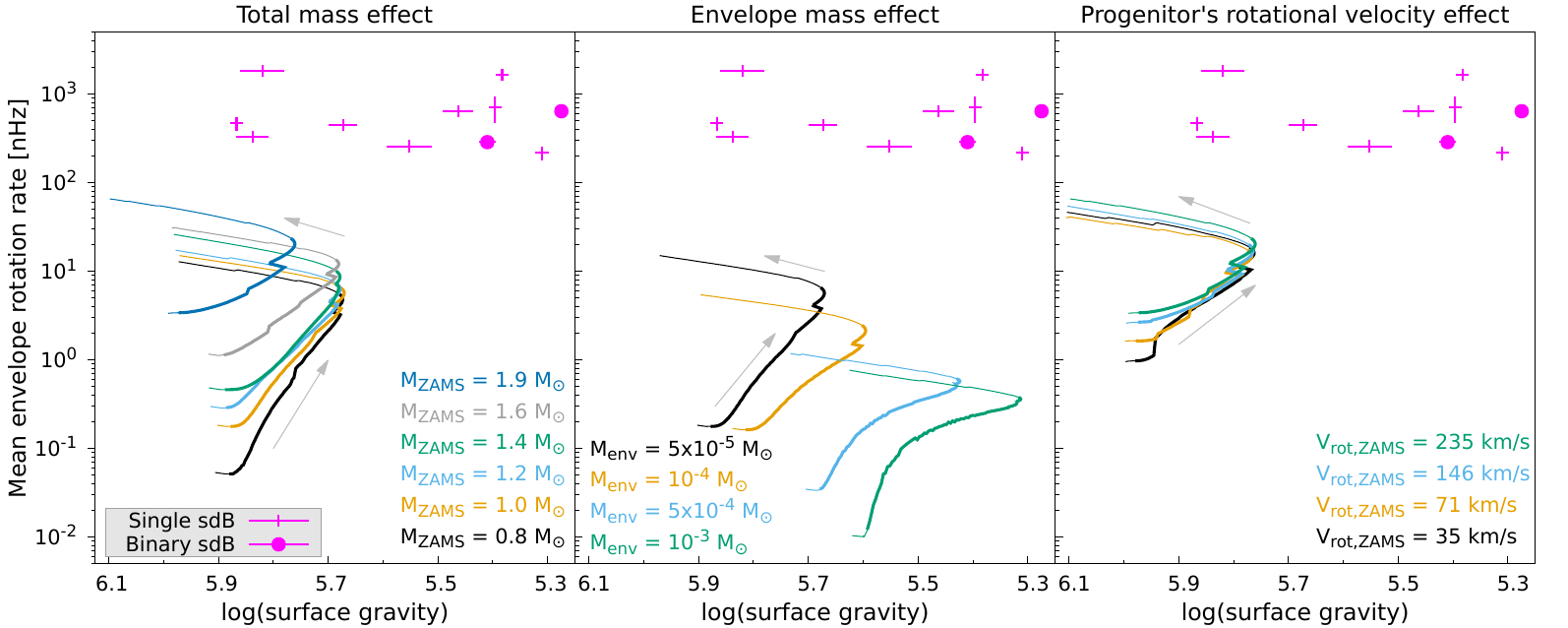}
    \caption{Mean envelope rotation rates as a function of the surface gravity.
      The data points are mean envelope rotation rates as given by the splitting of p-modes, taken from the literature (see Tables \ref{table_singlesdb} and \ref{table_binarysdb}).
      Both single (crosses) and binary (circles) sdB stars in unsynchronised short orbital period systems ($P_{\rm orb} \gtrsim 1$ day) are shown.
      Similarly to Fig. \ref{omegac_gsurf_sdb}, each panel shows the effect of the total mass, envelope mass, and rotational velocity of the sdB's progenitor at the ZAMS.
      The evolutionary sense is indicated by gray arrows.
      The models in the leftmost panel have an envelope mass of $M_{\rm env}= 5 \times 10^{-5} M_{\odot}$ and an initial rotation rate of $\Omega/2\pi= 30 \mu$Hz (or equivalently an initial rotational period of $P_{\rm rot} = 0.39$ days) for models with $M_{\rm ZAMS} > 1.3 M_{\odot}$ and initial rotation periods of $P_{\rm rot}= 0.3$ days for models with $M_{\rm ZAMS} < 1.3 M_{\odot}$.
      The initial mass of the models in the middle panel is $M_{\rm ZAMS}=1 M_{\odot}$ and has an initial rotation period of $P_{\rm rot}=0.3$ days.
      The models of the rightmost panel have an initial mass of $M_{\rm ZAMS}=1.9 M_{\odot}$ and an envelope mass of $M_{\rm env}= 5 \times 10^{-5} M_{\odot}$.
      All sdB models are shown only during the core-helium burning phase.
             In the leftmost panel, starting from the lowest $M_{\rm ZAMS}$ value, the mass of the sdB models are: 0.4753, 0.4737, 0.4735, 0.4728, 0.4649, and 0.4165 $M_{\odot}$.
     }
     \label{omegas_gsurf_sdb}
   \end{figure*}

   The asteroseismic envelope rotation rates as provided by p-modes are in the same range as their core rotation rates in the sample of pulsating sdB stars with asteroseismic measurements to date (see Fig. \ref{omegac_omegas_data}).
   In Fig. \ref{omegas_gsurf_sdb} we compare the asteroseismic envelope rotation rates of observed sdB stars to those of our models (i.e. $\omp/2\pi$, see Sect. \ref{method_omp}) in a few representative cases only, for the sake of clarity.
   The values of $\omp/2\pi$ in these models disagree by 2 to 5 orders of magnitude with the observational sample.
   The smallest disagreement is obtained in models with higher-mass progenitors and less massive envelopes, while the largest one is obtained in models with lower-mass progenitors and more massive hydrogen-rich envelopes.
   A higher initial velocity of the models at the ZAMS can mildly increase $\omp/2\pi$ of our sdB models, but it cannot fully account for the observations.

   In our full range of models, the progenitor's mass at the ZAMS and the envelope mass of the sdB have the most significant impact on the $\omp$ of sdB models.
   While the progenitor's mass changes $\omp/2\pi$ by $\simeq 2$ dex, the envelope mass can change it by $\simeq 1$ dex.
   The change in $\omp$ in sdB models with different progenitor masses (i.e. different $M_{\rm ZAMS}$)  occurs because higher-mass stars at the RGB tip have higher specific AM in the layers just above the hydrogen-burning shell, which later form the envelope of the sdB.
   This occurs because our lower-mass models reach larger radii at the RGB tip than the higher-mass ones (around 200 $R_{\odot}$ for our 0.8 $M_{\odot}$ model and around 90 $R_{\odot}$ for our 1.9 M$_{\odot}$ model), which in combination to the internal AM transport leads to lower specific AM in the radiative regions above the hydrogen-burning shell.   
   In the lower-mass range of our models ($M_{\rm ZAMS} < 1.3 M_{\odot}$), low specific AM in these layers is further imposed by the AM loss due to magnetic braking on their main sequence.
   The decrease of $\omp$ in sdB models with higher envelope mass occurs because the extension over which the AM is redistributed is larger due to the larger radii obtained in sdB models with more massive hydrogen-rich envelopes; the opposite is also valid for sdB models with less massive envelopes.
   Additionally, sdB models with faster-rotating progenitors have higher $\omp$ simply because they had a higher content of total AM at the RGB tip, which follows from their higher initial AM content at the ZAMS.
   
   As for their evolutionary behaviour, despite the sdB expansion during most of its life, the values of $\omp$ increase.
   Since no external sources nor sinks of AM are included in these models, the expansion of the envelope would lead to a decrease of the surface rotation rate under local conservation of AM; this indeed is true for the outermost layers of our models (see Fig. \ref{profile_rotation}).
   However, since $\omp$ is computed over the p-mode cavity, the combined action of the AM transport by internal magnetic fields and the contraction of internal layers (hence spin up) contribute to increasing the envelope rotation rate as sensed by p-modes.
   This can be seen in Fig. \ref{profile_rotation} where the rotation rate increases progressively outwards as the central helium mass fraction decreases.
   This explains the main trends seen in the change of $\omp$ with the physical properties of both the sdB stars and their progenitors, as well as their evolutionary behaviour.

 %
 %

   %------------------------------------------------------------------------------------------
   \subsection{Possible reasons for the disagreement between the rotation rates of models and asteroseismic measurements}
   The disagreement in the envelope rotation rates could be due to several reasons, such as the modelling of the red giants, the need for additional processes during the ejection of the red giants' envelope, or an external source of AM during the sdB phase, among others.
   As for the modelling of the sdB progenitors during the RGB, a higher content of AM in the regions above the hydrogen-burning shell would lead to higher envelope rotation rates of the sdB models.
   This would require a physical treatment of the AM transport different to the one we take in this work.
   However, our treatment of internal AM transport is supported by the asteroseismic rotation rates of the evolutionary phases previous to the RGB tip (see Fig. \ref{omegac_gsurf_zamstosdb}).
   Nevertheless, there is one possibility that we do not address in this work, which is having differential rotation in the convective envelope.
   In our grids of models we assumed that only convection transports AM in the convective zones, with an efficiency equal to that for chemical elements which essentially leads to solid-body rotation in the convective envelope of red giants.
   There are other theories of efficient AM transport that allow for differential rotation in extended convective zones \citep{Takahashi2021,Kissin2015}.
   These theories are supported by multidimensional numerical simulations \citep{Brun2009}, and in some specific cases by observations \citep{Tayar2022}, despite the lack of general observational evidence.
   Differential rotation in the convective envelope of our red giants' models would likely lead to a higher content of AM at the base of the convective envelope (and so higher angular velocities) which can eventually lead to higher envelope rotation rates in the sdB models.
   Although this is a good opportunity to test mechanisms of internal AM transport in upper RGB stars where asteroseismology cannot probe their interior, we do not explore it in detail in this work.
   
   Other AM transport processes operating in radiative regions of red giants could affect the values of $\omp$ in sdB stars, but only if the dominating process is of non-diffusive nature (i.e. it should not erase gradients of rotation).
   That is, the AM in the red giants' radiative regions above the helium core should increase, without significantly slowing down the core (which is already in disagreement with the sdB core rotation rates).
   To achieve this the AM from the convective envelope should be extracted and put into these layers, which is only possible for non-diffusive processes.
   One such process that could operate efficiently in upper RGB stars is the transport by mixed-modes themselves \citep{Belkacem2015a,Belkacem2015b,Bordadagua2025}.
   However, it was shown that at least in the lower RGB this process does not spin up the regions above the hydrogen-burning shell \citep{Bordadagua2025}, although its efficiency could change in the upper RGB above the RGB bump.
   Other non-diffusive AM transport phenomena in post-main sequence stars include Alfv\'en waves in stars with large-scale magnetic fields \citep{Takahashi2021,Takahashi2025}, or internal gravity waves \citep{Rogers2013,Fuller2014,Pincon2016,Pincon2017,Rogers2025}.
   Nonetheless, internal gravity waves are expected to be efficiently damped in red giants \citep{Pincon2017} and large-scale magnetic fields do not likely represent the majority of sdB progenitors, given the low incidence of surface magnetic fields detected in sdB stars \citep{Pelisoli2022,Dorsch2022}; although asteroseismic studies often found red giants' cores to be strongly magnetised \citep{Stello2016a,Stello2016b,Li2022,Deheuvels2023,Hatt2024} \textbfake{, as well as the cores of their main sequence progenitors \citep{Takata2025}}.
   Therefore, we argue that the disagreement in envelope rotation rates is unlikely to be fully addressed by invoking alternative scenarios for the internal AM transport in red giants.

   Another possibility to explain the disagreement in envelope rotation rates would be the transfer of AM during the formation of the sdB stars.
   A sdB in a binary system with short orbital period ($P_{\rm orb} \lesssim 10$ days) most likely forms through a common envelope ejection event when its red giant progenitor approached the tip of the RGB \citep{Han2002,Han2003}.
   When the binary system enters into the common envelope phase, the companion of the red giant spirals into the regions close to the red giant's helium core, releasing orbital energy that contributes to the energy necessary to overcome its binding energy and thus eject the envelope to form the sdB \citep{Ropke2023}.
   The timescales during this phase are short enough that any change in the AM distribution due to internal processes operating during the secular stellar evolution can be neglected.
   However, given the dynamical nature of the common-envelope phase, we speculate that the spiraling-in companion can spin up the gas that would later form the envelope of the sdB.
   This could be possible through drag forces aided by the gas' intrinsic viscosity and instabilities of the Kelvin-Helmholtz type, such as witnessed in planet-engulfment simulations of red giants \citep{Staff2016,Lau2025}.
   Or possibly aided by the amplification of magnetic fields during common envelope evolution \citep{Gagnier2024}.
   Currently, there is no physical prescription that we could implement in one-dimensional stellar evolution codes to simulate these kinds of processes, and no detailed study thereof.
   \textbfake{It is also possible that external torques due to interactions of the sdB with circumstellar matter may play a role; we discuss this scenario in the next subsection.}

%NEW DISCUSSION ON CIRCUMBINARY MATTER AROUND SDB BINARIES
%------------------------------------------------------------------------------------------
\textbfake{\subsection{Circumstellar matter around sdB stars as a source of angular momentum}}
\textbfake{An intermediate plausible step in the formation of sdB stars in close binary systems is the formation of a stable circumstellar environment such as a circumbinary disk.
  This is because sdB stars in close binary systems most likely underwent a common envelope ejection event with a red giant companion, which leads to the ejection of the red giants' envelope.
  But it is likely that a fraction of the matter ejected remains gravitationally bound to the newly formed binary system \citep[e.g.][]{Kashi2011}, if the ejected matter does not reach velocities above the escape velocity of the system.
     That is to say, such matter would still be removed and further ejected from the regions close to the outer boundary of the helium core in their progenitors (i.e.\ RGB tip stars).
     But then this matter could fall back onto the central binary system and possibly form a stable circumbinary disk if it has enough AM \citep[][]{Kashi2011,Ropke2023}.}

\textbfake{In general, circumbinary disks can be accreted onto the stars of the inner binary system (i.e.\ inner to the circumbinary disk) either through accretion streams or via the formation of smaller accretion disks around the individual stars \citep[e.g.\ review by][]{Lai2023}; we speculate that such scenario is also realisable in the formation of sdB stars that underwent a common envelope ejection event.
  The hypothesis of accretion from a circumbinary disk onto its inner binary system is largely supported by direct multi-dimensional numerical simulations \citep[e.g.][]{Shi2012,D'Orazio2013,Miranda2017,Moody2019,Munoz2019,Munoz2020,Duffell2020}, which also demonstrated that the net effect of the interaction between a circumbinary disk and its inner binary system (i.e.\ the combined action of gravitational torques, mass accretion and viscous stresses) is a transfer of positive specific AM to the binary system \citep{Miranda2017,Munoz2019,Munoz2020,Duffell2020}.
  This implies that the components of the central binary system receive AM and thus can spin up, or that the binary system can gain orbital AM and thus widen.
  In our framework, this would imply that the sdB could gain AM through the net interaction with a circumbinary disk, and thus rotate faster as it interacts with the disk.
      }
      
\textbfake{Although this scenario is physically plausible from a theoretical point of view, there is also observational evidence that support this hypothesis.
  Recent observational works have pointed out the ubiquitous presence of circumstellar matter around sdB stars \citep{LiJiangdan2025}, based on the detection of circumstellar matter that exceeds the expectations of the interstellar environment, as inferred from the absorption of the CaII K line; this line is not affected by blending with H lines contrary to other CaII lines \citep[see Fig. 1 of ][]{LiJiangdan2025}.
  These authors also found that in most cases the radial velocity of the ejected matter does not deviate strongly from that of the host sdB star \citep[see Fig.\ 6 of][]{LiJiangdan2025}, indicating that the matter is gravitationally bound and could thus be part of a stable structure such as a circumbinary disk \citep[see also][]{LiJiangdan2022}.
  There is also tentative evidence that circumstellar disks could be present around single sdB stars, as pointed out by \citet{Vos2021}, who found an sdB star whose particular spectrum and infrared excess can be well explained by the presence of a circumstellar disk; this points at a rather general outcome of the formation of sdB stars.
  Since both theoretical and observational arguments support this scenario we then speculate that it may provide an additional source of AM to spin up the envelope of sdB stars once they evolve past the common-envelope phase, which we explore in the following section.}

   %==========================================================================================
   \section{sdB stars spun up by accretion}
   \subsection{Method}
   In this section we present a grid of models where we assume that the sdB can accrete matter from a disk around it.
   Since we assume that the matter is orbiting the sdB and thus forms a disk, the matter should have Keplerian specific AM.
   We then compute models of sdB stars with an accretion disk by assuming that they accrete AM with Keplerian specific AM, although we do not take into account the structural effect of the mass accreted (i.e. the models only accrete AM).
   We explain in Sect. \ref{mass_accreted} why this is a good approximation.
   We then assume that due to accretion a uniform torque per unit of mass is applied to the hydrogen-rich envelope of our sdB models, which leads to a torque of the form
   \begin{equation}
     \frac{{\rm d} j}{{\rm d}t} = \frac{j_{\rm Kep}}{M_{\rm env}} \frac{{\rm d} m}{{\rm d} t} \ ,
     \label{eq_jdot}
   \end{equation}
   where $j_{\rm Kep}=\sqrt{GMR}$ is the Keplerian specific AM evaluated at the surface of the sdB model, $M_{\rm env}$ is the mass of the hydrogen-rich envelope, and ${\rm d}m/{\rm d}t$ is the mass accretion rate which we take as $10^{-12} M_{\odot}$/yr and assume it is constant.
   We assume that the sdB accretes matter as soon as it begins the core-helium burning phase and is well located in the sdB regime in terms of surface gravity and effective temperature.
   \textbfake{That is, the sdB model already has a hydrogen-rich envelope when it begins accreting, and only accretes additional matter from a disk once it begins the core-helium burning phase (i.e. the hydrogen-rich envelope is not provided by the disk).}

   To compute our models of rotating sdB stars with accretion, we choose as initial models the sdB models that we computed without assuming any external source of AM nor additional processes during common-envelope evolution presented in Sect. \ref{sect_sdbmodels}.
   Then we compute sdB evolutionary models with accretion of AM using Eq. \ref{eq_jdot} until their $\omp/2\pi$ reaches 200 nHz or 2000 nHz, and $\omg$ reaches 100 nHz or 1000 nHz, that is, four different kinds of models.
   These particular values of $\omp/2\pi$ and $\omg/2\pi$ are the limits that we defined based on the observed range (see Fig. \ref{omegac_omegas_data}).
   Once the models reach these limiting values, we stop the accretion of AM.
   Using these models we can estimate the minimum and maximum mass that sdB stars need to accrete to match their asteroseismic rotation rates.
   We repeated this process for all of our sdB rotating models, with different initial progenitor masses, initial velocities and envelope masses, as presented in Sect. \ref{inputs_sdb}.

   \begin{figure}   \includegraphics[width=\columnwidth]{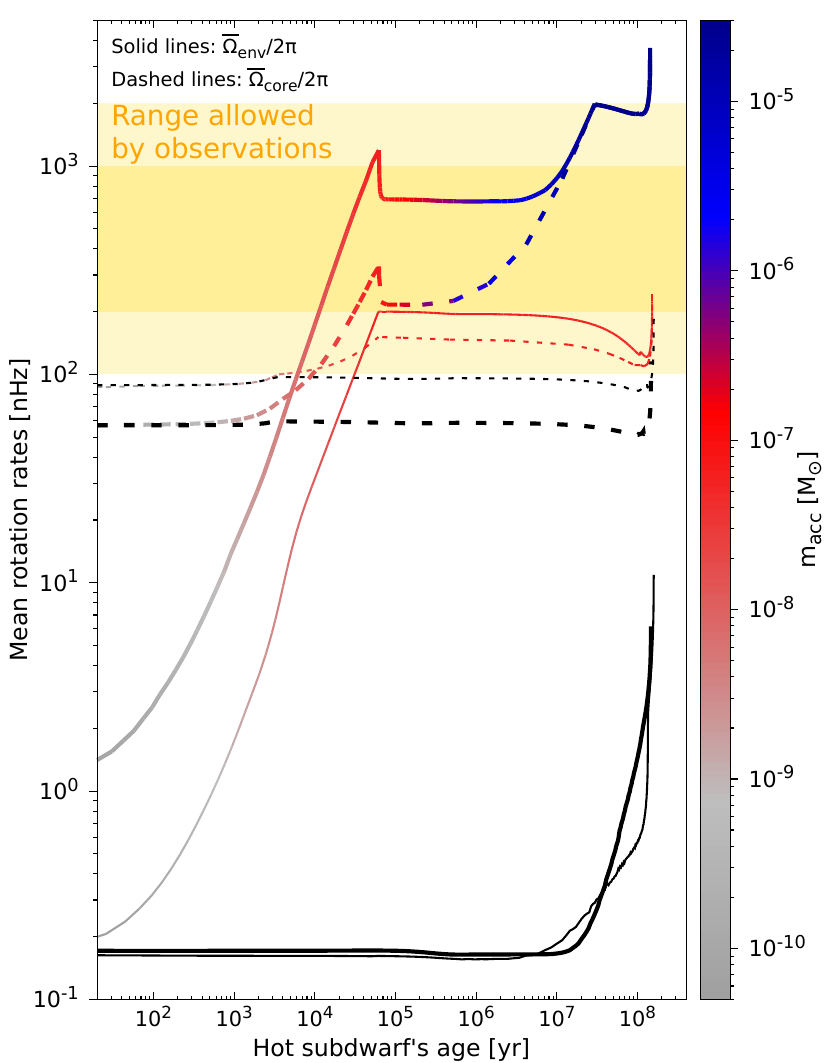}
     \caption{Mean core and envelope rotation rates as a function of the hot subdwarf's age in two accreting sdB models (colored lines) and two non-accreting ones (black lines).
       The colour bar shows the amount of AM accreted by the models in terms of mass ($m_{\rm acc}$) containing Keplerian specific AM.
       The yellow shaded bands show the range of values favoured by measurements of both core and envelope rotation rates (darker shade), or favoured either only by core or envelope rotation rates (lighter shade).
       The models have $M_{\rm ZAMS} = 1.6 M_{\odot}, M_{\rm env} = 10^{-3} M_{\odot}, P_{\rm rot,ZAMS} = 0.39$ days (thin lines), or $M_{\rm ZAMS} = 1 M_{\odot}, M_{\rm env} = 10^{-4} M_{\odot}, P_{\rm rot,ZAMS} = 1 $ day (thick lines).
     }
     \label{omegas_gsurf_accretion}
      \end{figure}
   \begin{figure*}
     \centering     \includegraphics[width=18cm]{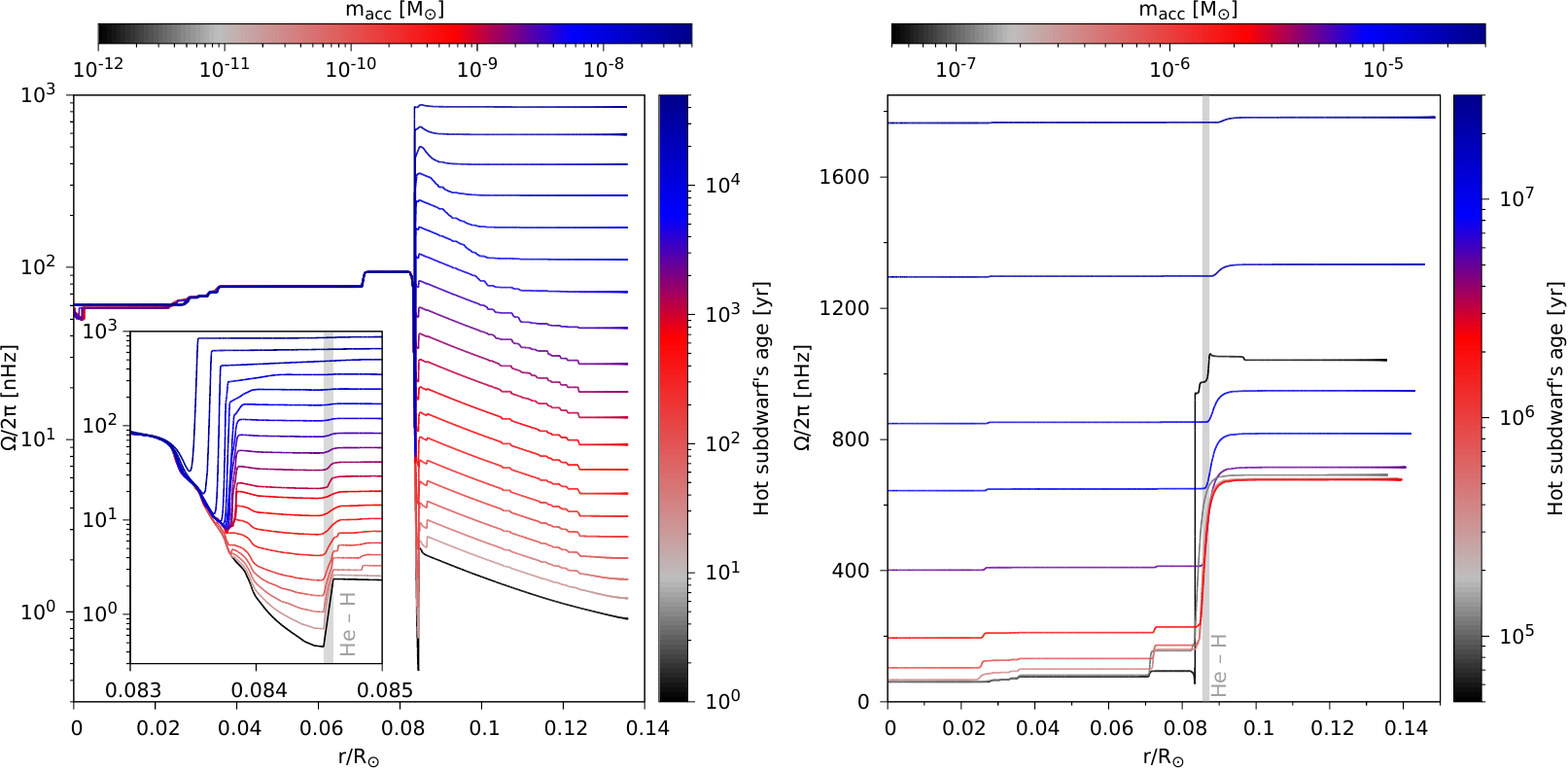}
     \caption{Rotation rate as a function of the radial coordinate at different times of a sdB model spun up by accretion.
       The rotation profiles correspond to the model spun up to $\omp/2\pi=2000$ nHz of Fig. \ref{omegas_gsurf_accretion}.
       The left panel shows the initial evolution as the star is spun up to $\omp/2\pi \simeq 1000$ nHz, while the panel on the right shows the evolution once the core begins spinning up and eventually the whole sdB rotates rigidly.
       The inset on the left figure shows a zoomed view on the transition from the helium-rich radiative interior to the hydrogen-rich envelope where most of the differential rotation develops.
       In both panels the colour bars show: the age of the sdB where the zero age is set at the beginning of the core-helium burning phase (colour bar on the right), and the mass with Keplerian specific AM that the star needs to accrete to be spun up (colour bar at the top).
       \textbfake{The grey band shows the mean location of the transition from the helium-rich radiative interior to the hydrogen-rich envelope}.
     }
     \label{profile_rotation_accretion}
   \end{figure*}
   %
   %

   %------------------------------------------------------------------------------------------
   \subsection{Mean core and envelope rotation rates of accreting sdB stars}
   In Fig. \ref{omegas_gsurf_accretion} we show an example of two accreting sdB models spun up until they reach $\omp/2\pi = 200$ nHz and $\omp/2\pi = 2000$ nHz, shown by thin and thick lines, respectively.
   In the models spun up to $200$ nHz, its $\omp$  increases monotonically with $m_{\rm acc}$, while in the models spun up to $2000$ nHz the internal transport of AM by magnetic fields briefly slows down the envelope as soon as it reaches $\omp/2\pi \simeq 1000$ nHz.
   This occurs because at higher rotation rates the magnetic fields can transport AM more efficiently, and thus transport the AM from the hydrogen-rich envelope into the helium-rich radiative regions.
   After this brief phase, the sdB continues accreting until it reaches 2000 nHz, at which point we stop the accretion.
   
   Once the sdB accreting models reach either 200 or 2000 nHz, they do not spin up or down significantly, changing their spin rates by at most a factor two.
   However, the evolution of their $\omp$ is different from that in non-accreting models (black lines in Fig. \ref{omegas_gsurf_accretion}) because by the time they stop accreting their envelopes rotate rigidly due to the AM transport being more efficient at higher rotation rates.
   Then, the rotation of their envelopes simply responds to the change in the size of the sdB model by local conservation of AM, that is, they spin down when they expand and vice versa.
   
   In our accreting sdB models, the core rotation rate as sensed by g-modes can also increase due to the accretion, as is shown in Fig. \ref{omegas_gsurf_accretion} by the dashed lines.
   This occurs primarily because in our models, $\omg$ as computed using Eq. \ref{eq_omg} can be affected by the rotation rates above the core, mainly by that at the He-H transition and secondarily by the rotation rate in the hydrogen-rich envelope if their rotation rates are high enough to overcome the weight functions (see Fig. \ref{weight_functions}), which can occur if the rotation rates in these regions are above $\simeq 100$ nHz.
   This is why in the model spun up to $\omp/2\pi = 200$ nHz the values of $\omg$ increase with respect to those of the non-accreting sdB model.
   In the case where the sdB accretes a higher amount of AM, the rotation rate in the envelope can be so high that the TSF dynamo can overcome the buoyancy force at the He-H transition and thus transport the AM from the hydrogen-rich envelope to the helium-rich radiative interior, eventually leading to solid-body rotation.
   This is why $\omg$ gradually reaches $\omp$ in the model spun up to $\omp/2\pi = 2000$ nHz in Fig. \ref{omegas_gsurf_accretion}.
   
   \textbfake{Models of accreting sdB stars without internal magnetic fields do not reach solid-body rotation (see App. \ref{app_nonmag}) because the AM transport efficiency by hydrodynamical processes is not high enough.
     These models behave slightly different from the magnetic models, in particular, they can develop a much higher core-to-envelope radial differential rotation because the AM deposited in the envelope through accretion cannot be transported into the helium-rich radiative regions; thus the central regions cannot be spun up.
     This is a fundamental difference with respect to stellar models including internal magnetic fields, which further reinforces the idea that internal magnetic fields may develop in all kinds of stars, including post-common envelope binaries.
     This is so because of the rather small core-to-envelope radial differential rotation observed in sdB stars, which in some cases display rigid rotation (see Tables \ref{table_singlesdb} \& \ref{table_binarysdb}).
     We also note that if the rotational splitting of g-modes of observed sdB stars probe mainly the deep helium-rich radiative regions (i.e. it is not affected by the rotation in the hydrogen-rich envelope) then non-magnetic accreting sdB models would always overestimate the amplitude of core-to-envelope radial differential rotation (see right panel of Fig. \ref{profile_rotation_accretion_nonmag}).
     Nevertheless, this depends on other aspects that we do not explore in this work, such as whether the g-modes are trapped or confined \citep[e.g.][]{Charpinet2000}.
   }

   %------------------------------------------------------------------------------------------
   \subsection{Interior and differential rotation of accreting sdB stars}
   The internal AM redistribution in our accreting sdB models is illustrated in Fig. \ref{profile_rotation_accretion} where we show the evolution of the rotation rates in the stellar interior during the whole spin-up phase of the sdB model that reaches quasi-rigid rotation.
   Initially, the rotation rate scales like $\Omega \propto r^{-2}$ in the hydrogen-rich envelope (whose boundary is at $r/R_{\odot} \simeq 0.0845$ in Fig.\ \ref{profile_rotation_accretion}, \textbfake{see also Fig. \ref{profile_rotation_accretion_withchems}}) because we assume the accretion leads to a uniform torque in all these regions; this is equivalent to assuming that the AM accreted is uniformly redistributed instantaneously in these regions.
   As evolution proceeds and the star accretes more AM the hydrogen-rich envelope is progressively spun up and the rotation profile flattens out from the surface inwards because the AM transport efficiency by the magnetic fields scales with both the rotation rate and the radial coordinate (Eq. \ref{eq_magvisc}). 
   In the regions close to the He-H transition (see inset of Fig. \ref{profile_rotation_accretion}), differential rotation can develop first because no torque is applied below the He-H transition, and second because the magnetic dynamo is inhibited by the strong chemical composition gradient present in these regions.
   This leads first to a sudden spin up of the innermost hydrogen-rich layers (at $r/R_{\odot} \simeq 0.0845$ in Fig. \ref{profile_rotation_accretion}) due to accretion, and then to a gradual spin up of the regions just below it due to the AM transport by internal magnetic fields.
   
   However, the rotation rate in the helium-rich radiative regions remains largely unchanged during this initial spin-up phase, mainly because there is not enough AM nor time for the internal magnetic fields to overcome the buoyancy forces at the He-H transition.
   At later times (see right panel of Fig. \ref{profile_rotation_accretion}) the magnetic dynamo can gradually overcome these forces and thus the AM can be transported from the hydrogen-rich envelope to the helium-rich radiative interior, spinning it up and thus increasing $\omg$.
   If the sdB accretes enough mass ($\simeq 10^{-5} M_{\odot}$), the magnetic fields become efficient enough to couple the whole star, as is shown in the last rotation profile of Fig. \ref{profile_rotation_accretion} (right panel) where the model reaches quasi-rigid rotation, with slight differential rotation at $r/R_{\odot} \simeq 0.09$ because the model is still accreting AM.
   All the accreting sdB models spun up to $\omp/2\pi = 2000$ nHz reach solid-body rotation, and the core cannot rotate faster than the envelope in any of them.

   \begin{figure}  \includegraphics[width=\columnwidth]{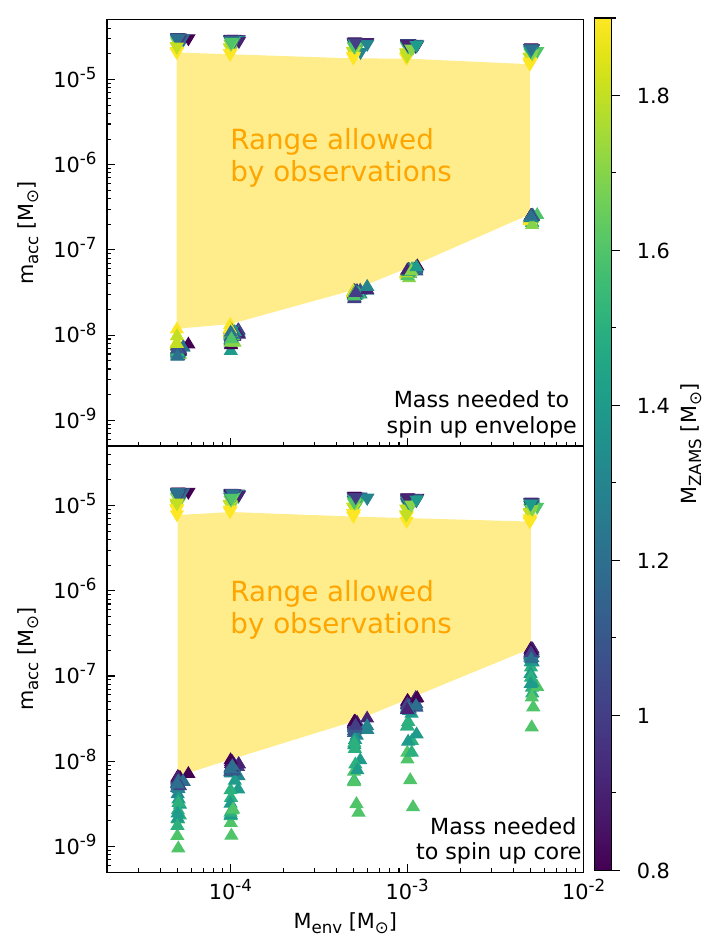}
     \caption{Mass needed to spin up the envelope (top panel) or core (bottom panel) of sdB stars to their observed range of asteroseismic rotation rates, as a function of the envelope mass of each model.
       The shaded regions indicate the range of values allowed by measurements of both core and envelope rotation rates for any progenitor mass.
       The two branches on each panel show the mass needed to increase the rotation rates to either $\omg/2\pi = 100 \rm{nHz}, \omp/2\pi= 200 \rm{nHz}$ for the lower branch (upward triangles), or $\omg/2\pi = 1000 \rm{nHz}, \omp/2\pi= 2000 \rm{nHz}$ for the upper branch (downward triangles).
       The colour bar shows the mass of the sdB's progenitor at the ZAMS.       
     }
     \label{macc_menv}
      \end{figure}
   %
   %
%------------------------------------------------------------------------------------------
   \subsection{Mass accreted}
   \label{mass_accreted}

   To spin up the sdB stars such that they can match the asteroseismic rotation rates, they need to accrete at least $\simeq 10^{-9} M_{\odot}$ but no more than $\simeq 10^{-5} M_{\odot}$ (see Fig. \ref{macc_menv}).
   To increase $\omp$ and $\omg$ in our sdB models to the lower boundaries set by the observations (either $\omg/2\pi = 100$ nHz or $\omp/2\pi= 200$ nHz) they need to accrete more mass if their hydrogen-rich envelope is more massive.
   This leads to a difference of a factor 10 to 100 of mass accreted in the range of envelope masses of our sdB models (Fig. \ref{macc_menv}).
   This is because in an sdB a more massive hydrogen-rich envelope has a larger moment of inertia mainly due to its increased extension, so it requires a larger amount of AM to spin it up.
   However, to increase both $\omp$ and $\omg$ to the upper boundaries (i.e. $\omp/2\pi=2000$ nHz or $\omg/2\pi=1000$ nHz), slightly less accreted mass is needed for models with more massive envelopes.
   This is because the specific AM accreted is evaluated at the surface of the sdB and thus can change with its total mass and radius.
   This makes the models with less massive envelopes accrete less AM for the same amount of mass because we recall the specific AM accreted is $j_{\rm Kep}=\sqrt{GMR}$.
   In fact the models with $M_{\rm env} = 5 \times 10^{-5} M_{\odot}$ accrete on average matter with a specific AM content of $j_{\rm Kep} \simeq 7 \times 10^{17}$ cm$^2$/s while those with $M_{\rm env} = 5 \times 10^{-3} M_{\odot}$ accrete on average matter with $j_{\rm Kep} \simeq 10^{18}$ cm$^2$/s.
   Because of this the sdB models with more massive envelopes, despite having a larger moment of inertia, need to accrete less mass than the sdB models with less massive envelopes.
   
   There is also a trend between the mass needed to increase $\omg$ and the mass of the sdB's progenitor at the ZAMS as shown by the colour bar in Fig. \ref{macc_menv}.
   Those sdB stars with more massive progenitors need to accrete less mass.
   This occurs because in the base non-accreting models used to compute the accreting ones, $\omg$ is higher in models with higher ZAMS progenitor masses (see left panel of Fig. \ref{omegac_gsurf_sdb}).
   In fact, the sdB models computed from the progenitors with $M_{\rm ZAMS}=1.7, 1.8, 1.9 M_{\odot}$ do not need to accrete any AM because their $\omg/2\pi$ is already around 100 nHz in the base models (see left panel of Fig. \ref{omegac_gsurf_sdb}); this is why they do not appear in Fig. \ref{macc_menv}.
   As for the envelope rotation rates there is not any trend with the mass of the progenitors, and there is no trend with the initial rotational velocities either.
   
   Regarding the fact that we do not include the effects on the structure of the sdB due to mass accretion (i.e. the models only accrete AM), we showed with our previous discussion that these effects can be neglected.
   Because they will only be relevant for the sdB models with the lowest envelope mass ($M_{\rm env}= 5 \times 10^{-5} M_{\odot}$), since at most the mass needed to reproduce the asteroseismic rotation rates is $3 \times 10^{-5} M_{\odot}$.
   This would represent a $60 \%$ increase in the mass of the envelope of these models, but we show in Fig. \ref{macc_menv} that the relation between the maximum mass needed to spin up the envelope and the envelope mass of the sdB is quite weak, leading to a difference of at most a factor two.
   Thus, the structural effect of increasing the envelope mass of the accreting sdB models will have a minor effect on the mass needed.

   %------------------------------------------------------------------------------------------
   \subsection{Core-to-envelope radial differential rotation observed in sdB stars}
   These accreting sdB models can reproduce both the range of mean core and envelope rotation rates, as well as the relatively low amplitude of radial differential rotation inferred in some sdB stars \citep[e.g. LT CnC; ][]{Reed2025}.
   Our models that accrete enough mass to spin up to $\omp/2\pi = 2000$ nHz allow for a maximum ratio of mean envelope-to-core rotation contrast of $\omp/\omg \simeq 4$ during the accretion, after which the models couple and reach rigid rotation (see also Fig. \ref{omegas_gsurf_accretion}), while those that are spun up to $\omp/2\pi = 200$ nHz only allow for $\omp/\omg \simeq 2$ at most.
   These are the maximum allowed values of envelope-to-core radial differential rotation as probed by the mean asterosesismic core and envelope rotation rates (using Eqs. \ref{eq_omg} \& \ref{eq_omp}).
   These values result from the combined action of accretion and internal AM transport by internal magnetic fields.
   Our accreting sdB models can also explain why the envelope of unsynchronised post common-envelope sdB stars in binary systems rotate faster than their core.
   Currently there are only two of such stars, where their core and envelope rotation rates were measured via the rotational splittings of g- and p-modes \citep[KIC11558725 and LT CnC;][]{Telting2012,Reed2025}.
   For these two systems in particular, we estimate that they should accrete roughly $5 \times 10^{-7} M_{\odot}$ (LT CnC) and $4 \times 10^{-8} M_{\odot}$ (KIC11558725) to reproduce their amplitude of radial differential rotation.

%==========================================================================================
   \section{Conclusions}
   We computed grids of sdB stellar evolution models with rotation, internal magnetic fields and mass accretion, based on models of red giants that can reproduce the range of asteroseismic core and envelope rotation rates of the evolutionary phases previous to the tip of the RGB.
   In particular, our sdB models correspond to those sdB stars formed through the common-envelope channel near the tip of the RGB.
   If sdB stars do not gain any angular momentum during their formation or initial evolution their asteroseismic rotation rates as measured by the rotation splittings of g- and p-modes cannot be explained.
   In particular, the asteroseismic envelope rotation rates disagree by two to five orders of magnitude if the sdB is not spun up by any external torque.

   We then showed that if sdB stars can accrete matter from a surrounding disk, the combined action of mass accretion and internal magnetic fields can spin up both their core and envelope.
   If they accrete $\sim 10^{-9} - 10^{-5} M_{\odot}$ from a disk, then the whole range of asteroseismic rotation rates can be reproduced.
   Moreover, this kind of models also allows for an envelope rotating faster than its core, as witnessed in some unsynchronised sdB stars in binary systems \citep{Telting2012,Reed2025}.
   However, we note that measurements of asteroseismic rotation rates of sdB stars are still scarce, and that our work relies strongly on the current measurements available in the literature.
   Future asteroseismic studies on already identified sdB pulsators \citep{Baran2023,Baran2024,Uzundag2024} should be able to confirm the observational trends and whether the hypothesis of spun-up sdB stars by accretion can be either further supported or rejected.
   \textbfake{In this regard, the upcoming PLATO space-mission will deliver unprecedented photometric precision, long uninterrupted time baselines, and wide sky coverage for bright pulsating compact stars.
     Simulations demonstrate that PLATO will be capable of detecting pulsation amplitudes down to $\simeq 0.1$ mma, enabling the measurement of rotational multiplets in pulsating hot subdwarfs \citep{Uzundag2025}.}
   
   From the theoretical perspective, although we propose a framework that can roughly reproduce the bulk asteroseismic rotational measurements of sdB stars, wide ample room is left to explore alternative scenarios such as: transfer of angular momentum during common envelope evolution \citep[e.g. via amplification of large-scale magnetic fields;][]{Gagnier2024}, engulfment of substellar objects \citep{Lau2025}, or uncertainties related to various physical phenomena affecting the internal AM transport of upper RGB stars \citep[e.g.][]{Moyano2023a,Meduri2024,Skoutnev2025,Bordadagua2025}.
   Moreover, in this work we only study sdB stars that formed through a common-envelope ejection from low-mass progenitors in binary systems, but their higher-mass counterparts as well as those formed through stable mass transfer remain largely unexplored, and deserve further study.

%==========================================================================================   
   \section*{Acknowledgements}
   We warmly thank the referee for their constructive and useful suggestions which helped improving the presentation of this work.
   M. U. gratefully acknowledges funding from the Research Foundation Flanders (FWO) by means of a junior postdoctoral fellowship (grant agreement No. 1247624N).
\textbfake{BB acknowledges support from the ERC Consolidator Grant DipolarSound (grant agreement no. 101000296).
  This project is supported by the National Natural Science Foundation of China (NSFC Nos. 12288102, 12090043, 12125303, 12525304), the National Key R\&D Program of China (No. 2021YFA1600403), the Strategic Priority Reserch Program of the Chinese Academy of Sciences (grant No. XDB1160201), the Yunnan Revitalization Talent Support Program - Science \& Technology Champion Project (No. 202305AB35003), the Yunnan Fundamental Research Projects (No. 202401BC070007), and the International Centre of Supernovae, Yunnan Key Laboratory (No. 202302AN36001).}
%%%%%%%%%%%%%%%%%%%%%%%%%%%%%%%%%%%%%%%%%%%%%%%%%%
\section*{Data Availability}
The models presented in this work, as well as the necessary extensions to reproduce them with {\mesa } are available at \url{https://zenodo.org/records/17332652}.
All the measurements of asteroseismic rotation rates were compiled from the literature and referenced either in the main body of the text or available in Tables \ref{table_singlesdb} and \ref{table_binarysdb}.



\bibliographystyle{mnras}
\bibliography{references} % references.bib

@ARTICLE{vanSaders2013,
       author = {{van Saders}, Jennifer L. and {Pinsonneault}, Marc H.},
        title = "{Fast Star, Slow Star; Old Star, Young Star: Subgiant Rotation as a Population and Stellar Physics Diagnostic}",
      journal = {\apj},
         year = 2013,
        month = oct,
       volume = {776},
       number = {2},
          eid = {67},
        pages = {67},
          doi = {10.1088/0004-637X/776/2/67},
archivePrefix = {arXiv},
       eprint = {1306.3701},
 primaryClass = {astro-ph.SR},
       adsurl = {https://ui.adsabs.harvard.edu/abs/2013ApJ...776...67V}
}

@ARTICLE{Hermes2015,
       author = {{Hermes}, J.~J. and {G{\"a}nsicke}, B.~T. and {Bischoff-Kim}, A. and et al.},
        title = "{Insights into internal effects of common-envelope evolution using the extended Kepler mission}",
      journal = {\mnras},
         year = 2015,
        month = aug,
       volume = {451},
       number = {2},
        pages = {1701-1712},
          doi = {10.1093/mnras/stv1053},
archivePrefix = {arXiv},
       eprint = {1505.01848},
 primaryClass = {astro-ph.SR},
       adsurl = {https://ui.adsabs.harvard.edu/abs/2015MNRAS.451.1701H}
}

@ARTICLE{Uzundag2025,
       author = {{Uzundag}, Murat and {Corsico}, Alejandro H. and {Jannsen}, Nicholas and et al.},
        title = "{Observing bright pulsating white dwarfs with PLATO: A new window into the late stages of stellar evolution}",
      journal = {arXiv e-prints},
         year = 2025,
        month = nov,
          eid = {arXiv:2511.19196},
        pages = {arXiv:2511.19196},
          doi = {10.48550/arXiv.2511.19196},
archivePrefix = {arXiv},
       eprint = {2511.19196},
 primaryClass = {astro-ph.SR},
       adsurl = {https://ui.adsabs.harvard.edu/abs/2025arXiv251119196U}
}

@ARTICLE{Reed2018,
       author = {{Reed}, M.~D. and {Armbrecht}, E.~L. and {Telting}, J.~H. and et al.},
        title = "{K2 Campaign 5 observations of pulsating subdwarf B stars: binaries and super-Nyquist frequencies}",
      journal = {\mnras},
         year = 2018,
        month = mar,
       volume = {474},
       number = {4},
        pages = {5186-5198},
          doi = {10.1093/mnras/stx3133},
       adsurl = {https://ui.adsabs.harvard.edu/abs/2018MNRAS.474.5186R}
}

@ARTICLE{Reed2016,
       author = {{Reed}, M.~D. and {Baran}, A.~S. and {{\O}stensen}, R.~H. and et al.},
        title = "{A pulsation analysis of K2 observations of the subdwarf B star PG 1142-037 during Campaign 1: A subsynchronously rotating ellipsoidal variable}",
      journal = {\mnras},
         year = 2016,
        month = may,
       volume = {458},
       number = {2},
        pages = {1417-1426},
          doi = {10.1093/mnras/stw348},
archivePrefix = {arXiv},
       eprint = {1602.06412},
 primaryClass = {astro-ph.SR},
       adsurl = {https://ui.adsabs.harvard.edu/abs/2016MNRAS.458.1417R}
}

@ARTICLE{Ketzer2017,
       author = {{Ketzer}, L. and {Reed}, M.~D. and {Baran}, A.~S. and et al.},
        title = "{K2 observations of pulsating subdwarf B stars: Analysis of EPIC 203948264 observed during Campaign 2}",
      journal = {\mnras},
         year = 2017,
        month = may,
       volume = {467},
       number = {1},
        pages = {461-468},
          doi = {10.1093/mnras/stx104},
       adsurl = {https://ui.adsabs.harvard.edu/abs/2017MNRAS.467..461K}
}

@ARTICLE{Bloemen2014,
       author = {{Bloemen}, S. and {Hu}, H. and {Aerts}, C. and et al.},
        title = "{The blue-edge problem of the V1093 Herculis instability strip revisited using evolutionary models with atomic diffusion}",
      journal = {\aap},
         year = 2014,
        month = sep,
       volume = {569},
          eid = {A123},
        pages = {A123},
          doi = {10.1051/0004-6361/201323309},
archivePrefix = {arXiv},
       eprint = {1409.1604},
 primaryClass = {astro-ph.SR},
       adsurl = {https://ui.adsabs.harvard.edu/abs/2014A&A...569A.123B}
}

@ARTICLE{Charpinet2001,
       author = {{Charpinet}, St{\'e}phane and {Fontaine}, G. and {Brassard}, P.},
        title = "{A Theoretical Exploration of the Pulsational Stability of Subdwarf B Stars}",
      journal = {\pasp},
         year = 2001,
        month = jul,
       volume = {113},
       number = {785},
        pages = {775-788},
          doi = {10.1086/322139},
       adsurl = {https://ui.adsabs.harvard.edu/abs/2001PASP..113..775C}
}

@ARTICLE{Charpinet2002a,
       author = {{Charpinet}, S. and {Fontaine}, G. and {Brassard}, P. and et al.},
        title = "{Adiabatic Survey of Subdwarf B Star Oscillations. II. Effects of Model Parameters on Pulsation Modes}",
      journal = {\apjs},
         year = 2002,
        month = apr,
       volume = {139},
       number = {2},
        pages = {487-537},
          doi = {10.1086/338822},
       adsurl = {https://ui.adsabs.harvard.edu/abs/2002ApJS..139..487C}
}

@ARTICLE{Charpinet2002b,
       author = {{Charpinet}, S. and {Fontaine}, G. and {Brassard}, P. and et al.},
        title = "{Adiabatic Survey of Subdwarf B Star Oscillations. III. Effects of Extreme Horizontal Branch Stellar Evolution on Pulsation Modes}",
      journal = {\apjs},
         year = 2002,
        month = jun,
       volume = {140},
       number = {2},
        pages = {469-561},
          doi = {10.1086/339707},
       adsurl = {https://ui.adsabs.harvard.edu/abs/2002ApJS..140..469C}
}

@ARTICLE{Charpinet1996,
       author = {{Charpinet}, S. and {Fontaine}, G. and {Brassard}, P. and et al.},
        title = "{The Potential of Asteroseismology for Hot, Subdwarf B Stars: A New Class of Pulsating Stars?}",
      journal = {\apjl},
         year = 1996,
        month = nov,
       volume = {471},
        pages = {L103},
          doi = {10.1086/310335},
archivePrefix = {arXiv},
       eprint = {astro-ph/9606069},
 primaryClass = {astro-ph},
       adsurl = {https://ui.adsabs.harvard.edu/abs/1996ApJ...471L.103C}
}

@ARTICLE{Corsico2011,
       author = {{C{\'o}rsico}, A.~H. and {Althaus}, L.~G. and {Kawaler}, S.~D. and et al.},
        title = "{Probing the internal rotation of pre-white dwarf stars with asteroseismology: the case of PG 0122+200}",
      journal = {\mnras},
         year = 2011,
        month = dec,
       volume = {418},
       number = {4},
        pages = {2519-2526},
          doi = {10.1111/j.1365-2966.2011.19642.x},
archivePrefix = {arXiv},
       eprint = {1108.3359},
 primaryClass = {astro-ph.SR},
       adsurl = {https://ui.adsabs.harvard.edu/abs/2011MNRAS.418.2519C}
}

@ARTICLE{Calcaferro2023,
       author = {{Calcaferro}, Leila M. and {C{\'o}rsico}, Alejandro H. and {Althaus}, Leandro G. and et al.},
        title = "{Exploring the internal rotation of the extremely low-mass He-core white dwarf GD 278 with TESS asteroseismology}",
      journal = {\aap},
         year = 2023,
        month = may,
       volume = {673},
          eid = {A135},
        pages = {A135},
          doi = {10.1051/0004-6361/202346007},
archivePrefix = {arXiv},
       eprint = {2303.15962},
 primaryClass = {astro-ph.SR},
       adsurl = {https://ui.adsabs.harvard.edu/abs/2023A&A...673A.135C}
}

@ARTICLE{Romero2022,
       author = {{Romero}, Alejandra D. and {Kepler}, S.~O. and {Hermes}, J.~J. and et al.},
        title = "{Discovery of 74 new bright ZZ Ceti stars in the first three years of TESS}",
      journal = {\mnras},
         year = 2022,
        month = apr,
       volume = {511},
       number = {2},
        pages = {1574-1590},
          doi = {10.1093/mnras/stac093},
archivePrefix = {arXiv},
       eprint = {2201.04158},
 primaryClass = {astro-ph.SR},
       adsurl = {https://ui.adsabs.harvard.edu/abs/2022MNRAS.511.1574R}
}

@ARTICLE{Romero2025,
       author = {{Romero}, Alejandra D. and {Kepler}, S.~O. and {Oliveira da Rosa}, Gabriela and et al.},
        title = "{Thirty-two New Bright ZZ Ceti Stars from TESS: Adding Cycles 4 and 5}",
      journal = {\apj},
         year = 2025,
        month = may,
       volume = {984},
       number = {2},
          eid = {112},
        pages = {112},
          doi = {10.3847/1538-4357/adc113},
archivePrefix = {arXiv},
       eprint = {2407.07260},
 primaryClass = {astro-ph.SR},
       adsurl = {https://ui.adsabs.harvard.edu/abs/2025ApJ...984..112R}
}

@ARTICLE{OliveiradaRosa2024,
       author = {{Oliveira da Rosa}, Gabriela and {Kepler}, S.~O. and {Soethe}, L.~T.~T. and et al.},
        title = "{Photometric White Dwarf Rotation}",
      journal = {\apj},
         year = 2024,
        month = oct,
       volume = {974},
       number = {2},
          eid = {314},
        pages = {314},
          doi = {10.3847/1538-4357/ad6987},
archivePrefix = {arXiv},
       eprint = {2407.05214},
 primaryClass = {astro-ph.SR},
       adsurl = {https://ui.adsabs.harvard.edu/abs/2024ApJ...974..314O}
}

@ARTICLE{Vos2021,
       author = {{Vos}, Joris and {Pelisoli}, Ingrid and {Budaj}, Jan and et al.},
        title = "{Looking into the cradle of the grave: J22564-5910, a potential young post-merger hot subdwarf}",
      journal = {\aap},
         year = 2021,
        month = nov,
       volume = {655},
          eid = {A43},
        pages = {A43},
          doi = {10.1051/0004-6361/202140391},
archivePrefix = {arXiv},
       eprint = {2106.03363},
 primaryClass = {astro-ph.SR},
       adsurl = {https://ui.adsabs.harvard.edu/abs/2021A&A...655A..43V}
}

@ARTICLE{Lai2023,
       author = {{Lai}, Dong and {Mu{\~n}oz}, Diego J.},
        title = "{Circumbinary Accretion: From Binary Stars to Massive Binary Black Holes}",
      journal = {\araa},
         year = 2023,
        month = aug,
       volume = {61},
        pages = {517-560},
          doi = {10.1146/annurev-astro-052622-022933},
archivePrefix = {arXiv},
       eprint = {2211.00028},
 primaryClass = {astro-ph.HE},
       adsurl = {https://ui.adsabs.harvard.edu/abs/2023ARA&A..61..517L}
}

@ARTICLE{Moody2019,
       author = {{Moody}, Mackenzie S.~L. and {Shi}, Ji-Ming and {Stone}, James M.},
        title = "{Hydrodynamic Torques in Circumbinary Accretion Disks}",
      journal = {\apj},
         year = 2019,
        month = apr,
       volume = {875},
       number = {1},
          eid = {66},
        pages = {66},
          doi = {10.3847/1538-4357/ab09ee},
archivePrefix = {arXiv},
       eprint = {1903.00008},
 primaryClass = {astro-ph.HE},
       adsurl = {https://ui.adsabs.harvard.edu/abs/2019ApJ...875...66M}
}

@ARTICLE{Munoz2020,
       author = {{Mu{\~n}oz}, Diego J. and {Lai}, Dong and {Kratter}, Kaitlin and et al.},
        title = "{Circumbinary Accretion from Finite and Infinite Disks}",
      journal = {\apj},
         year = 2020,
        month = feb,
       volume = {889},
       number = {2},
          eid = {114},
        pages = {114},
          doi = {10.3847/1538-4357/ab5d33},
archivePrefix = {arXiv},
       eprint = {1910.04763},
 primaryClass = {astro-ph.HE},
       adsurl = {https://ui.adsabs.harvard.edu/abs/2020ApJ...889..114M}
}

@ARTICLE{Duffell2020,
       author = {{Duffell}, Paul C. and {D'Orazio}, Daniel and {Derdzinski}, Andrea and et al.},
        title = "{Circumbinary Disks: Accretion and Torque as a Function of Mass Ratio and Disk Viscosity}",
      journal = {\apj},
         year = 2020,
        month = sep,
       volume = {901},
       number = {1},
          eid = {25},
        pages = {25},
          doi = {10.3847/1538-4357/abab95},
archivePrefix = {arXiv},
       eprint = {1911.05506},
 primaryClass = {astro-ph.SR},
       adsurl = {https://ui.adsabs.harvard.edu/abs/2020ApJ...901...25D}
}

@ARTICLE{Miranda2017,
       author = {{Miranda}, Ryan and {Mu{\~n}oz}, Diego J. and {Lai}, Dong},
        title = "{Viscous hydrodynamics simulations of circumbinary accretion discs: variability, quasi-steady state and angular momentum transfer}",
      journal = {\mnras},
         year = 2017,
        month = apr,
       volume = {466},
       number = {1},
        pages = {1170-1191},
          doi = {10.1093/mnras/stw3189},
archivePrefix = {arXiv},
       eprint = {1610.07263},
 primaryClass = {astro-ph.SR},
       adsurl = {https://ui.adsabs.harvard.edu/abs/2017MNRAS.466.1170M}
}

@ARTICLE{Munoz2019,
       author = {{Mu{\~n}oz}, Diego J. and {Miranda}, Ryan and {Lai}, Dong},
        title = "{Hydrodynamics of Circumbinary Accretion: Angular Momentum Transfer and Binary Orbital Evolution}",
      journal = {\apj},
         year = 2019,
        month = jan,
       volume = {871},
       number = {1},
          eid = {84},
        pages = {84},
          doi = {10.3847/1538-4357/aaf867},
archivePrefix = {arXiv},
       eprint = {1810.04676},
 primaryClass = {astro-ph.HE},
       adsurl = {https://ui.adsabs.harvard.edu/abs/2019ApJ...871...84M}
}

@ARTICLE{Shi2012,
       author = {{Shi}, Ji-Ming and {Krolik}, Julian H. and {Lubow}, Stephen H. and et al.},
        title = "{Three-dimensional Magnetohydrodynamic Simulations of Circumbinary Accretion Disks: Disk Structures and Angular Momentum Transport}",
      journal = {\apj},
         year = 2012,
        month = apr,
       volume = {749},
       number = {2},
          eid = {118},
        pages = {118},
          doi = {10.1088/0004-637X/749/2/118},
archivePrefix = {arXiv},
       eprint = {1110.4866},
 primaryClass = {astro-ph.HE},
       adsurl = {https://ui.adsabs.harvard.edu/abs/2012ApJ...749..118S}
}

@ARTICLE{Kashi2011,
       author = {{Kashi}, Amit and {Soker}, Noam},
        title = "{A circumbinary disc in the final stages of common envelope and the core-degenerate scenario for Type Ia supernovae}",
      journal = {\mnras},
         year = 2011,
        month = oct,
       volume = {417},
       number = {2},
        pages = {1466-1479},
          doi = {10.1111/j.1365-2966.2011.19361.x},
archivePrefix = {arXiv},
       eprint = {1105.5698},
 primaryClass = {astro-ph.SR},
       adsurl = {https://ui.adsabs.harvard.edu/abs/2011MNRAS.417.1466K}
}

@ARTICLE{LiJiangdan2025,
       author = {{Li}, Jiangdan and {Wolf}, Christian and {Li}, Jiao and {Luo}, Yangping and {Zhao}, Jingkun and {Chen}, Bingqiu and {Zhang}, Lin and {Jia}, Shi and {Chen}, Xuefei and {Han}, Zhanwen},
        title = "{Detection of ubiquitous circumbinary matter in hot subdwarfs formed from common-envelope ejections}",
      journal = {\mnras},
         year = 2025,
        month = feb,
       volume = {537},
       number = {2},
        pages = {1950-1962},
          doi = {10.1093/mnras/staf165},
archivePrefix = {arXiv},
       eprint = {2502.00822},
 primaryClass = {astro-ph.SR},
       adsurl = {https://ui.adsabs.harvard.edu/abs/2025MNRAS.537.1950L}
}

@ARTICLE{LiJiangdan2022,
       author = {{Li}, Jiangdan and {Onken}, Christopher A. and {Wolf}, Christian and et al.},
        title = "{A Roche lobe-filling hot subdwarf and white dwarf binary: possible detection of an ejected common envelope}",
      journal = {\mnras},
         year = 2022,
        month = sep,
       volume = {515},
       number = {3},
        pages = {3370-3382},
          doi = {10.1093/mnras/stac1768},
archivePrefix = {arXiv},
       eprint = {2208.01253},
 primaryClass = {astro-ph.SR},
       adsurl = {https://ui.adsabs.harvard.edu/abs/2022MNRAS.515.3370L}
}

@ARTICLE{Reed2011,
       author = {{Reed}, M.~D. and {Baran}, A. and {Quint}, A.~C. and et al.},
        title = "{First Kepler results on compact pulsators - VIII. Mode identifications via period spacings in g-mode pulsating subdwarf B stars}",
      journal = {\mnras},
         year = 2011,
        month = jul,
       volume = {414},
       number = {4},
        pages = {2885-2892},
          doi = {10.1111/j.1365-2966.2011.18532.x},
archivePrefix = {arXiv},
       eprint = {1102.4286},
 primaryClass = {astro-ph.SR},
       adsurl = {https://ui.adsabs.harvard.edu/abs/2011MNRAS.414.2885R}
}

@ARTICLE{Borucki2010,
       author = {{Borucki}, William J. and {Koch}, David and {Basri}, Gibor and {Batalha}, Natalie and {Brown}, Timothy and {Caldwell}, Douglas and {Caldwell}, John and {Christensen-Dalsgaard}, J{\o}rgen and {Cochran}, William D. and {DeVore}, Edna and {Dunham}, Edward W. and {Dupree}, Andrea K. and {Gautier}, Thomas N. and {Geary}, John C. and {Gilliland}, Ronald and {Gould}, Alan and {Howell}, Steve B. and {Jenkins}, Jon M. and {Kondo}, Yoji and {Latham}, David W. and {Marcy}, Geoffrey W. and {Meibom}, S{\o}ren and {Kjeldsen}, Hans and {Lissauer}, Jack J. and {Monet}, David G. and {Morrison}, David and {Sasselov}, Dimitar and {Tarter}, Jill and {Boss}, Alan and {Brownlee}, Don and {Owen}, Toby and {Buzasi}, Derek and {Charbonneau}, David and {Doyle}, Laurance and {Fortney}, Jonathan and {Ford}, Eric B. and {Holman}, Matthew J. and {Seager}, Sara and {Steffen}, Jason H. and {Welsh}, William F. and {Rowe}, Jason and {Anderson}, Howard and {Buchhave}, Lars and {Ciardi}, David and {Walkowicz}, Lucianne and {Sherry}, William and {Horch}, Elliott and {Isaacson}, Howard and {Everett}, Mark E. and {Fischer}, Debra and {Torres}, Guillermo and {Johnson}, John Asher and {Endl}, Michael and {MacQueen}, Phillip and {Bryson}, Stephen T. and {Dotson}, Jessie and {Haas}, Michael and {Kolodziejczak}, Jeffrey and {Van Cleve}, Jeffrey and {Chandrasekaran}, Hema and {Twicken}, Joseph D. and {Quintana}, Elisa V. and {Clarke}, Bruce D. and {Allen}, Christopher and {Li}, Jie and {Wu}, Haley and {Tenenbaum}, Peter and {Verner}, Ekaterina and {Bruhweiler}, Frederick and {Barnes}, Jason and {Prsa}, Andrej},
        title = "{Kepler Planet-Detection Mission: Introduction and First Results}",
      journal = {Science},
         year = 2010,
        month = feb,
       volume = {327},
       number = {5968},
        pages = {977},
          doi = {10.1126/science.1185402},
       adsurl = {https://ui.adsabs.harvard.edu/abs/2010Sci...327..977B}
}

@ARTICLE{Ricker2015,
       author = {{Ricker}, George R. and {Winn}, Joshua N. and {Vanderspek}, Roland and {Latham}, David W. and {Bakos}, G{\'a}sp{\'a}r {\'A}. and {Bean}, Jacob L. and {Berta-Thompson}, Zachory K. and {Brown}, Timothy M. and {Buchhave}, Lars and {Butler}, Nathaniel R. and {Butler}, R. Paul and {Chaplin}, William J. and {Charbonneau}, David and {Christensen-Dalsgaard}, J{\o}rgen and {Clampin}, Mark and {Deming}, Drake and {Doty}, John and {De Lee}, Nathan and {Dressing}, Courtney and {Dunham}, Edward W. and {Endl}, Michael and {Fressin}, Francois and {Ge}, Jian and {Henning}, Thomas and {Holman}, Matthew J. and {Howard}, Andrew W. and {Ida}, Shigeru and {Jenkins}, Jon M. and {Jernigan}, Garrett and {Johnson}, John Asher and {Kaltenegger}, Lisa and {Kawai}, Nobuyuki and {Kjeldsen}, Hans and {Laughlin}, Gregory and {Levine}, Alan M. and {Lin}, Douglas and {Lissauer}, Jack J. and {MacQueen}, Phillip and {Marcy}, Geoffrey and {McCullough}, Peter R. and {Morton}, Timothy D. and {Narita}, Norio and {Paegert}, Martin and {Palle}, Enric and {Pepe}, Francesco and {Pepper}, Joshua and {Quirrenbach}, Andreas and {Rinehart}, Stephen A. and {Sasselov}, Dimitar and {Sato}, Bun'ei and {Seager}, Sara and {Sozzetti}, Alessandro and {Stassun}, Keivan G. and {Sullivan}, Peter and {Szentgyorgyi}, Andrew and {Torres}, Guillermo and {Udry}, Stephane and {Villasenor}, Joel},
        title = "{Transiting Exoplanet Survey Satellite (TESS)}",
      journal = {Journal of Astronomical Telescopes, Instruments, and Systems},
         year = 2015,
        month = jan,
       volume = {1},
          eid = {014003},
        pages = {014003},
          doi = {10.1117/1.JATIS.1.1.014003},
       adsurl = {https://ui.adsabs.harvard.edu/abs/2015JATIS...1a4003R}
}

@ARTICLE{Uzundag2024,
       author = {{Uzundag}, Murat and {Krzesinski}, Jurek and {Pelisoli}, Ingrid and et al.},
        title = "{A comprehensive search for hot subdwarf stars using Gaia and TESS. I. Pulsating hot subdwarf B stars}",
      journal = {\aap},
         year = 2024,
        month = apr,
       volume = {684},
          eid = {A118},
        pages = {A118},
          doi = {10.1051/0004-6361/202348829},
archivePrefix = {arXiv},
       eprint = {2401.17707},
 primaryClass = {astro-ph.SR},
       adsurl = {https://ui.adsabs.harvard.edu/abs/2024A&A...684A.118U}
}

@ARTICLE{Holdsworth2017,
       author = {{Holdsworth}, Daniel L. and {{\O}stensen}, Roy H. and {Smalley}, Barry and et al.},
        title = "{Three new pulsating sdB stars discovered with SuperWASP}",
      journal = {\mnras},
         year = 2017,
        month = apr,
       volume = {466},
       number = {4},
        pages = {5020-5032},
          doi = {10.1093/mnras/stx077},
archivePrefix = {arXiv},
       eprint = {1701.02761},
 primaryClass = {astro-ph.SR},
       adsurl = {https://ui.adsabs.harvard.edu/abs/2017MNRAS.466.5020H}
}

@ARTICLE{Ostensen2010,
       author = {{{\O}stensen}, R.~H. and {Oreiro}, R. and {Solheim}, J.-E. and et al.},
        title = "{A survey for pulsating subdwarf B stars with the Nordic Optical Telescope}",
      journal = {\aap},
         year = 2010,
        month = apr,
       volume = {513},
          eid = {A6},
        pages = {A6},
          doi = {10.1051/0004-6361/200913480},
archivePrefix = {arXiv},
       eprint = {1001.3657},
 primaryClass = {astro-ph.SR},
       adsurl = {https://ui.adsabs.harvard.edu/abs/2010A&A...513A...6O}
}

@ARTICLE{Takata2025,
       author = {{Takata}, Masao and {Murphy}, Simon J. and {Kurtz}, Donald W. and et al.},
        title = "{Asteroseismic detection of a predominantly toroidal magnetic field in the deep interior of the main-sequence F star KIC 9244992}",
      journal = {\mnras},
         year = 2025,
        month = dec,
          doi = {10.1093/mnras/staf2153},
archivePrefix = {arXiv},
       eprint = {2512.00786},
 primaryClass = {astro-ph.SR},
       adsurl = {https://ui.adsabs.harvard.edu/abs/2025MNRAS.tmp.2020T}
}

@ARTICLE{Arancibia-Rojas2024,
       author = {{Arancibia-Rojas}, Eduardo and {Zorotovic}, Monica and {Vu{\v{c}}kovi{\'c}}, Maja and et al.},
        title = "{The mass range of hot subdwarf B stars from MESA simulations}",
      journal = {\mnras},
         year = 2024,
        month = feb,
       volume = {527},
       number = {4},
        pages = {11184-11197},
          doi = {10.1093/mnras/stad3891},
archivePrefix = {arXiv},
       eprint = {2312.09920},
 primaryClass = {astro-ph.SR},
       adsurl = {https://ui.adsabs.harvard.edu/abs/2024MNRAS.52711184A}
}

@ARTICLE{Grosjean2014,
       author = {{Grosjean}, M. and {Dupret}, M.-A. and {Belkacem}, K. and et al.},
        title = "{Theoretical power spectra of mixed modes in low-mass red giant stars}",
      journal = {\aap},
         year = 2014,
        month = dec,
       volume = {572},
          eid = {A11},
        pages = {A11},
          doi = {10.1051/0004-6361/201423827},
archivePrefix = {arXiv},
       eprint = {1409.6121},
 primaryClass = {astro-ph.SR},
       adsurl = {https://ui.adsabs.harvard.edu/abs/2014A&A...572A..11G}
}

@ARTICLE{Clausen2012,
       author = {{Clausen}, Drew and {Wade}, Richard A. and {Kopparapu}, Ravi Kumar and et al.},
        title = "{Population Synthesis of Hot Subdwarfs: A Parameter Study}",
      journal = {\apj},
         year = 2012,
        month = feb,
       volume = {746},
       number = {2},
          eid = {186},
        pages = {186},
          doi = {10.1088/0004-637X/746/2/186},
archivePrefix = {arXiv},
       eprint = {1201.0012},
 primaryClass = {astro-ph.SR},
       adsurl = {https://ui.adsabs.harvard.edu/abs/2012ApJ...746..186C}
}

@ARTICLE{Nelemans2010,
       author = {{Nelemans}, G.},
        title = "{Population synthesis of Galactic subdwarf B stars}",
      journal = {\apss},
         year = 2010,
        month = oct,
       volume = {329},
       number = {1-2},
        pages = {25-31},
          doi = {10.1007/s10509-010-0392-0},
archivePrefix = {arXiv},
       eprint = {1302.0135},
 primaryClass = {astro-ph.SR},
       adsurl = {https://ui.adsabs.harvard.edu/abs/2010Ap&SS.329...25N}
}

@ARTICLE{Rodriguez-Segovia2025b,
       author = {{Rodr{\'\i}guez-Segovia}, Nicol{\'a}s and {Ruiter}, Ashley J. and {Seitenzahl}, Ivo R.},
        title = "{Population synthesis of hot-subdwarf B stars with COMPAS: Parameter variations and a prescription for hydrogen-rich shells}",
      journal = {\pasa},
         year = 2025,
        month = dec,
       volume = {42},
          eid = {e012},
        pages = {e012},
          doi = {10.1017/pasa.2024.135},
archivePrefix = {arXiv},
       eprint = {2412.11712},
 primaryClass = {astro-ph.SR},
       adsurl = {https://ui.adsabs.harvard.edu/abs/2025PASA...42...12R}
}

@ARTICLE{Rodriguez-Segovia2025a,
       author = {{Rodr{\'\i}guez-Segovia}, Nicol{\'a}s and {Ruiter}, Ashley J.},
        title = "{Population synthesis of hot subdwarf B stars with COMPAS: on the observed Galactic population}",
      journal = {\mnras},
         year = 2025,
        month = jun,
       volume = {539},
       number = {4},
        pages = {3273-3284},
          doi = {10.1093/mnras/staf710},
archivePrefix = {arXiv},
       eprint = {2505.05791},
 primaryClass = {astro-ph.SR},
       adsurl = {https://ui.adsabs.harvard.edu/abs/2025MNRAS.539.3273R}
}

@ARTICLE{Ge2022,
       author = {{Ge}, Hongwei and {Tout}, Christopher A. and {Chen}, Xuefei and et al.},
        title = "{The Common Envelope Evolution Outcome-A Case Study on Hot Subdwarf B Stars}",
      journal = {\apj},
         year = 2022,
        month = jul,
       volume = {933},
       number = {2},
          eid = {137},
        pages = {137},
          doi = {10.3847/1538-4357/ac75d3},
archivePrefix = {arXiv},
       eprint = {2205.14256},
 primaryClass = {astro-ph.SR},
       adsurl = {https://ui.adsabs.harvard.edu/abs/2022ApJ...933..137G}
}

@ARTICLE{Ge2024,
       author = {{Ge}, Hongwei and {Tout}, Christopher A. and {Webbink}, Ronald F. and et al.},
        title = "{The Common Envelope Evolution Outcome. II. Short-orbital-period Hot Subdwarf B Binaries Reveal a Clear Picture}",
      journal = {\apj},
         year = 2024,
        month = feb,
       volume = {961},
       number = {2},
          eid = {202},
        pages = {202},
          doi = {10.3847/1538-4357/ad158e},
archivePrefix = {arXiv},
       eprint = {2311.17304},
 primaryClass = {astro-ph.SR},
       adsurl = {https://ui.adsabs.harvard.edu/abs/2024ApJ...961..202G}
}

@ARTICLE{Saio2021,
       author = {{Saio}, Hideyuki and {Takata}, Masao and {Lee}, Umin and et al.},
        title = "{Rotation of the convective core in {\ensuremath{\gamma}} Dor stars measured by dips in period spacings of g modes coupled with inertial modes}",
      journal = {\mnras},
         year = 2021,
        month = apr,
       volume = {502},
       number = {4},
        pages = {5856-5874},
          doi = {10.1093/mnras/stab482},
archivePrefix = {arXiv},
       eprint = {2102.08548},
 primaryClass = {astro-ph.SR},
       adsurl = {https://ui.adsabs.harvard.edu/abs/2021MNRAS.502.5856S}
}

@ARTICLE{Santos2021,
       author = {{Santos}, A.~R.~G. and {Breton}, S.~N. and {Mathur}, S. and {Garc{\'\i}a}, R.~A.},
        title = "{Surface Rotation and Photometric Activity for Kepler Targets. II. G and F Main-sequence Stars and Cool Subgiant Stars}",
      journal = {\apjs},
         year = 2021,
        month = jul,
       volume = {255},
       number = {1},
          eid = {17},
        pages = {17},
          doi = {10.3847/1538-4365/ac033f},
archivePrefix = {arXiv},
       eprint = {2107.02217},
 primaryClass = {astro-ph.SR},
       adsurl = {https://ui.adsabs.harvard.edu/abs/2021ApJS..255...17S}
}

@article{Charpinet2008,
   author = {S. Charpinet and V. Van Grootel and D. Reese and G. Fontaine and E. M. Green and P. Brassard and P. Chayer},
   doi = {10.1051/0004-6361:200809907},
   issn = {00046361},
   issue = {1},
   journal = {Astronomy and Astrophysics},
   pages = {377-394},
   title = {Testing the forward modeling approach in asteroseismology II. Structure and internal dynamics of the hot B subdwarf component in the close eclipsing binary system PG 1336-018},
   volume = {489},
   year = {2008}
}

@article{Pablo2011,
   author = {Herbert Pablo and Steven D. Kawaler and Elizabeth M. Green},
   doi = {10.1088/2041-8205/740/2/L47},
   issn = {20418205},
   issue = {2},
   journal = {Astrophysical Journal Letters},
   pages = {4-10},
   title = {Exploring B4: A pulsating sdB star, in a binary, in the open cluster NGC 6791},
   volume = {740},
   year = {2011}
}

@article{Pablo2012,
   author = {Herbert Pablo and Steven D. Kawaler and M. D. Reed and S. Bloemen and S. Charpinet and H. Hu and J. Telting and R. H. Østensen and A. S. Baran and E. M. Green and J. J. Hermes and T. Barclay and S. J. O'Toole and Fergal Mullally and D. W. Kurtz and J. Christensen-Dalsgaard and Douglas A. Caldwell and Jessie L. Christiansen and K. Kinemuchi},
   doi = {10.1111/j.1365-2966.2012.20707.x},
   issn = {00358711},
   issue = {2},
   journal = {Monthly Notices of the Royal Astronomical Society},
   pages = {1343-1351},
   title = {Seismic evidence for non-synchronization in two close sdb+dM binaries from Kepler photometry},
   volume = {422},
   year = {2012}
}

@article{Baran2016,
   author = {A. S. Baran and J. H. Telting and P. Németh and R. H. Østensen and M. D. Reed and F. Kiaeerad},
   doi = {10.1051/0004-6361/201527182},
   issn = {14320746},
   journal = {Astronomy and Astrophysics},
   title = {A subsynchronously rotating pulsating subdwarf B star in a short-period binary with a white dwarf companion},
   volume = {585},
   year = {2016}
}

@article{Telting2012,
   author = {J. H. Telting and R. H. Østensen and A. S. Baran and S. Bloemen and M. D. Reed and R. Oreiro and L. Farris and T. A. Ottosen and C. Aerts and S. D. Kawaler and U. Heber and S. Prins and E. M. Green and B. Kalomeni and S. J. O'Toole and F. Mullally and D. T. Sanderfer and J. C. Smith and H. Kjeldsen},
   doi = {10.1051/0004-6361/201219458},
   issn = {00046361},
   journal = {Astronomy and Astrophysics},
   pages = {1-16},
   title = {Three ways to solve the orbit of KIC 11558725: A 10-day beaming sdB+WD binary with a pulsating subdwarf},
   volume = {544},
   year = {2012}
}

@article{Telting2014,
   author = {J. H. Telting and A. S. Baran and P. Nemeth and R. H. Østensen and T. Kupfer and S. MacFarlane and U. Heber and C. Aerts and S. Geier},
   doi = {10.1051/0004-6361/201424169},
   issn = {14320746},
   journal = {Astronomy and Astrophysics},
   pages = {1-18},
   title = {KIC 7668647: A 14 day beaming sdB+WD binary with a pulsating subdwarf},
   volume = {570},
   year = {2014}
}

@article{Foster2015,
   author = {H. M. Foster and M. D. Reed and J. H. Telting and R. H. Østensen and A. S. Baran},
   doi = {10.1088/0004-637X/805/2/94},
   issn = {15384357},
   issue = {2},
   journal = {Astrophysical Journal},
   pages = {1-12},
   publisher = {IOP Publishing},
   title = {The discovery of differential radial rotation in the pulsating subdwarf B star KIC 3527751},
   volume = {805},
   url = {http://dx.doi.org/10.1088/0004-637X/805/2/94},
   year = {2015}
}

@article{Krzesinski2015,
   author = {J. Krzesinski},
   doi = {10.1051/0004-6361/201526346},
   issn = {14320746},
   journal = {Astronomy and Astrophysics},
   pages = {1-7},
   title = {Planetary candidates around the pulsating sdB star KIC 5807616 considered doubtful},
   volume = {581},
   year = {2015}
}

@article{VanGrootel2010,
doi = {10.1088/2041-8205/718/2/L97},
url = {https://doi.org/10.1088/2041-8205/718/2/L97},
year = {2010},
month = {jul},
publisher = {The American Astronomical Society},
volume = {718},
number = {2},
pages = {L97},
author = {Van Grootel, V. and Charpinet, S. and Fontaine, G. and Brassard, P. and Green, E. M. and Randall, S. K. and Silvotti, R. and Østensen, R. H. and Kjeldsen, H. and Christensen-Dalsgaard, J. and Borucki, W. J. and Koch, D.},
title = {EARLY ASTEROSEISMIC RESULTS FROM KEPLER: STRUCTURAL AND CORE PARAMETERS OF THE HOT B SUBDWARF KPD 1943+4058 AS INFERRED FROM g-MODE OSCILLATIONS},
journal = {The Astrophysical Journal Letters}
}

@ARTICLE{Baran2012a,
       author = {{Baran}, A.~S. and {Reed}, M.~D. and {Stello}, D. and et al.},
        title = "{A pulsation zoo in the hot subdwarf B star KIC 10139564 observed by Kepler}",
      journal = {\mnras},
         year = 2012,
        month = aug,
       volume = {424},
       number = {4},
        pages = {2686-2700},
          doi = {10.1111/j.1365-2966.2012.21355.x},
archivePrefix = {arXiv},
       eprint = {1206.3841},
 primaryClass = {astro-ph.SR},
       adsurl = {https://ui.adsabs.harvard.edu/abs/2012MNRAS.424.2686B}
}

@article{Reed2014,
   author = {M. D. Reed and H. Foster and J. H. Telting and R. H. Østensen and L. H. Farris and R. Oreiro and A. S. Baran},
   doi = {10.1093/mnras/stu412},
   issn = {13652966},
   issue = {4},
   journal = {Monthly Notices of the Royal Astronomical Society},
   pages = {3809-3824},
   title = {Analysis of the rich frequency spectrum of KIC 10670103 revealing the most slowly rotating subdwarf B star in the kepler field},
   volume = {440},
   year = {2014}
}

@article{Kern2017,
   author = {J. W. Kern and M. D. Reed and A. S. Baran and R. H. Østensen and J. H. Telting},
   doi = {10.1093/mnras/stw2794},
   issn = {13652966},
   issue = {1},
   journal = {Monthly Notices of the Royal Astronomical Society},
   pages = {1057-1065},
   title = {Kepler observations of the pulsating subdwarf B star KIC 2697388: The detection of converging frequency multiplets in the full data set},
   volume = {465},
   year = {2017}
}

@article{Baran2017,
   author = {A. S. Baran and M. D. Reed and R. H. Østensen and J. H. Telting and C. S. Jeffery},
   doi = {10.1051/0004-6361/201629651},
   issn = {14320746},
   journal = {Astronomy and Astrophysics},
   pages = {1-12},
   title = {EPIC 211779126: A rare hybrid pulsating subdwarf B star richly pulsating in both pressure and gravity modes},
   volume = {597},
   year = {2017}
}

@article{Reed2023,
   author = {M. D. Reed and A. S. Baran and J. H. Telting and R. H. Østensen},
   doi = {10.1093/mnras/stad2392},
   issn = {13652966},
   issue = {1},
   journal = {Monthly Notices of the Royal Astronomical Society},
   pages = {1342-1352},
   title = {TESS photometry of the pulsating hot subdwarf star V585 Peg},
   volume = {525},
   year = {2023}
}

@article{Baran2019,
   author = {A. S. Baran and J. H. Telting and C. S. Jeffery and R. H. Østensen and J. Vos and M. D. Reed and M. Vǔcković},
   doi = {10.1093/mnras/stz2209},
   issn = {13652966},
   issue = {2},
   journal = {Monthly Notices of the Royal Astronomical Society},
   pages = {1556-1571},
   title = {K2 observations of the sdBV + dM/bd binaries PHL457 and EQPsc},
   volume = {489},
   year = {2019}
}

@article{Latour2019,
   author = {M. Latour and E. M. Green and G. Fontaine},
   doi = {10.1051/0004-6361/201935307},
   issn = {14320746},
   journal = {Astronomy and Astrophysics},
   pages = {10-15},
   title = {Discovery of a second pulsating intermediate helium-enriched sdOB star},
   volume = {623},
   year = {2019}
}

@article{Silvotti2022,
   author = {Roberto Silvotti and Péter Németh and John H. Telting and Andrzej S. Baran and Roy H. Østensen and Jakub Ostrowski and Sumanta K. Sahoo and Saskia Prins},
   doi = {10.1093/mnras/stac160},
   issn = {13652966},
   issue = {2},
   journal = {Monthly Notices of the Royal Astronomical Society},
   pages = {2201-2217},
   title = {Filling the gap between synchronized and non-synchronized sdBs in short-period sdBV+dM binaries with TESS: TIC 137608661, a new system with a well-defined rotational splitting},
   volume = {511},
   year = {2022}
}

@ARTICLE{Uzundag2023,
       author = {{Uzundag}, Murat and {Silvotti}, Roberto and {Baran}, Andrzej S. and et al.},
        title = "{Asteroseismology of hot subdwarf B stars observed with TESS: discovery of two new gravity mode pulsating stars}",
      journal = {Bulletin de la Societe Royale des Sciences de Liege},
         year = 2023,
        month = jan,
       volume = {92},
       number = {2},
          eid = {11294},
        pages = {11294},
          doi = {10.25518/0037-9565.11294},
archivePrefix = {arXiv},
       eprint = {2212.02950},
 primaryClass = {astro-ph.SR},
       adsurl = {https://ui.adsabs.harvard.edu/abs/2023BSRSL..9211294U}
}

@article{Baran2012b,
   author = {A. S. Baran and A. Winans},
   issn = {00015237},
   issue = {4},
   journal = {Acta Astronomica},
   pages = {343-355},
   title = {Mode identification of three pulsating subdwarf B stars via multiplets and period spacing},
   volume = {62},
   year = {2012}
}

@article{Reed2020,
   author = {M. D. Reed and M. Yeager and J. Vos and J. H. Telting and R. H. Østensen and A. Slayton and A. S. Baran and C. S. Jeffery},
   doi = {10.1093/MNRAS/STAA144},
   issn = {13652966},
   issue = {4},
   journal = {Monthly Notices of the Royal Astronomical Society},
   pages = {5202-5217},
   publisher = {Oxford University Press},
   title = {K2 observations of the pulsating subdwarf B stars UY Sex and V1405 Ori},
   volume = {492},
   year = {2020}
}

@article{Reed2021,
   author = {M. D. Reed and A. Slayton and A. S. Baran and J. H. Telting and R. H. Østensen and C. S. Jeffery and M. Uzundag and S. Sanjayan},
   doi = {10.1093/mnras/stab2405},
   issn = {13652966},
   issue = {3},
   journal = {Monthly Notices of the Royal Astronomical Society},
   month = {11},
   pages = {4178-4195},
   publisher = {Oxford University Press},
   title = {Pulsating subdwarf B stars observed with K2 during Campaign 7 and an examination of seismic group properties},
   volume = {507},
   year = {2021}
}

@article{Kern2018,
   author = {J. W. Kern and M. D. Reed and A. S. Baran and J. H. Telting and R. H. Østensen},
   doi = {10.1093/mnras/stx2893},
   issn = {13652966},
   issue = {4},
   journal = {Monthly Notices of the Royal Astronomical Society},
   month = {3},
   pages = {4709-4716},
   publisher = {Oxford University Press},
   title = {Asteroseismic analysis of the pulsating subdwarf B star KIC 11558725: An sdB+WD system with divergent frequency multiplets and mode trapping observed by Kepler},
   volume = {474},
   year = {2018}
}

@article{Pelisoli2021,
   author = {Ingrid Pelisoli and P. Neunteufel and S. Geier and T. Kupfer and U. Heber and A. Irrgang and D. Schneider and A. Bastian and J. van Roestel and V. Schaffenroth and B. N. Barlow},
   doi = {10.1038/s41550-021-01413-0},
   issn = {23973366},
   issue = {10},
   journal = {Nature Astronomy},
   month = {10},
   pages = {1052-1061},
   publisher = {Nature Research},
   title = {A hot subdwarf–white dwarf super-Chandrasekhar candidate supernova Ia progenitor},
   volume = {5},
   year = {2021}
}

@ARTICLE{VanGrootel2008,
       author = {{Van Grootel}, V. and {Charpinet}, S. and {Fontaine}, G. and et al.},
        title = "{Asteroseismology in action: a test of spin-orbit synchronism in the close binary system Feige 48}",
      journal = {\aap},
         year = 2008,
        month = jun,
       volume = {483},
       number = {3},
        pages = {875-885},
          doi = {10.1051/0004-6361:200809554},
       adsurl = {https://ui.adsabs.harvard.edu/abs/2008A&A...483..875V}
}

@ARTICLE{Paxton2011,
       author = {{Paxton}, Bill and {Bildsten}, Lars and {Dotter}, Aaron and et al.},
        title = "{Modules for Experiments in Stellar Astrophysics (MESA)}",
      journal = {\apjs},
         year = 2011,
        month = jan,
       volume = {192},
       number = {1},
          eid = {3},
        pages = {3},
          doi = {10.1088/0067-0049/192/1/3},
archivePrefix = {arXiv},
       eprint = {1009.1622},
 primaryClass = {astro-ph.SR},
       adsurl = {https://ui.adsabs.harvard.edu/abs/2011ApJS..192....3P}
}

@ARTICLE{Paxton2015,
       author = {{Paxton}, Bill and {Marchant}, Pablo and {Schwab}, Josiah and et al.},
        title = "{Modules for Experiments in Stellar Astrophysics (MESA): Binaries, Pulsations, and Explosions}",
      journal = {\apjs},
         year = 2015,
        month = sep,
       volume = {220},
       number = {1},
          eid = {15},
        pages = {15},
          doi = {10.1088/0067-0049/220/1/15},
archivePrefix = {arXiv},
       eprint = {1506.03146},
 primaryClass = {astro-ph.SR},
       adsurl = {https://ui.adsabs.harvard.edu/abs/2015ApJS..220...15P}
}

@ARTICLE{Paxton2013,
       author = {{Paxton}, Bill and {Cantiello}, Matteo and {Arras}, Phil and et al.},
        title = "{Modules for Experiments in Stellar Astrophysics (MESA): Planets, Oscillations, Rotation, and Massive Stars}",
      journal = {\apjs},
         year = 2013,
        month = sep,
       volume = {208},
       number = {1},
          eid = {4},
        pages = {4},
          doi = {10.1088/0067-0049/208/1/4},
archivePrefix = {arXiv},
       eprint = {1301.0319},
 primaryClass = {astro-ph.SR},
       adsurl = {https://ui.adsabs.harvard.edu/abs/2013ApJS..208....4P}
}

@ARTICLE{Paxton2018,
       author = {{Paxton}, Bill and {Schwab}, Josiah and {Bauer}, Evan B. and et al.},
        title = "{Modules for Experiments in Stellar Astrophysics (MESA): Convective Boundaries, Element Diffusion, and Massive Star Explosions}",
      journal = {\apjs},
         year = 2018,
        month = feb,
       volume = {234},
       number = {2},
          eid = {34},
        pages = {34},
          doi = {10.3847/1538-4365/aaa5a8},
archivePrefix = {arXiv},
       eprint = {1710.08424},
 primaryClass = {astro-ph.SR},
       adsurl = {https://ui.adsabs.harvard.edu/abs/2018ApJS..234...34P}
}

@ARTICLE{Jermyn2023,
       author = {{Jermyn}, Adam S. and {Bauer}, Evan B. and {Schwab}, Josiah and et al.},
        title = "{Modules for Experiments in Stellar Astrophysics (MESA): Time-dependent Convection, Energy Conservation, Automatic Differentiation, and Infrastructure}",
      journal = {\apjs},
         year = 2023,
        month = mar,
       volume = {265},
       number = {1},
          eid = {15},
        pages = {15},
          doi = {10.3847/1538-4365/acae8d},
archivePrefix = {arXiv},
       eprint = {2208.03651},
 primaryClass = {astro-ph.SR},
       adsurl = {https://ui.adsabs.harvard.edu/abs/2023ApJS..265...15J}
}

@ARTICLE{Heger2000,
       author = {{Heger}, A. and {Langer}, N. and {Woosley}, S.~E.},
        title = "{Presupernova Evolution of Rotating Massive Stars. I. Numerical Method and Evolution of the Internal Stellar Structure}",
      journal = {\apj},
         year = 2000,
        month = jan,
       volume = {528},
       number = {1},
        pages = {368-396},
          doi = {10.1086/308158},
archivePrefix = {arXiv},
       eprint = {astro-ph/9904132},
 primaryClass = {astro-ph},
       adsurl = {https://ui.adsabs.harvard.edu/abs/2000ApJ...528..368H}
}

@ARTICLE{Han2003,
       author = {{Han}, Z. and {Podsiadlowski}, Ph. and {Maxted}, P.~F.~L. and et al.},
        title = "{The origin of subdwarf B stars - II}",
      journal = {\mnras},
         year = 2003,
        month = may,
       volume = {341},
       number = {2},
        pages = {669-691},
          doi = {10.1046/j.1365-8711.2003.06451.x},
archivePrefix = {arXiv},
       eprint = {astro-ph/0301380},
 primaryClass = {astro-ph},
       adsurl = {https://ui.adsabs.harvard.edu/abs/2003MNRAS.341..669H}
}

@ARTICLE{Heber2016,
       author = {{Heber}, U.},
        title = "{Hot Subluminous Stars}",
      journal = {\pasp},
         year = 2016,
        month = aug,
       volume = {128},
       number = {966},
        pages = {082001},
          doi = {10.1088/1538-3873/128/966/082001},
archivePrefix = {arXiv},
       eprint = {1604.07749},
 primaryClass = {astro-ph.SR},
       adsurl = {https://ui.adsabs.harvard.edu/abs/2016PASP..128h2001H}
}

@ARTICLE{Spruit2002,
       author = {{Spruit}, H.~C.},
        title = "{Dynamo action by differential rotation in a stably stratified stellar interior}",
      journal = {\aap},
         year = 2002,
        month = jan,
       volume = {381},
        pages = {923-932},
          doi = {10.1051/0004-6361:20011465},
archivePrefix = {arXiv},
       eprint = {astro-ph/0108207},
 primaryClass = {astro-ph},
       adsurl = {https://ui.adsabs.harvard.edu/abs/2002A&A...381..923S}
}

@ARTICLE{Han2002,
       author = {{Han}, Z. and {Podsiadlowski}, Ph. and {Maxted}, P.~F.~L. and et al.},
        title = "{The origin of subdwarf B stars - I. The formation channels}",
      journal = {\mnras},
         year = 2002,
        month = oct,
       volume = {336},
       number = {2},
        pages = {449-466},
          doi = {10.1046/j.1365-8711.2002.05752.x},
archivePrefix = {arXiv},
       eprint = {astro-ph/0206130},
 primaryClass = {astro-ph},
       adsurl = {https://ui.adsabs.harvard.edu/abs/2002MNRAS.336..449H}
}

@ARTICLE{Aerts2019,
       author = {{Aerts}, Conny and {Mathis}, St{\'e}phane and {Rogers}, Tamara M.},
        title = "{Angular Momentum Transport in Stellar Interiors}",
      journal = {\araa},
         year = 2019,
        month = aug,
       volume = {57},
        pages = {35-78},
          doi = {10.1146/annurev-astro-091918-104359},
archivePrefix = {arXiv},
       eprint = {1809.07779},
 primaryClass = {astro-ph.SR},
       adsurl = {https://ui.adsabs.harvard.edu/abs/2019ARA&A..57...35A}
}

@ARTICLE{Ropke2023,
       author = {{R{\"o}pke}, Friedrich K. and {De Marco}, Orsola},
        title = "{Simulations of common-envelope evolution in binary stellar systems: physical models and numerical techniques}",
      journal = {Living Reviews in Computational Astrophysics},
         year = 2023,
        month = dec,
       volume = {9},
       number = {1},
          eid = {2},
        pages = {2},
          doi = {10.1007/s41115-023-00017-x},
archivePrefix = {arXiv},
       eprint = {2212.07308},
 primaryClass = {astro-ph.SR},
       adsurl = {https://ui.adsabs.harvard.edu/abs/2023LRCA....9....2R}
}

@ARTICLE{Fuller2019,
       author = {{Fuller}, Jim and {Piro}, Anthony L. and {Jermyn}, Adam S.},
        title = "{Slowing the spins of stellar cores}",
      journal = {\mnras},
         year = 2019,
        month = may,
       volume = {485},
       number = {3},
        pages = {3661-3680},
          doi = {10.1093/mnras/stz514},
archivePrefix = {arXiv},
       eprint = {1902.08227},
 primaryClass = {astro-ph.SR},
       adsurl = {https://ui.adsabs.harvard.edu/abs/2019MNRAS.485.3661F}
}

@ARTICLE{Charpinet1997,
       author = {{Charpinet}, S. and {Fontaine}, G. and {Brassard}, P. and et al.},
        title = "{A Driving Mechanism for the Newly Discovered Class of Pulsating Subdwarf B Stars}",
      journal = {\apjl},
         year = 1997,
        month = jul,
       volume = {483},
       number = {2},
        pages = {L123-L126},
          doi = {10.1086/310741},
       adsurl = {https://ui.adsabs.harvard.edu/abs/1997ApJ...483L.123C}
}

@ARTICLE{Li2024,
       author = {{Li}, Gang and {Deheuvels}, S{\'e}bastien and {Ballot}, J{\'e}r{\^o}me},
        title = "{Asteroseismic measurement of core and envelope rotation rates for 2006 red giant branch stars}",
      journal = {\aap},
         year = 2024,
        month = aug,
       volume = {688},
          eid = {A184},
        pages = {A184},
          doi = {10.1051/0004-6361/202449882},
archivePrefix = {arXiv},
       eprint = {2405.12116},
 primaryClass = {astro-ph.SR},
       adsurl = {https://ui.adsabs.harvard.edu/abs/2024A&A...688A.184L}
}

@ARTICLE{Tayler1973,
       author = {{Tayler}, R.~J.},
        title = "{The adiabatic stability of stars containing magnetic fields-I.Toroidal fields}",
      journal = {\mnras},
         year = 1973,
        month = jan,
       volume = {161},
        pages = {365},
          doi = {10.1093/mnras/161.4.365},
       adsurl = {https://ui.adsabs.harvard.edu/abs/1973MNRAS.161..365T}
}

@ARTICLE{Freytag1996,
       author = {{Freytag}, B. and {Ludwig}, H. -G. and {Steffen}, M.},
        title = "{Hydrodynamical models of stellar convection. The role of overshoot in DA white dwarfs, A-type stars, and the Sun.}",
      journal = {\aap},
         year = 1996,
        month = sep,
       volume = {313},
        pages = {497-516},
       adsurl = {https://ui.adsabs.harvard.edu/abs/1996A&A...313..497F}
}

@ARTICLE{Ma2024,
       author = {{Ma}, Linhao and {Fuller}, Jim},
        title = "{Tidal Spin-up of Subdwarf B Stars}",
      journal = {\apj},
         year = 2024,
        month = nov,
       volume = {975},
       number = {1},
          eid = {1},
        pages = {1},
          doi = {10.3847/1538-4357/ad7788},
archivePrefix = {arXiv},
       eprint = {2408.16158},
 primaryClass = {astro-ph.SR},
       adsurl = {https://ui.adsabs.harvard.edu/abs/2024ApJ...975....1M}
}

@ARTICLE{Matt2015,
       author = {{Matt}, Sean P. and {Brun}, A. Sacha and {Baraffe}, Isabelle and et al.},
        title = "{The Mass-dependence of Angular Momentum Evolution in Sun-like Stars}",
      journal = {\apjl},
         year = 2015,
        month = jan,
       volume = {799},
       number = {2},
          eid = {L23},
        pages = {L23},
          doi = {10.1088/2041-8205/799/2/L23},
archivePrefix = {arXiv},
       eprint = {1412.4786},
 primaryClass = {astro-ph.SR},
       adsurl = {https://ui.adsabs.harvard.edu/abs/2015ApJ...799L..23M}
}

@ARTICLE{Herwig1997,
       author = {{Herwig}, F. and {Bloecker}, T. and {Schoenberner}, D. and et al.},
        title = "{Stellar evolution of low and intermediate-mass stars. IV. Hydrodynamically-based overshoot and nucleosynthesis in AGB stars.}",
      journal = {\aap},
         year = 1997,
        month = aug,
       volume = {324},
        pages = {L81-L84},
          doi = {10.48550/arXiv.astro-ph/9706122},
archivePrefix = {arXiv},
       eprint = {astro-ph/9706122},
 primaryClass = {astro-ph},
       adsurl = {https://ui.adsabs.harvard.edu/abs/1997A&A...324L..81H}
}

@ARTICLE{Marques2013,
       author = {{Marques}, J.~P. and {Goupil}, M.~J. and {Lebreton}, Y. and et al.},
        title = "{Seismic diagnostics for transport of angular momentum in stars. I. Rotational splittings from the pre-main sequence to the red-giant branch}",
      journal = {\aap},
         year = 2013,
        month = jan,
       volume = {549},
          eid = {A74},
        pages = {A74},
          doi = {10.1051/0004-6361/201220211},
archivePrefix = {arXiv},
       eprint = {1211.1271},
 primaryClass = {astro-ph.SR},
       adsurl = {https://ui.adsabs.harvard.edu/abs/2013A&A...549A..74M}
}

@ARTICLE{Bordadagua2025,
       author = {{Bordad{\'a}gua}, B. and {Ahlborn}, F. and {Copp{\'e}e}, Q. and et al.},
        title = "{The efficiency of mixed modes for angular momentum transport}",
      journal = {\aap},
         year = 2025,
        month = jul,
       volume = {699},
          eid = {A310},
        pages = {A310},
          doi = {10.1051/0004-6361/202555405},
archivePrefix = {arXiv},
       eprint = {2506.13283},
 primaryClass = {astro-ph.SR},
       adsurl = {https://ui.adsabs.harvard.edu/abs/2025A&A...699A.310B}
}

@ARTICLE{Skoutnev2025,
       author = {{Skoutnev}, Valentin A. and {Beloborodov}, Andrei M.},
        title = "{Magnetic Webs in Stellar Radiative Zones}",
      journal = {\apjl},
         year = 2025,
        month = aug,
       volume = {989},
       number = {1},
          eid = {L4},
        pages = {L4},
          doi = {10.3847/2041-8213/adefda},
archivePrefix = {arXiv},
       eprint = {2504.07223},
 primaryClass = {astro-ph.SR},
       adsurl = {https://ui.adsabs.harvard.edu/abs/2025ApJ...989L...4S}
}

@ARTICLE{Aerts2025,
       author = {{Aerts}, Conny and {Van Reeth}, Timothy and {Mombarg}, Joey S.~G. and et al.},
        title = "{Evolution of the near-core rotation frequency of 2497 intermediate-mass stars from their dominant gravito-inertial mode}",
      journal = {\aap},
         year = 2025,
        month = mar,
       volume = {695},
          eid = {A214},
        pages = {A214},
          doi = {10.1051/0004-6361/202452691},
archivePrefix = {arXiv},
       eprint = {2502.17692},
 primaryClass = {astro-ph.SR},
       adsurl = {https://ui.adsabs.harvard.edu/abs/2025A&A...695A.214A}
}

@ARTICLE{Li2020,
       author = {{Li}, Gang and {Van Reeth}, Timothy and {Bedding}, Timothy R. and et al.},
        title = "{Gravity-mode period spacings and near-core rotation rates of 611 {\ensuremath{\gamma}} Doradus stars with Kepler}",
      journal = {\mnras},
         year = 2020,
        month = jan,
       volume = {491},
       number = {3},
        pages = {3586-3605},
          doi = {10.1093/mnras/stz2906},
archivePrefix = {arXiv},
       eprint = {1910.06634},
 primaryClass = {astro-ph.SR},
       adsurl = {https://ui.adsabs.harvard.edu/abs/2020MNRAS.491.3586L}
}

@ARTICLE{Deheuvels2014,
       author = {{Deheuvels}, S. and {Do{\u{g}}an}, G. and {Goupil}, M.~J. and et al.},
        title = "{Seismic constraints on the radial dependence of the internal rotation profiles of six Kepler subgiants and young red giants}",
      journal = {\aap},
         year = 2014,
        month = apr,
       volume = {564},
          eid = {A27},
        pages = {A27},
          doi = {10.1051/0004-6361/201322779},
archivePrefix = {arXiv},
       eprint = {1401.3096},
 primaryClass = {astro-ph.SR},
       adsurl = {https://ui.adsabs.harvard.edu/abs/2014A&A...564A..27D}
}

@ARTICLE{Rogers2013,
       author = {{Rogers}, T.~M. and {Lin}, D.~N.~C. and {McElwaine}, J.~N. and et al.},
        title = "{Internal Gravity Waves in Massive Stars: Angular Momentum Transport}",
      journal = {\apj},
         year = 2013,
        month = jul,
       volume = {772},
       number = {1},
          eid = {21},
        pages = {21},
          doi = {10.1088/0004-637X/772/1/21},
archivePrefix = {arXiv},
       eprint = {1306.3262},
 primaryClass = {astro-ph.SR},
       adsurl = {https://ui.adsabs.harvard.edu/abs/2013ApJ...772...21R}
}

@ARTICLE{Eggenberger2022,
       author = {{Eggenberger}, P. and {Moyano}, F.~D. and {den Hartogh}, J.~W.},
        title = "{Rotation in stellar interiors: General formulation and an asteroseismic-calibrated transport by the Tayler instability}",
      journal = {\aap},
         year = 2022,
        month = aug,
       volume = {664},
          eid = {L16},
        pages = {L16},
          doi = {10.1051/0004-6361/202243781},
archivePrefix = {arXiv},
       eprint = {2309.17396},
 primaryClass = {astro-ph.SR},
       adsurl = {https://ui.adsabs.harvard.edu/abs/2022A&A...664L..16E}
}

@ARTICLE{Cantiello2014,
       author = {{Cantiello}, Matteo and {Mankovich}, Christopher and {Bildsten}, Lars and et al.},
        title = "{Angular Momentum Transport within Evolved Low-mass Stars}",
      journal = {\apj},
         year = 2014,
        month = jun,
       volume = {788},
       number = {1},
          eid = {93},
        pages = {93},
          doi = {10.1088/0004-637X/788/1/93},
archivePrefix = {arXiv},
       eprint = {1405.1419},
 primaryClass = {astro-ph.SR},
       adsurl = {https://ui.adsabs.harvard.edu/abs/2014ApJ...788...93C}
}

@ARTICLE{Su2024,
       author = {{Su}, Wenchao and {Charpinet}, St{\'e}phane and {Latour}, Marilyn and et al.},
        title = "{TIC 441725813: A new bright hybrid hot B subdwarf pulsator with differential core versus envelope rotation}",
      journal = {\aap},
         year = 2024,
        month = oct,
       volume = {690},
          eid = {A36},
        pages = {A36},
          doi = {10.1051/0004-6361/202450020},
archivePrefix = {arXiv},
       eprint = {2407.17887},
 primaryClass = {astro-ph.SR},
       adsurl = {https://ui.adsabs.harvard.edu/abs/2024A&A...690A..36S}
}

@ARTICLE{Baran2024,
       author = {{Baran}, A.~S. and {Charpinet}, S. and {{\O}stensen}, R.~H. and et al.},
        title = "{Short-period pulsating hot subdwarf stars observed by TESS. II. Northern ecliptic hemisphere}",
      journal = {\aap},
         year = 2024,
        month = jun,
       volume = {686},
          eid = {A65},
        pages = {A65},
          doi = {10.1051/0004-6361/202348571},
archivePrefix = {arXiv},
       eprint = {2403.02384},
 primaryClass = {astro-ph.SR},
       adsurl = {https://ui.adsabs.harvard.edu/abs/2024A&A...686A..65B}
}

@ARTICLE{Eggenberger2012,
       author = {{Eggenberger}, P. and {Montalb{\'a}n}, J. and {Miglio}, A.},
        title = "{Angular momentum transport in stellar interiors constrained by rotational splittings of mixed modes in red giants}",
      journal = {\aap},
         year = 2012,
        month = aug,
       volume = {544},
          eid = {L4},
        pages = {L4},
          doi = {10.1051/0004-6361/201219729},
archivePrefix = {arXiv},
       eprint = {1207.1023},
 primaryClass = {astro-ph.SR},
       adsurl = {https://ui.adsabs.harvard.edu/abs/2012A&A...544L...4E}
}

@ARTICLE{Goupil2013,
       author = {{Goupil}, M.~J. and {Mosser}, B. and {Marques}, J.~P. and et al.},
        title = "{Seismic diagnostics for transport of angular momentum in stars. II. Interpreting observed rotational splittings of slowly rotating red giant stars}",
      journal = {\aap},
         year = 2013,
        month = jan,
       volume = {549},
          eid = {A75},
        pages = {A75},
          doi = {10.1051/0004-6361/201220266},
archivePrefix = {arXiv},
       eprint = {1211.1546},
 primaryClass = {astro-ph.SR},
       adsurl = {https://ui.adsabs.harvard.edu/abs/2013A&A...549A..75G}
}

@ARTICLE{Moyano2024,
       author = {{Moyano}, F.~D. and {Eggenberger}, P. and {Salmon}, S.~J.~A.~J.},
        title = "{Angular momentum transport near convective-core boundaries of Gamma Doradus stars}",
      journal = {\aap},
         year = 2024,
        month = jan,
       volume = {681},
          eid = {L16},
        pages = {L16},
          doi = {10.1051/0004-6361/202348704},
archivePrefix = {arXiv},
       eprint = {2401.05543},
 primaryClass = {astro-ph.SR},
       adsurl = {https://ui.adsabs.harvard.edu/abs/2024A&A...681L..16M}
}

@ARTICLE{Moyano2023b,
       author = {{Moyano}, F.~D. and {Eggenberger}, P. and {Salmon}, S.~J.~A.~J. and et al.},
        title = "{Angular momentum transport by magnetic fields in main-sequence stars with Gamma Doradus pulsators}",
      journal = {\aap},
         year = 2023,
        month = sep,
       volume = {677},
          eid = {A6},
        pages = {A6},
          doi = {10.1051/0004-6361/202346548},
archivePrefix = {arXiv},
       eprint = {2304.00674},
 primaryClass = {astro-ph.SR},
       adsurl = {https://ui.adsabs.harvard.edu/abs/2023A&A...677A...6M}
}

@ARTICLE{Gossage2023,
       author = {{Gossage}, Seth and {Kalogera}, Vicky and {Sun}, Meng},
        title = "{Magnetic Braking with MESA Evolutionary Models in the Single Star and Low-mass X-Ray Binary Regimes}",
      journal = {\apj},
         year = 2023,
        month = jun,
       volume = {950},
       number = {1},
          eid = {27},
        pages = {27},
          doi = {10.3847/1538-4357/acc86e},
archivePrefix = {arXiv},
       eprint = {2212.12037},
 primaryClass = {astro-ph.SR},
       adsurl = {https://ui.adsabs.harvard.edu/abs/2023ApJ...950...27G}
}

@ARTICLE{Fuller2014,
       author = {{Fuller}, Jim and {Lecoanet}, Daniel and {Cantiello}, Matteo and et al.},
        title = "{Angular Momentum Transport via Internal Gravity Waves in Evolving Stars}",
      journal = {\apj},
         year = 2014,
        month = nov,
       volume = {796},
       number = {1},
          eid = {17},
        pages = {17},
          doi = {10.1088/0004-637X/796/1/17},
archivePrefix = {arXiv},
       eprint = {1409.6835},
 primaryClass = {astro-ph.SR},
       adsurl = {https://ui.adsabs.harvard.edu/abs/2014ApJ...796...17F}
}

@ARTICLE{Langer2014,
       author = {{Langer}, N. and {Kudritzki}, R.~P.},
        title = "{The spectroscopic Hertzsprung-Russell diagram}",
      journal = {\aap},
         year = 2014,
        month = apr,
       volume = {564},
          eid = {A52},
        pages = {A52},
          doi = {10.1051/0004-6361/201423374},
archivePrefix = {arXiv},
       eprint = {1403.2212},
 primaryClass = {astro-ph.SR},
       adsurl = {https://ui.adsabs.harvard.edu/abs/2014A&A...564A..52L}
}

@ARTICLE{Jin2024,
       author = {{Jin}, Harim and {Langer}, Norbert and {Lennon}, Daniel J. and et al.},
        title = "{Boron depletion in Galactic early B-type stars reveals two different main sequence star populations}",
      journal = {\aap},
         year = 2024,
        month = oct,
       volume = {690},
          eid = {A135},
        pages = {A135},
          doi = {10.1051/0004-6361/202450896},
archivePrefix = {arXiv},
       eprint = {2405.18266},
 primaryClass = {astro-ph.SR},
       adsurl = {https://ui.adsabs.harvard.edu/abs/2024A&A...690A.135J}
}

@ARTICLE{Kjeldsen2011,
       author = {{Kjeldsen}, H. and {Bedding}, T.~R.},
        title = "{Amplitudes of solar-like oscillations: a new scaling relation}",
      journal = {\aap},
         year = 2011,
        month = may,
       volume = {529},
          eid = {L8},
        pages = {L8},
          doi = {10.1051/0004-6361/201116789},
archivePrefix = {arXiv},
       eprint = {1104.1659},
 primaryClass = {astro-ph.SR},
       adsurl = {https://ui.adsabs.harvard.edu/abs/2011A&A...529L...8K}
}

@ARTICLE{Tayar2022,
       author = {{Tayar}, Jamie and {Moyano}, Facundo D. and {Soares-Furtado}, Melinda and et al.},
        title = "{Spinning up the Surface: Evidence for Planetary Engulfment or Unexpected Angular Momentum Transport?}",
      journal = {\apj},
         year = 2022,
        month = nov,
       volume = {940},
       number = {1},
          eid = {23},
        pages = {23},
          doi = {10.3847/1538-4357/ac9312},
archivePrefix = {arXiv},
       eprint = {2208.01678},
 primaryClass = {astro-ph.SR},
       adsurl = {https://ui.adsabs.harvard.edu/abs/2022ApJ...940...23T}
}

@ARTICLE{Baran2023,
       author = {{Baran}, A.~S. and {Van Grootel}, V. and {{\O}stensen}, R.~H. and et al.},
        title = "{Short-period pulsating hot-subdwarf stars observed by TESS. I. Southern ecliptic hemisphere}",
      journal = {\aap},
         year = 2023,
        month = jan,
       volume = {669},
          eid = {A48},
        pages = {A48},
          doi = {10.1051/0004-6361/202244888},
archivePrefix = {arXiv},
       eprint = {2211.09137},
 primaryClass = {astro-ph.SR},
       adsurl = {https://ui.adsabs.harvard.edu/abs/2023A&A...669A..48B}
}

@ARTICLE{Takahashi2021,
       author = {{Takahashi}, K. and {Langer}, N.},
        title = "{Modeling of magneto-rotational stellar evolution. I. Method and first applications}",
      journal = {\aap},
         year = 2021,
        month = feb,
       volume = {646},
          eid = {A19},
        pages = {A19},
          doi = {10.1051/0004-6361/202039253},
archivePrefix = {arXiv},
       eprint = {2010.13909},
 primaryClass = {astro-ph.SR},
       adsurl = {https://ui.adsabs.harvard.edu/abs/2021A&A...646A..19T}
}

@ARTICLE{Lau2025,
       author = {{Lau}, Mike Y.~M. and {Cantiello}, Matteo and {Jermyn}, Adam S. and et al.},
        title = "{Hot Jupiter engulfment by an early red giant in 3D hydrodynamics}",
      journal = {\aap},
         year = 2025,
        month = feb,
       volume = {694},
          eid = {A264},
        pages = {A264},
          doi = {10.1051/0004-6361/202452081},
archivePrefix = {arXiv},
       eprint = {2210.15848},
 primaryClass = {astro-ph.SR},
       adsurl = {https://ui.adsabs.harvard.edu/abs/2025A&A...694A.264L}
}

@ARTICLE{Pelisoli2022,
       author = {{Pelisoli}, Ingrid and {Dorsch}, M. and {Heber}, U. and et al.},
        title = "{Discovery and analysis of three magnetic hot subdwarf stars: evidence for merger-induced magnetic fields}",
      journal = {\mnras},
         year = 2022,
        month = sep,
       volume = {515},
       number = {2},
        pages = {2496-2510},
          doi = {10.1093/mnras/stac1069},
archivePrefix = {arXiv},
       eprint = {2204.06575},
 primaryClass = {astro-ph.SR},
       adsurl = {https://ui.adsabs.harvard.edu/abs/2022MNRAS.515.2496P}
}

@ARTICLE{Ma2023,
       author = {{Ma}, X. -Y. and {Zong}, W. and {Fu}, J. -N. and et al.},
        title = "{Amplitude and frequency variations in PG 0101+039 from K2 photometry. A pulsating hot B subdwarf star in an unsynchronized binary system}",
      journal = {\aap},
         year = 2023,
        month = dec,
       volume = {680},
          eid = {A11},
        pages = {A11},
          doi = {10.1051/0004-6361/202347410},
archivePrefix = {arXiv},
       eprint = {2309.11532},
 primaryClass = {astro-ph.SR},
       adsurl = {https://ui.adsabs.harvard.edu/abs/2023A&A...680A..11M}
}

@ARTICLE{Moyano2023a,
       author = {{Moyano}, F.~D. and {Eggenberger}, P. and {Mosser}, B. and et al.},
        title = "{Asteroseismology of evolved stars to constrain the internal transport of angular momentum. VI. Testing a parametric formulation for the azimuthal magneto-rotational instability}",
      journal = {\aap},
         year = 2023,
        month = may,
       volume = {673},
          eid = {A110},
        pages = {A110},
          doi = {10.1051/0004-6361/202245519},
archivePrefix = {arXiv},
       eprint = {2302.07811},
 primaryClass = {astro-ph.SR},
       adsurl = {https://ui.adsabs.harvard.edu/abs/2023A&A...673A.110M}
}

@ARTICLE{Takahashi2025,
       author = {{Takahashi}, Koh and {Langer}, Norbert},
        title = "{Standing torsional Alfv{\'e}n waves as the source of the rotational period variation in magnetic early-type stars}",
      journal = {\aap},
         year = 2025,
        month = apr,
       volume = {696},
          eid = {A129},
        pages = {A129},
          doi = {10.1051/0004-6361/202452850},
archivePrefix = {arXiv},
       eprint = {2411.01193},
 primaryClass = {astro-ph.SR},
       adsurl = {https://ui.adsabs.harvard.edu/abs/2025A&A...696A.129T}
}

@INPROCEEDINGS{Heber2026,
       author = {{Heber}, Ulrich},
        title = "{Hot subdwarf stars}",
    booktitle = {Encyclopedia of Astrophysics},
         year = 2026,
       volume = {2},
        month = jan,
        pages = {488-507},
          doi = {10.1016/B978-0-443-21439-4.00043-2},
       adsurl = {https://ui.adsabs.harvard.edu/abs/2026enap....2..488H}
}

@ARTICLE{Claret2016,
       author = {{Claret}, A. and {Torres}, G.},
        title = "{The dependence of convective core overshooting on stellar mass}",
      journal = {\aap},
         year = 2016,
        month = jul,
       volume = {592},
          eid = {A15},
        pages = {A15},
          doi = {10.1051/0004-6361/201628779},
       adsurl = {https://ui.adsabs.harvard.edu/abs/2016A&A...592A..15C}
}

@ARTICLE{Claret2018,
       author = {{Claret}, Antonio and {Torres}, Guillermo},
        title = "{The Dependence of Convective Core Overshooting on Stellar Mass: Additional Binary Systems and Improved Calibration}",
      journal = {\apj},
         year = 2018,
        month = jun,
       volume = {859},
       number = {2},
          eid = {100},
        pages = {100},
          doi = {10.3847/1538-4357/aabd35},
archivePrefix = {arXiv},
       eprint = {1804.03148},
 primaryClass = {astro-ph.SR},
       adsurl = {https://ui.adsabs.harvard.edu/abs/2018ApJ...859..100C}
}

@ARTICLE{Staff2016,
       author = {{Staff}, Jan E. and {De Marco}, Orsola and {Wood}, Peter and et al.},
        title = "{Hydrodynamic simulations of the interaction between giant stars and planets}",
      journal = {\mnras},
         year = 2016,
        month = may,
       volume = {458},
       number = {1},
        pages = {832-844},
          doi = {10.1093/mnras/stw331},
archivePrefix = {arXiv},
       eprint = {1602.03130},
 primaryClass = {astro-ph.SR},
       adsurl = {https://ui.adsabs.harvard.edu/abs/2016MNRAS.458..832S}
}

@ARTICLE{Belkacem2015b,
       author = {{Belkacem}, K. and {Marques}, J.~P. and {Goupil}, M.~J. and et al.},
        title = "{Angular momentum redistribution by mixed modes in evolved low-mass stars. II. Spin-down of the core of red giants induced by mixed modes}",
      journal = {\aap},
         year = 2015,
        month = jul,
       volume = {579},
          eid = {A31},
        pages = {A31},
          doi = {10.1051/0004-6361/201526043},
archivePrefix = {arXiv},
       eprint = {1505.05452},
 primaryClass = {astro-ph.SR},
       adsurl = {https://ui.adsabs.harvard.edu/abs/2015A&A...579A..31B}
}

@ARTICLE{Dorsch2022,
       author = {{Dorsch}, M. and {Reindl}, N. and {Pelisoli}, I. and et al.},
        title = "{Discovery of a highly magnetic He-sdO star from a double-degenerate binary merger}",
      journal = {\aap},
         year = 2022,
        month = feb,
       volume = {658},
          eid = {L9},
        pages = {L9},
          doi = {10.1051/0004-6361/202142880},
archivePrefix = {arXiv},
       eprint = {2201.08146},
 primaryClass = {astro-ph.SR},
       adsurl = {https://ui.adsabs.harvard.edu/abs/2022A&A...658L...9D}
}

@ARTICLE{Rogers2025,
       author = {{Rogers}, T.~M. and {Ratnasingam}, R.~P.},
        title = "{Angular Momentum Transport by Internal Gravity Waves across Age}",
      journal = {\apjl},
         year = 2025,
        month = apr,
       volume = {983},
       number = {2},
          eid = {L38},
        pages = {L38},
          doi = {10.3847/2041-8213/adc45a},
archivePrefix = {arXiv},
       eprint = {2504.03827},
 primaryClass = {astro-ph.SR},
       adsurl = {https://ui.adsabs.harvard.edu/abs/2025ApJ...983L..38R}
}

@ARTICLE{Charpinet2000,
       author = {{Charpinet}, S. and {Fontaine}, G. and {Brassard}, P. and et al.},
        title = "{Adiabatic Survey of Subdwarf B Star Oscillations. I. Pulsation Properties of a Representative Evolutionary Model}",
      journal = {\apjs},
         year = 2000,
        month = nov,
       volume = {131},
       number = {1},
        pages = {223-247},
          doi = {10.1086/317359},
       adsurl = {https://ui.adsabs.harvard.edu/abs/2000ApJS..131..223C}
}

@ARTICLE{Claret2019,
       author = {{Claret}, Antonio and {Torres}, Guillermo},
        title = "{The Dependence of Convective Core Overshooting on Stellar Mass: Reality Check and Additional Evidence}",
      journal = {\apj},
         year = 2019,
        month = may,
       volume = {876},
       number = {2},
          eid = {134},
        pages = {134},
          doi = {10.3847/1538-4357/ab1589},
archivePrefix = {arXiv},
       eprint = {1904.02714},
 primaryClass = {astro-ph.SR},
       adsurl = {https://ui.adsabs.harvard.edu/abs/2019ApJ...876..134C}
}

@ARTICLE{Meduri2024,
       author = {{Meduri}, Domenico G. and {Jouve}, Laur{\`e}ne and {Ligni{\`e}res}, Fran{\c{c}}ois},
        title = "{Angular momentum and chemical transport by azimuthal magnetorotational instability in radiative stellar interiors}",
      journal = {\aap},
         year = 2024,
        month = mar,
       volume = {683},
          eid = {A12},
        pages = {A12},
          doi = {10.1051/0004-6361/202347672},
       adsurl = {https://ui.adsabs.harvard.edu/abs/2024A&A...683A..12M}
}

@ARTICLE{Gossage2021,
       author = {{Gossage}, Seth and {Dotter}, Aaron and {Garraffo}, Cecilia and et al.},
        title = "{MESA Models with Magnetic Braking}",
      journal = {\apj},
         year = 2021,
        month = may,
       volume = {912},
       number = {1},
          eid = {65},
        pages = {65},
          doi = {10.3847/1538-4357/abebdf},
archivePrefix = {arXiv},
       eprint = {2011.02470},
 primaryClass = {astro-ph.SR},
       adsurl = {https://ui.adsabs.harvard.edu/abs/2021ApJ...912...65G}
}

@ARTICLE{Brun2009,
       author = {{Brun}, A.~S. and {Palacios}, A.},
        title = "{Numerical Simulations of a Rotating Red Giant Star. I. Three-dimensional Models of Turbulent Convection and Associated Mean Flows}",
      journal = {\apj},
         year = 2009,
        month = sep,
       volume = {702},
       number = {2},
        pages = {1078-1097},
          doi = {10.1088/0004-637X/702/2/1078},
       adsurl = {https://ui.adsabs.harvard.edu/abs/2009ApJ...702.1078B}
}

@ARTICLE{Claret2017,
       author = {{Claret}, Antonio and {Torres}, Guillermo},
        title = "{The Dependence of Convective Core Overshooting on Stellar Mass: A Semi-empirical Determination Using the Diffusive Approach with Two Different Element Mixtures}",
      journal = {\apj},
         year = 2017,
        month = nov,
       volume = {849},
       number = {1},
          eid = {18},
        pages = {18},
          doi = {10.3847/1538-4357/aa8770},
archivePrefix = {arXiv},
       eprint = {1710.08417},
 primaryClass = {astro-ph.SR},
       adsurl = {https://ui.adsabs.harvard.edu/abs/2017ApJ...849...18C}
}

@ARTICLE{Uzundag2021,
       author = {{Uzundag}, Murat and {Vu{\v{c}}kovi{\'c}}, Maja and {N{\'e}meth}, P{\'e}ter and et al.},
        title = "{Asteroseismic analysis of variable hot subdwarf stars observed with TESS. I. The mean g-mode period spacings in hot subdwarf B stars}",
      journal = {\aap},
         year = 2021,
        month = jul,
       volume = {651},
          eid = {A121},
        pages = {A121},
          doi = {10.1051/0004-6361/202140961},
archivePrefix = {arXiv},
       eprint = {2105.15137},
 primaryClass = {astro-ph.SR},
       adsurl = {https://ui.adsabs.harvard.edu/abs/2021A&A...651A.121U}
}

@ARTICLE{Belkacem2015a,
       author = {{Belkacem}, K. and {Marques}, J.~P. and {Goupil}, M.~J. and et al.},
        title = "{Angular momentum redistribution by mixed modes in evolved low-mass stars. I. Theoretical formalism}",
      journal = {\aap},
         year = 2015,
        month = jul,
       volume = {579},
          eid = {A30},
        pages = {A30},
          doi = {10.1051/0004-6361/201526042},
archivePrefix = {arXiv},
       eprint = {1505.05447},
 primaryClass = {astro-ph.SR},
       adsurl = {https://ui.adsabs.harvard.edu/abs/2015A&A...579A..30B}
}

@ARTICLE{Pincon2016,
       author = {{Pin{\c{c}}on}, C. and {Belkacem}, K. and {Goupil}, M.~J.},
        title = "{Generation of internal gravity waves by penetrative convection}",
      journal = {\aap},
         year = 2016,
        month = apr,
       volume = {588},
          eid = {A122},
        pages = {A122},
          doi = {10.1051/0004-6361/201527663},
archivePrefix = {arXiv},
       eprint = {1512.07028},
 primaryClass = {astro-ph.SR},
       adsurl = {https://ui.adsabs.harvard.edu/abs/2016A&A...588A.122P}
}

@ARTICLE{Pincon2017,
       author = {{Pin{\c{c}}on}, C. and {Belkacem}, K. and {Goupil}, M.~J. and et al.},
        title = "{Can plume-induced internal gravity waves regulate the core rotation of subgiant stars?}",
      journal = {\aap},
         year = 2017,
        month = sep,
       volume = {605},
          eid = {A31},
        pages = {A31},
          doi = {10.1051/0004-6361/201730998},
archivePrefix = {arXiv},
       eprint = {1705.10101},
 primaryClass = {astro-ph.SR},
       adsurl = {https://ui.adsabs.harvard.edu/abs/2017A&A...605A..31P}
}

@ARTICLE{Schwab2018,
       author = {{Schwab}, Josiah},
        title = "{Hot subdwarfs formed from the merger of two He white dwarfs}",
      journal = {\mnras},
         year = 2018,
        month = jun,
       volume = {476},
       number = {4},
        pages = {5303-5311},
          doi = {10.1093/mnras/sty586},
archivePrefix = {arXiv},
       eprint = {1803.00576},
 primaryClass = {astro-ph.SR},
       adsurl = {https://ui.adsabs.harvard.edu/abs/2018MNRAS.476.5303S}
}

@ARTICLE{Charpinet2018,
       author = {{Charpinet}, Stephane and {Giammichele}, Noemi and {Zong}, Weikai and et al.},
        title = "{Rotation in sdB stars as revealed by stellar oscillations}",
      journal = {Open Astronomy},
         year = 2018,
        month = jul,
       volume = {27},
       number = {1},
        pages = {112-119},
          doi = {10.1515/astro-2018-0012},
       adsurl = {https://ui.adsabs.harvard.edu/abs/2018OAst...27..112C}
}

@ARTICLE{Kissin2015,
       author = {{Kissin}, Yevgeni and {Thompson}, Christopher},
        title = "{Rotation of Giant Stars}",
      journal = {\apj},
         year = 2015,
        month = jul,
       volume = {808},
       number = {1},
          eid = {35},
        pages = {35},
          doi = {10.1088/0004-637X/808/1/35},
archivePrefix = {arXiv},
       eprint = {1501.07217},
 primaryClass = {astro-ph.SR},
       adsurl = {https://ui.adsabs.harvard.edu/abs/2015ApJ...808...35K}
}

@ARTICLE{Reed2019,
       author = {{Reed}, M.~D. and {Telting}, J.~H. and {Ketzer}, L. and et al.},
        title = "{Two p-mode-dominated subdwarf B pulsators in binaries with F-star companions observed with K2}",
      journal = {\mnras},
         year = 2019,
        month = feb,
       volume = {483},
       number = {2},
        pages = {2282-2299},
          doi = {10.1093/mnras/sty3025},
       adsurl = {https://ui.adsabs.harvard.edu/abs/2019MNRAS.483.2282R}
}

@ARTICLE{MoniBidin2017,
       author = {{Moni Bidin}, C. and {Casetti-Dinescu}, D.~I. and {Girard}, T.~M. and et al.},
        title = "{Young stars in the periphery of the Large Magellanic Cloud}",
      journal = {\mnras},
         year = 2017,
        month = apr,
       volume = {466},
       number = {3},
        pages = {3077-3087},
          doi = {10.1093/mnras/stw3242},
archivePrefix = {arXiv},
       eprint = {1612.03072},
 primaryClass = {astro-ph.GA},
       adsurl = {https://ui.adsabs.harvard.edu/abs/2017MNRAS.466.3077M}
}

@ARTICLE{Geier2008,
       author = {{Geier}, S. and {Nesslinger}, S. and {Heber}, U. and et al.},
        title = "{Tidal synchronisation of the subdwarf B binary PG 0101+039}",
      journal = {\aap},
         year = 2008,
        month = jan,
       volume = {477},
       number = {2},
        pages = {L13-L16},
          doi = {10.1051/0004-6361:20078797},
       adsurl = {https://ui.adsabs.harvard.edu/abs/2008A&A...477L..13G}
}

@ARTICLE{Zong2018,
       author = {{Zong}, Weikai and {Charpinet}, St{\'e}phane and {Fu}, Jian-Ning and et al.},
        title = "{Oscillation Mode Variability in Evolved Compact Pulsators from Kepler Photometry. I. The Hot B Subdwarf Star KIC 3527751}",
      journal = {\apj},
         year = 2018,
        month = feb,
       volume = {853},
       number = {2},
          eid = {98},
        pages = {98},
          doi = {10.3847/1538-4357/aaa548},
       adsurl = {https://ui.adsabs.harvard.edu/abs/2018ApJ...853...98Z}
}

@ARTICLE{Ostensen2012,
       author = {{{\O}stensen}, R.~H. and {Degroote}, P. and {Telting}, J.~H. and et al.},
        title = "{KIC 1718290: A Helium-rich V1093-Her-like Pulsator on the Blue Horizontal Branch}",
      journal = {\apjl},
         year = 2012,
        month = jul,
       volume = {753},
       number = {1},
          eid = {L17},
        pages = {L17},
          doi = {10.1088/2041-8205/753/1/L17},
archivePrefix = {arXiv},
       eprint = {1206.0656},
 primaryClass = {astro-ph.SR},
       adsurl = {https://ui.adsabs.harvard.edu/abs/2012ApJ...753L..17O}
}

@ARTICLE{VanReeth2018,
       author = {{Van Reeth}, T. and {Mombarg}, J.~S.~G. and {Mathis}, S. and et al.},
        title = "{Sensitivity of gravito-inertial modes to differential rotation in intermediate-mass main-sequence stars}",
      journal = {\aap},
         year = 2018,
        month = oct,
       volume = {618},
          eid = {A24},
        pages = {A24},
          doi = {10.1051/0004-6361/201832718},
archivePrefix = {arXiv},
       eprint = {1806.03586},
 primaryClass = {astro-ph.SR},
       adsurl = {https://ui.adsabs.harvard.edu/abs/2018A&A...618A..24V}
}

@ARTICLE{VanGrootel2018,
       author = {{Van Grootel}, Val{\'e}rie and {P{\'e}ters}, Marie-Julie and {Green}, Elizabeth M. and et al.},
        title = "{New observations and asteroseismic analysis of the subdwarf B pulsator PG 1219+534}",
      journal = {Open Astronomy},
         year = 2018,
        month = mar,
       volume = {27},
       number = {1},
        pages = {44-56},
          doi = {10.1515/astro-2018-0014},
       adsurl = {https://ui.adsabs.harvard.edu/abs/2018OAst...27...44V}
}

@ARTICLE{OstensenReed2014,
       author = {{{\O}stensen}, R.~H. and {Reed}, M.~D. and {Baran}, A.~S. and et al.},
        title = "{Stochastic pulsations in the subdwarf-B star KIC 2991276}",
      journal = {\aap},
         year = 2014,
        month = apr,
       volume = {564},
          eid = {L14},
        pages = {L14},
          doi = {10.1051/0004-6361/201423734},
       adsurl = {https://ui.adsabs.harvard.edu/abs/2014A&A...564L..14O}
}

@ARTICLE{Stello2016b,
       author = {{Stello}, Dennis and {Cantiello}, Matteo and {Fuller}, Jim and et al.},
        title = "{Suppression of Quadrupole and Octupole Modes in Red Giants Observed by Kepler *}",
      journal = {\pasa},
         year = 2016,
        month = mar,
       volume = {33},
          eid = {e011},
        pages = {e011},
          doi = {10.1017/pasa.2016.9},
archivePrefix = {arXiv},
       eprint = {1602.05193},
 primaryClass = {astro-ph.SR},
       adsurl = {https://ui.adsabs.harvard.edu/abs/2016PASA...33...11S}
}

@ARTICLE{Stello2016a,
       author = {{Stello}, Dennis and {Cantiello}, Matteo and {Fuller}, Jim and et al.},
        title = "{A prevalence of dynamo-generated magnetic fields in the cores of intermediate-mass stars}",
      journal = {\nat},
         year = 2016,
        month = jan,
       volume = {529},
       number = {7586},
        pages = {364-367},
          doi = {10.1038/nature16171},
archivePrefix = {arXiv},
       eprint = {1601.00004},
 primaryClass = {astro-ph.SR},
       adsurl = {https://ui.adsabs.harvard.edu/abs/2016Natur.529..364S}
}

@ARTICLE{Li2022,
       author = {{Li}, Gang and {Deheuvels}, S{\'e}bastien and {Ballot}, J{\'e}r{\^o}me and et al.},
        title = "{Magnetic fields of 30 to 100 kG in the cores of red giant stars}",
      journal = {\nat},
         year = 2022,
        month = oct,
       volume = {610},
       number = {7930},
        pages = {43-46},
          doi = {10.1038/s41586-022-05176-0},
archivePrefix = {arXiv},
       eprint = {2208.09487},
 primaryClass = {astro-ph.SR},
       adsurl = {https://ui.adsabs.harvard.edu/abs/2022Natur.610...43L}
}

@ARTICLE{Deheuvels2023,
       author = {{Deheuvels}, S. and {Li}, G. and {Ballot}, J. and et al.},
        title = "{Strong magnetic fields detected in the cores of 11 red giant stars using gravity-mode period spacings}",
      journal = {\aap},
         year = 2023,
        month = feb,
       volume = {670},
          eid = {L16},
        pages = {L16},
          doi = {10.1051/0004-6361/202245282},
archivePrefix = {arXiv},
       eprint = {2301.01308},
 primaryClass = {astro-ph.SR},
       adsurl = {https://ui.adsabs.harvard.edu/abs/2023A&A...670L..16D}
}

@ARTICLE{Hatt2024,
       author = {{Hatt}, Emily J. and {Ong}, J.~M. Joel and {Nielsen}, Martin B. and et al.},
        title = "{Asteroseismic signatures of core magnetism and rotation in hundreds of low-luminosity red giants}",
      journal = {\mnras},
         year = 2024,
        month = oct,
       volume = {534},
       number = {2},
        pages = {1060-1076},
          doi = {10.1093/mnras/stae2053},
archivePrefix = {arXiv},
       eprint = {2409.01157},
 primaryClass = {astro-ph.SR},
       adsurl = {https://ui.adsabs.harvard.edu/abs/2024MNRAS.534.1060H}
}

@ARTICLE{Gagnier2024,
       author = {{Gagnier}, Damien and {Pejcha}, Ond{\v{r}}ej},
        title = "{Post-dynamical inspiral phase of common envelope evolution. The role of magnetic fields}",
      journal = {\aap},
         year = 2024,
        month = mar,
       volume = {683},
          eid = {A4},
        pages = {A4},
          doi = {10.1051/0004-6361/202348383},
archivePrefix = {arXiv},
       eprint = {2310.16880},
 primaryClass = {astro-ph.SR},
       adsurl = {https://ui.adsabs.harvard.edu/abs/2024A&A...683A...4G}
}

@ARTICLE{Reed2025,
       author = {{Reed}, M.~D. and {Baran}, A.~S. and {Telting}, J.~H. and et al.},
        title = "{K2 observations of five pulsating subdwarf-B stars with white dwarf companions}",
      journal = {\mnras},
         year = 2025,
        month = jul,
       volume = {540},
       number = {4},
        pages = {3725-3752},
          doi = {10.1093/mnras/staf917},
archivePrefix = {arXiv},
       eprint = {2506.05033},
 primaryClass = {astro-ph.SR},
       adsurl = {https://ui.adsabs.harvard.edu/abs/2025MNRAS.540.3725R}
}

@ARTICLE{OstensenTelting2014,
       author = {{{\O}stensen}, R.~H. and {Telting}, J.~H. and {Reed}, M.~D. and et al.},
        title = "{Asteroseismology revealing trapped modes in KIC 10553698A}",
      journal = {\aap},
         year = 2014,
        month = sep,
       volume = {569},
          eid = {A15},
        pages = {A15},
          doi = {10.1051/0004-6361/201423611},
archivePrefix = {arXiv},
       eprint = {1406.6941},
 primaryClass = {astro-ph.SR},
       adsurl = {https://ui.adsabs.harvard.edu/abs/2014A&A...569A..15O}
}

@ARTICLE{Kupfer2015,
       author = {{Kupfer}, T. and {Geier}, S. and {Heber}, U. and et al.},
        title = "{Hot subdwarf binaries from the MUCHFUSS project. Analysis of 12 new systems and a study of the short-period binary population}",
      journal = {\aap},
         year = 2015,
        month = apr,
       volume = {576},
          eid = {A44},
        pages = {A44},
          doi = {10.1051/0004-6361/201425213},
archivePrefix = {arXiv},
       eprint = {1501.03692},
 primaryClass = {astro-ph.SR},
       adsurl = {https://ui.adsabs.harvard.edu/abs/2015A&A...576A..44K}
}

@ARTICLE{Aerts2021,
       author = {{Aerts}, C.},
        title = "{Probing the interior physics of stars through asteroseismology}",
      journal = {Reviews of Modern Physics},
         year = 2021,
        month = jan,
       volume = {93},
       number = {1},
          eid = {015001},
        pages = {015001},
          doi = {10.1103/RevModPhys.93.015001},
archivePrefix = {arXiv},
       eprint = {1912.12300},
 primaryClass = {astro-ph.SR},
       adsurl = {https://ui.adsabs.harvard.edu/abs/2021RvMP...93a5001A}
}

@ARTICLE{Ouazzani2019,
       author = {{Ouazzani}, R. -M. and {Marques}, J.~P. and {Goupil}, M. -J. and et al.},
        title = "{{\ensuremath{\gamma}} Doradus stars as a test of angular momentum transport models}",
      journal = {\aap},
         year = 2019,
        month = jun,
       volume = {626},
          eid = {A121},
        pages = {A121},
          doi = {10.1051/0004-6361/201832607},
archivePrefix = {arXiv},
       eprint = {1801.09228},
 primaryClass = {astro-ph.SR},
       adsurl = {https://ui.adsabs.harvard.edu/abs/2019A&A...626A.121O}
}

@ARTICLE{Pedersen2022,
       author = {{Pedersen}, May G.},
        title = "{Internal Rotation and Inclinations of Slowly Pulsating B Stars: Evidence of Interior Angular Momentum Transport}",
      journal = {\apj},
         year = 2022,
        month = nov,
       volume = {940},
       number = {1},
          eid = {49},
        pages = {49},
          doi = {10.3847/1538-4357/ac947f},
archivePrefix = {arXiv},
       eprint = {2208.14497},
 primaryClass = {astro-ph.SR},
       adsurl = {https://ui.adsabs.harvard.edu/abs/2022ApJ...940...49P}
}

@ARTICLE{Deheuvels2020,
       author = {{Deheuvels}, S. and {Ballot}, J. and {Eggenberger}, P. and et al.},
        title = "{Seismic evidence for near solid-body rotation in two Kepler subgiants and implications for angular momentum transport}",
      journal = {\aap},
         year = 2020,
        month = sep,
       volume = {641},
          eid = {A117},
        pages = {A117},
          doi = {10.1051/0004-6361/202038578},
archivePrefix = {arXiv},
       eprint = {2007.02585},
 primaryClass = {astro-ph.SR},
       adsurl = {https://ui.adsabs.harvard.edu/abs/2020A&A...641A.117D}
}

@ARTICLE{Mosser2012,
       author = {{Mosser}, B. and {Goupil}, M.~J. and {Belkacem}, K. and et al.},
        title = "{Spin down of the core rotation in red giants}",
      journal = {\aap},
         year = 2012,
        month = dec,
       volume = {548},
          eid = {A10},
        pages = {A10},
          doi = {10.1051/0004-6361/201220106},
archivePrefix = {arXiv},
       eprint = {1209.3336},
 primaryClass = {astro-ph.SR},
       adsurl = {https://ui.adsabs.harvard.edu/abs/2012A&A...548A..10M}
}

@ARTICLE{Gehan2018,
       author = {{Gehan}, C. and {Mosser}, B. and {Michel}, E. and et al.},
        title = "{Core rotation braking on the red giant branch for various mass ranges}",
      journal = {\aap},
         year = 2018,
        month = aug,
       volume = {616},
          eid = {A24},
        pages = {A24},
          doi = {10.1051/0004-6361/201832822},
archivePrefix = {arXiv},
       eprint = {1802.04558},
 primaryClass = {astro-ph.SR},
       adsurl = {https://ui.adsabs.harvard.edu/abs/2018A&A...616A..24G}
}

@ARTICLE{Dhanpal2025,
       author = {{Dhanpal}, Siddharth and {Benomar}, Othman and {Hanasoge}, Shravan and et al.},
        title = "{Anomalously Fast Core and Envelope Rotation in Red Giants}",
      journal = {\apj},
         year = 2025,
        month = aug,
       volume = {988},
       number = {2},
          eid = {224},
        pages = {224},
          doi = {10.3847/1538-4357/ade2d1},
archivePrefix = {arXiv},
       eprint = {2506.06415},
 primaryClass = {astro-ph.SR},
       adsurl = {https://ui.adsabs.harvard.edu/abs/2025ApJ...988..224D}
}

@ARTICLE{Mosser2024,
       author = {{Mosser}, B. and {Dr{\'e}au}, G. and {Pin{\c{c}}on}, C. and et al.},
        title = "{Locked differential rotation in core-helium burning red giants}",
      journal = {\aap},
         year = 2024,
        month = jan,
       volume = {681},
          eid = {L20},
        pages = {L20},
          doi = {10.1051/0004-6361/202348338},
archivePrefix = {arXiv},
       eprint = {2401.07161},
 primaryClass = {astro-ph.SR},
       adsurl = {https://ui.adsabs.harvard.edu/abs/2024A&A...681L..20M}
}

@INPROCEEDINGS{Kawaler2015,
       author = {{Kawaler}, Steven D.},
        title = "{Rotation of White Dwarf Stars}",
    booktitle = {19th European Workshop on White Dwarfs},
         year = 2015,
       editor = {{Dufour}, P. and {Bergeron}, P. and {Fontaine}, G.},
       series = {Astronomical Society of the Pacific Conference Series},
       volume = {493},
        month = jun,
        pages = {65},
          doi = {10.48550/arXiv.1410.6934},
archivePrefix = {arXiv},
       eprint = {1410.6934},
 primaryClass = {astro-ph.SR},
       adsurl = {https://ui.adsabs.harvard.edu/abs/2015ASPC..493...65K}
}

@ARTICLE{Hermes2017,
       author = {{Hermes}, J.~J. and {G{\"a}nsicke}, B.~T. and {Kawaler}, Steven D. and et al.},
        title = "{White Dwarf Rotation as a Function of Mass and a Dichotomy of Mode Line Widths: Kepler Observations of 27 Pulsating DA White Dwarfs through K2 Campaign 8}",
      journal = {\apjs},
         year = 2017,
        month = oct,
       volume = {232},
       number = {2},
          eid = {23},
        pages = {23},
          doi = {10.3847/1538-4365/aa8bb5},
archivePrefix = {arXiv},
       eprint = {1709.07004},
 primaryClass = {astro-ph.SR},
       adsurl = {https://ui.adsabs.harvard.edu/abs/2017ApJS..232...23H}
}

@ARTICLE{Kawaler2005,
       author = {{Kawaler}, Steven D. and {Hostler}, Shelbi R.},
        title = "{Internal Rotation of Subdwarf B Stars: Limiting Cases and Asteroseismological Consequences}",
      journal = {\apj},
         year = 2005,
        month = mar,
       volume = {621},
       number = {1},
        pages = {432-444},
          doi = {10.1086/427403},
archivePrefix = {arXiv},
       eprint = {astro-ph/0411314},
 primaryClass = {astro-ph},
       adsurl = {https://ui.adsabs.harvard.edu/abs/2005ApJ...621..432K}
}

@ARTICLE{Sills2000,
       author = {{Sills}, Alison and {Pinsonneault}, M.~H.},
        title = "{Rotation of Horizontal-Branch Stars in Globular Clusters}",
      journal = {\apj},
         year = 2000,
        month = sep,
       volume = {540},
       number = {1},
        pages = {489-503},
          doi = {10.1086/309306},
archivePrefix = {arXiv},
       eprint = {astro-ph/9911024},
 primaryClass = {astro-ph},
       adsurl = {https://ui.adsabs.harvard.edu/abs/2000ApJ...540..489S}
}

@ARTICLE{MillerBertolami2008,
       author = {{Miller Bertolami}, M.~M. and {Althaus}, L.~G. and {Unglaub}, K. and et al.},
        title = "{Modeling He-rich subdwarfs through the hot-flasher scenario}",
      journal = {\aap},
         year = 2008,
        month = nov,
       volume = {491},
       number = {1},
        pages = {253-265},
          doi = {10.1051/0004-6361:200810373},
archivePrefix = {arXiv},
       eprint = {0808.3580},
 primaryClass = {astro-ph},
       adsurl = {https://ui.adsabs.harvard.edu/abs/2008A&A...491..253M}
}

@ARTICLE{Lanz2004,
       author = {{Lanz}, Thierry and {Brown}, Thomas M. and {Sweigart}, Allen V. and et al.},
        title = "{Flash Mixing on the White Dwarf Cooling Curve: Far Ultraviolet Spectroscopic Explorer Observations of Three He-rich sdB Stars}",
      journal = {\apj},
         year = 2004,
        month = feb,
       volume = {602},
       number = {1},
        pages = {342-355},
          doi = {10.1086/380904},
archivePrefix = {arXiv},
       eprint = {astro-ph/0308440},
 primaryClass = {astro-ph},
       adsurl = {https://ui.adsabs.harvard.edu/abs/2004ApJ...602..342L}
}

%\begin{thebibliography}{99}
%\end{thebibliography}

%%%%%%%%%%%%%%%%%%%%%%%%%%%%%%%%%%%%%%%%%%%%%%%%%%

%%%%%%%%%%%%%%%%% APPENDICES %%%%%%%%%%%%%%%%%%%%%

\appendix
  \section{Asteroseismic rotation rates from the literature}
   \label{data}   
   %sdB stars
   We compiled measurements of core and envelope rotation rates of sdB stars from different publications in the literature (see Tables \ref{table_singlesdb} and \ref{table_binarysdb}).
   The mean core rotation rate ($\overline{\Omega}_{\rm core}/2\pi$) is the rotation as inferred from the splitting of g-modes, while the envelope rotation rate  ($\omp/2\pi$) is the one provided by the splitting of p-modes.
   These measurements are mean values of the rotation rates in the regions where the g- or p-modes propagate, where the g-modes are sensitive mainly to the helium-rich radiative regions below the hydrogen-rich envelope and the p-modes are sensitive to the hydrogen-rich envelope \citep{Charpinet2000}.
   The compilation provided in Tables \ref{table_singlesdb} and \ref{table_binarysdb} is the largest set of reliable asteroseismic measurements of mean core and envelope rotation rates of sdB stars available in the literature.
   Table \ref{table_singlesdb} contains single sdB stars, while Table \ref{table_binarysdb} contains sdB stars in binary systems.

%==========================================================================================
   \section{Accreting sdB models without internal magnetic fields}
   \label{app_nonmag}
\begin{figure}   \includegraphics[width=\columnwidth]{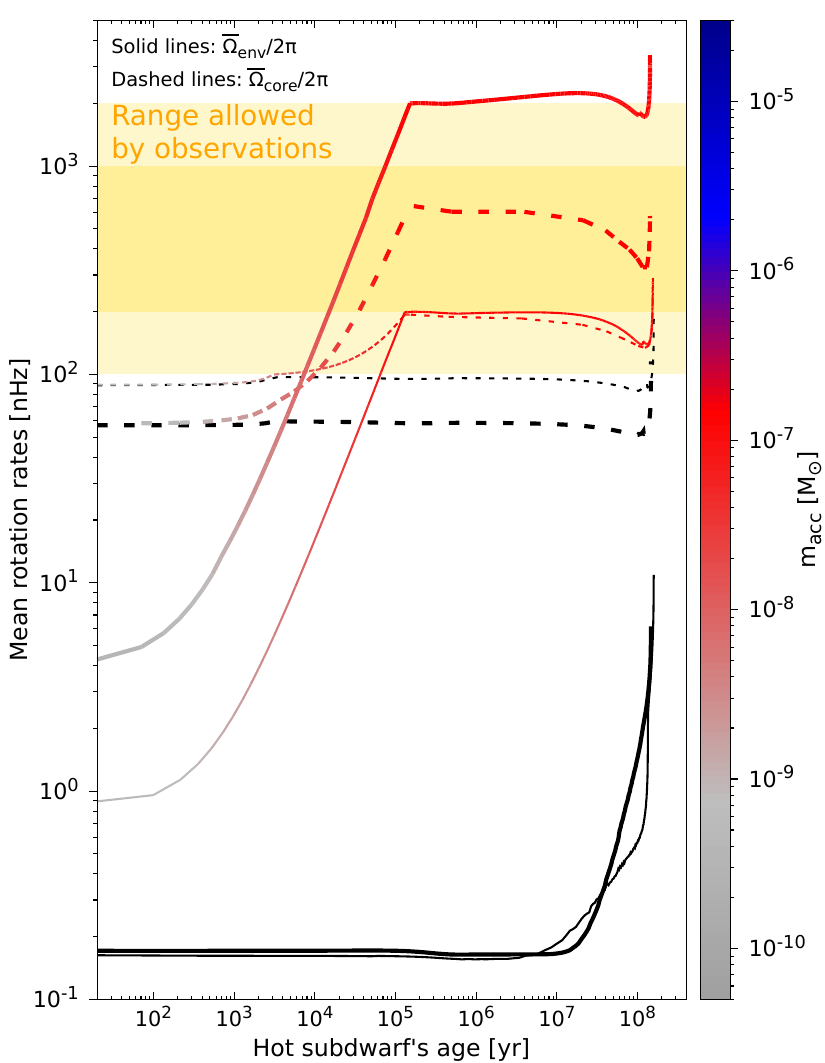}
    \caption{Similar to Fig.\ \ref{omegas_gsurf_accretion} but showing the effect of neglecting the internal magnetic fields.
Models shown with coloured lines are accreting sdB models computed without internal magnetic fields, while those in black are sdB models without accretion but including internal magnetic fields.
    }
     \label{omegas_gsurf_accretion_nonmag}
      \end{figure}

   \textbfake{Although we included internal magnetic fields driven by the TSF dynamo in all of the models presented in the main body of this work, in this section we explain how neglecting the internal magnetic fields in our accreting sdB models would impact our results.
   In Fig. \ref{omegas_gsurf_accretion_nonmag} we show the evolution of the mean rotation rates of the core and envelope in both sdB models with and without internal magnetic fields, with the same initial parameters (i.e.\ sdB mass, envelope mass, and initial velocity) as in Fig. \ref{omegas_gsurf_accretion}.
   While accreting sdB models with internal magnetic fields can reach solid-body rotation if they accrete enough AM, models without internal magnetic fields cannot transport the AM from the hydrogen-rich envelope to the helium-rich radiative regions and thus always rotate differentially.
   This is illustrated in Fig. \ref{omegas_gsurf_accretion_nonmag} by the models whose envelope is spun up to $\omp/2\pi = 2000$ nHz by accretion.
   The model without magnetic fields remains rotating differentially, with its envelope rotating always faster than the core.}

\begin{figure*}
     \centering     \includegraphics[width=18cm]{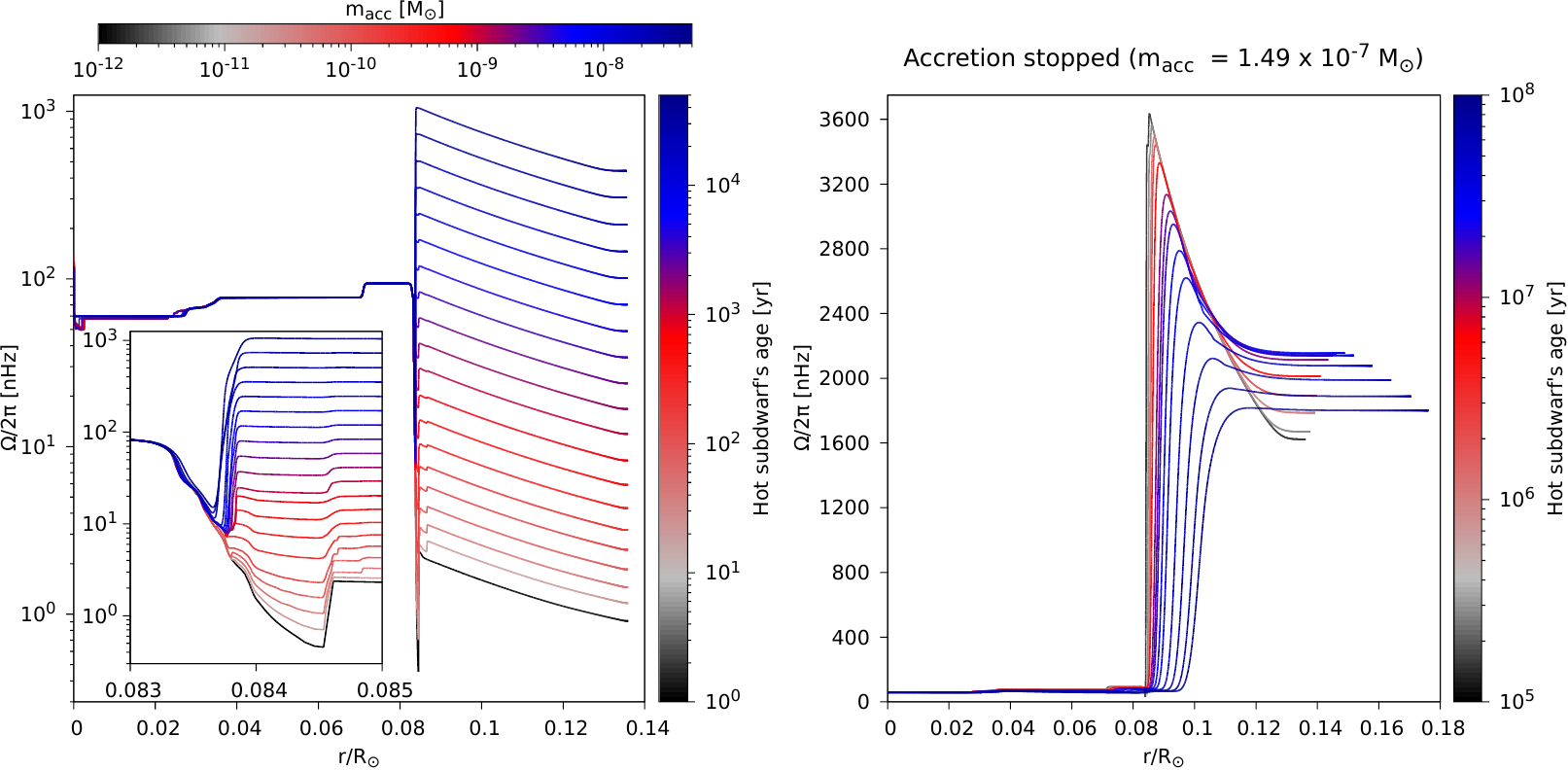}
     \caption{\textbfake{Similar to Fig.\ \ref{profile_rotation_accretion} but for accreting sdB models without internal magnetic fields.
       The panel on the left shows the spin-up phase due to accretion while the one on the right shows the evolution of the rotation profiles once the accretion stops, until the sdB model reaches an age of $10^8$ years since it started its core-helium burning phase (the core-helium burning lifetime of this sdB model is around $1.44 \times 10^{8}$ years), where the total mass accreted is indicated above the panel.}
     }
     \label{profile_rotation_accretion_nonmag}
   \end{figure*}

\textbfake{This behaviour is further illustrated in Fig. \ref{profile_rotation_accretion_nonmag}, where we show the rotation profiles of the non-magnetic model spun up to $\omp/2\pi = 2000$ nHz.
   The initial evolution of the rotation profiles is similar to that of the magnetic models, where the rotation rate in the envelope initially scales as $\Omega \propto 1/r^{2}$, because the specific AM from accretion is redistributed uniformly in the envelope.
   However, the behaviour after a few thousands years is different to that of the magnetic models (compare to left panel of Fig.\ \ref{profile_rotation_accretion}).
   The rotation profiles in the hydrogen-rich envelope do not become flat with time, because these models do not experience efficient AM transport, as is the case of models with internal magnetic fields.}

\textbfake{Afterwards, once the accretion stops (right panel of Fig.\ \ref{profile_rotation_accretion_nonmag}), the timescales are long enough for the hydrodynamical processes to diffuse the AM in the envelope.
   In this phase, initially the AM is diffused into a relatively small extension, spinning up the surface layers, which leads to higher mean envelope rotation rates as seen in the regions close to $\omp/2\pi = 2000$ nHz in the non-magnetic model in Fig.\ \ref{omegas_gsurf_accretion_nonmag}.
   This is partially due to the sensitivity of the weight function used to compute $\omp$ (see Fig. \ref{weight_functions} and Eq.\ \ref{eq_omp}), which are more sensitive to the surface layers than the deep ones.
   After this initial increase in $\omp$, the AM is still being diffused, but the sdB model is expanding as well which eventually leads to lower envelope rotation rates.}

  \textbfake{Notably, in this non-magnetic model the AM transport by hydrodynamical processes is efficient enough to diffuse all of the AM in the envelope, leading to almost rigid rotation in the envelope after $10^8$ years.
    However, the AM is never diffused into the core.
    We also note that the increasing $\omg$ seen in Fig. \ref{omegas_gsurf_accretion_nonmag} is purely due to the weight functions used to compute $\omg$ that are also affected by the rotation rate in the hydrogen-rich envelope.
   This is a fundamental difference with respect to the magnetic models, which illustrates the importance of the AM transport efficiency.
   Thus, non-magnetic accreting sdB models remain rotating differentially during their whole lifetime.}

   %
   %

%==========================================================================================   
  \section{Additional figures}
\begin{figure} \includegraphics[width=\columnwidth]{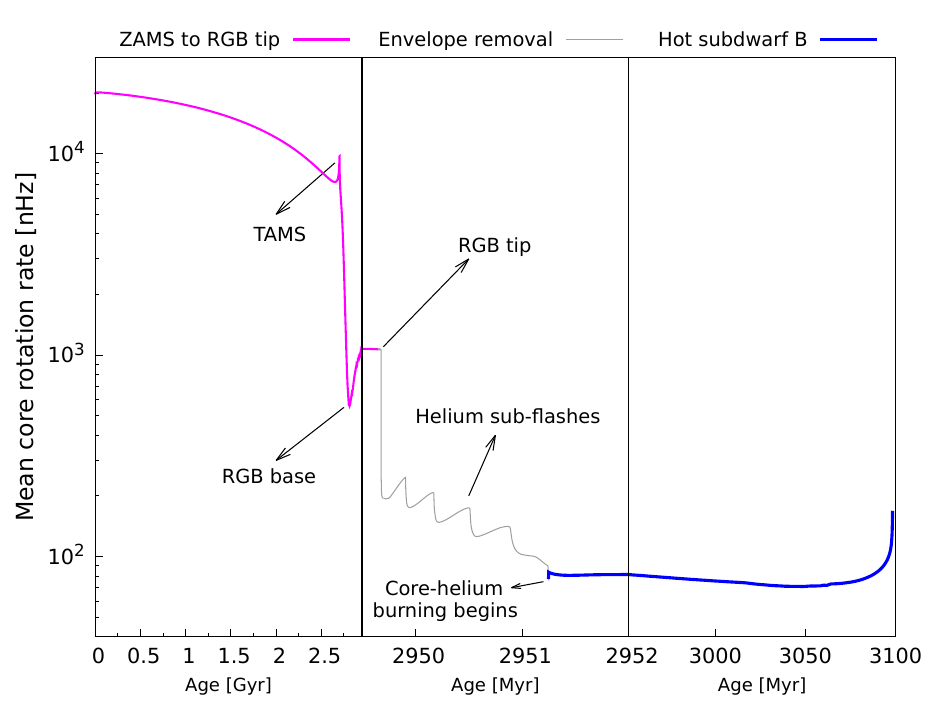}
     \caption{\textbfake{Mean core rotation rate as a function of age for the same model presented in Fig.\ \ref{omegac_gsurf_zamstosdb} (initial mass of 1.5 $M_{\odot}$).
       Each panel shows the evolution in different timescales, as indicated by the x-axis.
       Relevant evolutionary phases are indicated in the figure by arrows.
       The evolution of the sdB model ends at the end of the core-helium burning phase.}
     }
     \label{omegac_age}
      \end{figure}

\textbfake{Figure \ref{omegac_age} shows the mean core rotation rate as a function of the age of the model presented in Fig. \ref{omegac_gsurf_zamstosdb} to illustrate the different timescales that the model goes through from the ZAMS until the sdB phase.}

\begin{figure}   \includegraphics[width=\columnwidth]{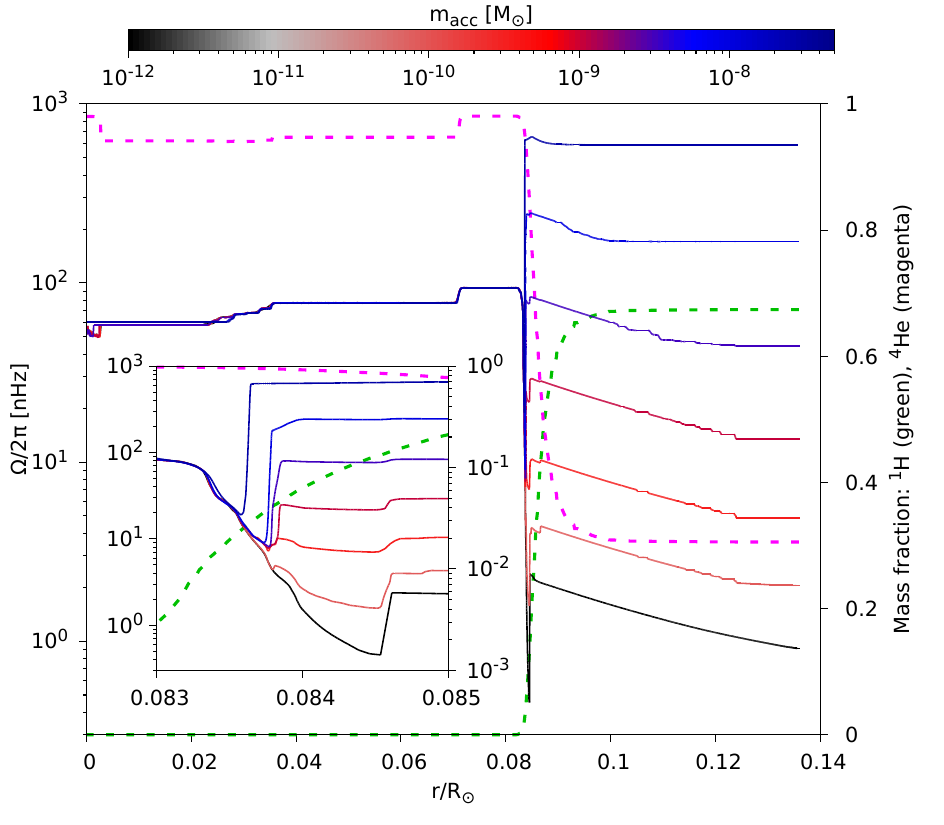}
     \caption{\textbfake{Rotation profiles as shown in Fig. \ref{profile_rotation_accretion} for the same model, including as well the chemical composition profiles of hydrogen and helium in mass fraction.
       The chemical composition profiles are shown for the first model only (i.e. the one with the lowest surface rotation rate), although they remain largely unchanged.}
     }
     \label{profile_rotation_accretion_withchems}
      \end{figure}

\textbfake{Figure \ref{profile_rotation_accretion_withchems} shows a subset of the rotation profiles shown in Fig. \ref{profile_rotation_accretion} along with the abundance profiles of hydrogen and helium, to better illustrate the change in chemical composition in the spun-up regions during accretion.}
   %
   %

%==========================================================================================   
   \begin{table*}
     \caption{Parameters of single sdB stars.
       From left to right, the parameters are: name, effective temperature ($T_{\rm eff})$, surface gravity ($\log g$), core rotation rate as sensed by g-modes ($\Omega_{\rm core}/2\pi$), envelope rotation rate as sensed by p-modes ($\Omega_{\rm env}/2\pi$), and period spacing of dipole modes ($\Delta\Pi_{1}$).
       The last column is a label for the source reference given as follows: 
       VG2010. \citet{VanGrootel2010},
       B2012.  \citet{Baran2012b},
       O2012.  \citet{Ostensen2012},
       O2014a. \citet{OstensenReed2014},
       R2014.  \citet{Reed2014},
       F2015.  \citet{Foster2015},
       Krz2015. \citet{Krzesinski2015},
       B2017. \citet{Baran2017},
       K2017. \citet{Kern2017},
       MB2017.  \citet{MoniBidin2017},
       VG2018.  \citet{VanGrootel2018},
       Z2018.   \citet{Zong2018},
       L2019. \citet{Latour2019},
       R2020. \citet{Reed2020},
       R2023. \citet{Reed2023},
       U2023. \citet{Uzundag2023},
       Su2024. \citet{Su2024}.
     }
     \resizebox{19cm}{!}{
     \begin{tabular}{lrrrrrrrrrl}
     Name &  T$_{\rm eff}$ [K] & $\log (g \rm [cm/s^{2}])$ & $\Omega_{\rm core}/2\pi$ [nHz] &$\Omega_{\rm env}/2\pi$ [nHz] & $\Delta\Pi_{1} $[s] & Ref. \\
     \hline
KIC3527751    & $27818 \pm 163 $ &  $5.35   \pm 0.03  $  & $ 271.69  \pm 20      $ &  $  $--$ $   & $266.4 \pm 0.2 $& F2015,Z2018\\
KIC10670103   & $21485 \pm 540 $ &  $5.14   \pm 0.05  $  & $ 131.52  \pm 12      $ &  $  $--$ $   & $251.6 \pm 0.2 $& R2014\\
KIC1718290    & $22350 \pm 200 $ &  $4.75   \pm 0.03  $  & $ 119.93  \pm 30      $ &  $  $--$ $   & $276.3 \pm 1   $& O2012\\
FEIGE46       & $36100 \pm 230 $ &  $5.93   \pm 0.04  $  & $ 264     \pm  46     $ &  $  $--$ $   & $ $--$           $& L2019\\
TIC033834484  & $24210 \pm 140 $ &  $5.28   \pm 0.03  $  & $ 181     \pm  80     $ &  $  $--$ $   & $263.84 \pm 1.4$& U2023,MB2017\\
TIC441725813  & $27827 \pm 177 $ &  $5.463  \pm 0.028 $  & $ 134     \pm  2      $ &  $ 647     \pm 26     $  & $267.93 \pm 1.63 $ & Su2024\\
KIC2697388    & $24165 \pm 77  $ &  $5.31   \pm 0.012 $  & $ 275.96  \pm  19     $ &  $ 218.38  \pm 38     $  & $240.06 \pm 0.19 $ & K2017\\
UYSEX         & $33030 \pm 220 $ &  $5.867  \pm 0.006 $  & $  $--$                 $ &  $ 470     \pm  80    $  & $ $--$             $ & R2020\\
KIC2991276    & $33900 \pm 200 $ &  $5.82   \pm 0.04  $  & $  $--$                 $ &  $ 1840    \pm  33    $  & $ $--$             $ & O2014a\\
V585PEG       & $28000 \pm 1200$ &  $5.383  \pm 0.004 $  & $  $--$                 $ &  $ 1662.94 \pm  68.7  $  & $ $--$             $ & R2023\\
PG1219+534    & $34258 \pm 170 $ &  $5.838  \pm 0.03  $  & $  $--$                 $ &  $ 331.54  \pm  10    $  & $ $--$             $ & VG2018\\
EPIC211779126 & $28557 \pm 82  $ &  $5.396  \pm 0.012 $  & $  $--$                 $ &  $ 707     \pm  230   $  & $256    \pm  5   $ & B2017\\
KIC5807616    & $27730 \pm 270 $ &  $5.552  \pm 0.041 $  & $  $--$                 $ &  $ 257.77  \pm  6     $  & $241.48 \pm  0.26$ & Krz2015,VG2010\\
KIC10139564   & $31859 \pm 126 $ &  $5.673  \pm 0.026 $  & $  $--$                 $ &  $ 452     \pm  30    $  & $ $--$             $ & B2012\\
     \hline
     \end{tabular}
     \label{table_singlesdb}
     }

   \end{table*}
%==========================================================================================

%==========================================================================================   
   \begin{table*}
     \caption{Same as table \ref{table_singlesdb} but for sdB stars in binary systems, including an additional column for the orbital period ($P_{\rm orb}$).
       The last column is a label for the source reference given as follows: 
       C2008.  \citet{Charpinet2008},
       G2008.  \citet{Geier2008},
       VG2008. \citet{VanGrootel2008},
       Pa2011.  \citet{Pablo2011},
       B2012.  \citet{Baran2012b},
       Pa2012.  \citet{Pablo2012},
       T2012.  \citet{Telting2012},
       O2014b. \citet{OstensenTelting2014},
       T2014.  \citet{Telting2014},
       Kup2015. \citet{Kupfer2015}
       B2016.   \citet{Baran2016},
       K2018. \citet{Kern2018},
       B2019. \citet{Baran2019},
       R2019. \citet{Reed2019},
       R2020. \citet{Reed2020},
       Pe2021. \citet{Pelisoli2021},
       Sil2022. \citet{Silvotti2022},
       Ma2023. \citet{Ma2023},
       R2025. \citet{Reed2025}.
     }
     \resizebox{19cm}{!}{
        \begin{tabular}{lrrrrrrrl}
     Name &  T$_{\rm eff}$ [K] & $\log (g \rm [cm/s^{2}])$ & $\Omega_{\rm core}/2\pi$ [nHz] &$\Omega_{\rm env}/2\pi$ [nHz]  & $P_{\rm orb}$ [d]  & $\Delta\Pi_{1} $ [s] & Ref. \\
     \hline
B3NGC6791     & $23540 \pm 210 $ &  $5.311  \pm 0.035 $  & $ 180.2    \pm 6.2     $ &  $  $--$ $   & $  $--$                 $ &$ 243.87 \pm 0.87 $ & Sil2022 \\
KIC2991403    & $27300 \pm 200 $ &  $5.43   \pm 0.03  $  & $ 1123.69  \pm 64      $ &  $  $--$ $   & $ 0.4431   \pm 6.25 \times 10^{-7}$  &$ 262.1  \pm 2.2  $ & Pa2012,B2012\\
KIC11179657   & $26000 \pm 800 $ &  $5.14   \pm 0.13  $  & $ 1564.06  \pm 84      $ &  $  $--$ $   & $ 0.3944   \pm 10^{-6} $ & $ 259.6 \pm 1.4 $ & Pa2012,B2012 \\
B4NGC6791     & $25290 \pm 300 $ &  $5.51   \pm 0.043 $  & $ 1258.6   \pm 9       $ &  $  $--$ $   & $ 0.3985              $ & $ 240.3 \pm 2.9$ & Pa2011,Sil2022,B2012\\
KIC7664467    & $27440 \pm 120 $ &  $5.38   \pm 0.02  $  & $ 330      \pm 50      $ &  $  $--$ $   & $ 1.5591   \pm 6 \times 10^{-5}  $ &$ 263 \pm 10$ & B2016\\
KIC7668647    & $27700 \pm 300 $ &  $5.5    \pm 0.03  $  & $ 246.26   \pm 5       $ &  $  $--$ $   & $ 14.1742  \pm 0.0042 $ &$ 248 \pm 10 $& T2014\\
PHL457        & $26690 \pm 60  $ &  $5.312  \pm 0.008 $  & $ 2500     \pm 16.28   $ &  $  $--$ $   & $ 0.31                $ &$ 259 \pm 2  $& B2019\\
EQPSC         & $28320 \pm 50  $ &  $5.632  \pm 0.015 $  & $ 1230     \pm 3.4     $ &  $  $--$ $   & $ 0.80088  \pm 9.7 \times 10^{-5} $  &$ 260    \pm 2    $& B2019\\
TIC137608661  & $27960 \pm 100 $ &  $5.42   \pm 0.04  $  & $ 2508     \pm 9.26    $ &  $  $--$ $   & $ 0.30058  \pm 1.6 \times 10^{-4} $  &$ 270.12 \pm 1.19 $& Sil2022\\
KIC10553698A  & $27750^{+130}_{-70}$ &  $5.452_{-0.008}^{+0.02}$ & $ 282.29 \pm 20.77 $   &  $  $--$ $   & $ 3.387    \pm 0.014   $ & $ 263.15 \pm 10$ & O2014b\\
PG0902+124    & $27287 \pm 77  $ & $5.465 \pm 0.01 $       & $ 1271.88 \pm 118.8   $ &  $  $--$ $                & $ 0.57739 \pm 1.5 \times 10^{-9}$ & $252.88 \pm 0.74$ & R2025 \\
PB6373        & $26968 \pm 125 $ & $5.404 \pm 0.018$       & $ 780 \pm 200         $ &  $ $--$  $                & $ 1.3 \pm  0.02 $                & $257.6  \pm 0.47$ & R2025,Kup2015 \\
PG0101+039    & $27500 \pm 500 $ &  $5.53   \pm 0.07  $  & $ 1322.75  \pm 19.65   $ &  $ 1345.82  \pm 25.03  $  & $ 0.569899 \pm 4 \times 10^{-7} $ &$ 249.2  \pm 1.5  $ & Ma2023,G2008,R2025 \\
KIC11558725   & $27910 \pm 210 $ &  $5.41   \pm 0.015 $  & $ 256.63   \pm 44.38   $ &  $ 287.91   \pm 23.63  $  & $ 10.0545  \pm 0.0048  $ & $ 244.45 \pm 0.32 $ & T2012,K2018\\
Feige48       & $29580 \pm 370 $ &  $5.462  \pm 0.006 $  & $ 30768.5  \pm 1635.9  $ &  $ 30768.5  \pm 1635.9 $  & $ 0.3762   \pm 0.003   $ & $  $--$  $ & VG2008\\
PG1315-123    & $36230_{-500}^{+710}$ &  $5.61 \pm 0.09  $  & $ 716      \pm 28      $ &  $ 731.15   \pm 9      $  & $ >100 $                 & $  $--$ $  & R2019\\
LTCnc         & $26032 \pm 83  $ & $5.275 \pm 0.011$       & $ 326.03  \pm 68.88   $ &  $ 643.0 \pm 96.45    $   & $ 6.122 \pm 0.004 $ & $259.77 \pm 0.51$ & R2025 \\
V1405Ori      & $31360 \pm 240 $ &  $5.573  \pm 0.044 $  & $  $--$                  $ &  $ 20850    \pm 1160 $  & $ 0.398023 \pm 3   \times 10^{-6}  $ &$  $--$ $ & R2020\\
HD265435      & $34300 \pm 400 $ &  $5.62   \pm 0.1   $  & $  $--$                  $ &  $ 168179   \pm 13   $  & $ 0.068818 \pm 3.2 \times 10^{-10} $ &$  $--$ $ & Pe2021\\
PG1336-018    & $32740 \pm 400 $ &  $5.739  \pm 0.002 $  & $  $--$                  $ &  $ 114576.83         $  & $ 0.1010      $            &$  $--$ $ & C2008\\
PG0048+091    & $32460_{-230}^{+290}$ &  $5.77_{-0.07}^{+0.05}$  & $  $--$                $ &  $ 2640     \pm 310  $   & $ >100 $ &$ 207.45 \pm 0.40 $& R2019\\
\hline
        \end{tabular}
        \label{table_binarysdb}
        }
   \end{table*}
%==========================================================================================
%%%%%%%%%%%%%%%%%%%%%%%%%%%%%%%%%%%%%%%%%%%%%%%%%%

% Don't change these lines
\bsp	% typesetting comment
\label{lastpage}
\end{document}